\documentclass{iopart}
\usepackage{iopams}  
\usepackage{graphicx}
\usepackage{balance}

\usepackage{changes}

\usepackage{tocloft}

\providecommand{\vect}[1]{{\boldsymbol{#1}}}
\renewcommand{\vec}[1]{{\bf {#1}}}

\newcommand{\fcs}{Fe$_{1-x}$Co$_{x}$Si}

\newcommand{\cso}{Cu$_{2}$OSeO$_{3}$}
\newcommand{\czm}{Co$_{8}$Zn$_8$Mn$_{4}$}

\begin{document}

\topical[The 2020 Skyrmionics Roadmap]{The 2020 Skyrmionics Roadmap}

%\ioptwocol

%\begin{flushleft}
\author{
{C. Back}$^{1}$,
{V. Cros}$^{2}$,
{H. Ebert}$^{3}$,
{K. Everschor-Sitte}$^{4}$,
{A. Fert}$^{2}$,
{M. Garst}$^{5}$,
{Tianping Ma}$^{6}$,
{S. Mankovsky}$^{3}$,
{T. L. Monchesky}$^{7}$,
{M. Mostovoy}$^{8}$,
{N. Nagaosa}$^{9,10}$,
{S.S.P. Parkin}$^{6}$,
{C. Pfleiderer}$^{1}$,
{N. Reyren}$^{2}$,
{A. Rosch}$^{11,12}$,
{Y. Taguchi}$^{9}$,
{Y. Tokura}$^{9,10}$,
{K. von Bergmann}$^{13}$,
{Jiadong Zang}$^{14}$
}
%\end{flushleft}

\address{
$^{1}${Physik-Department, Technical University of Munich, James-Franck-Str. 1, 85748 Garching, Germany},\\
$^{2}${Unit\'{e} Mixte de Physique CNRS/Thales (UMR137), 1 avenue A. Fresnel, 91767 Palaiseau Cedex, France},\\
$^{3}${LMU Munich, Department of Chemistry, Butenandtstrasse 11, D-81377 Munich, Germany},\\
$^{4}${Institute of Physics, Johannes Gutenberg University, 55128 Mainz, Germany},\\
$^{5}${Institut f\"ur Theoretische Festk\"orperphysik, Karlsruhe Institute of Technology, 76131 Karlsruhe, Germany},\\
$^{6}${Max Planck Institute for Microstructure Physics, Halle (Saale), Germany},\\
$^{7}${Department of Physics and Atmospheric Science, Dalhousie University, Halifax NS, Canada B3H 4R2},\\
$^{8}${Zernike Institute for Advanced Materials, University of Groningen, Nijenborgh 4, 9747 AG Groningen, The Netherlands},\\
$^{9}${RIKEN Center for Emergent Matter Science, Wako 351-0198, Japan},\\
$^{10}${Department of Applied Physics and Quantum Phase Electronics Center, University of Tokyo, Tokyo 113-8656, Japan},\\
$^{11}${Institute for Theoretical Physics, University of Cologne, Cologne, Germany},\\
$^{12}${Department of Physics, Harvard University, Cambridge MA 02138, USA},\\
$^{13}${Department of Physics, University of Hamburg, 20355 Hamburg, Germany},\\
$^{14}${Department of Physics and Astronomy, University of New Hampshire, Durham, New Hampshire 03824, USA}
}

%\eads{
%\mailto{christian.back@tum.de},
%\mailto{vincent.cros@cnrs-thales.fr},
%\mailto{hubert.ebert@cup.uni-muenchen.de},
%\mailto{kaeversc@uni-mainz.de},
%\mailto{albert.fert@cnrs-thales.fr},
%\mailto{markus.garst@kit.edu},
%\mailto{tianping.ma@mpi-halle.mpg.de},
%\mailto{manpc@cup.uni-muenchen.de},
%\mailto{theodore.monchesky@dal.ca},
%\mailto{m.mostovoy@rug.nl},
%\mailto{nagaosa@riken.jp},
%\mailto{stuart.parkin@mpi-halle.mpg.de},
%\mailto{christian.pfleiderer@tum.de},
%\mailto{nicolas.reyren@cnrs-thales.fr},
%\mailto{rosch@thp.uni-koeln.de},
%\mailto{y-taguchi@riken.jp},
%\mailto{tokura@riken.jp},
%\mailto{kbergman@physnet.uni-hamburg.de},
%\mailto{jiadong.zang@unh.edu}
%}

%%%%%%%%%%%%%%%%%%%%%%%%%%%%%%%%%%%%%%%%%%%%%%%%%

\vspace{10pt}
%\begin{indented}
%\item[]14. October 2018

%\end{indented}

%Please see the full \LaTeXe template for intructions on  using \LaTeXe\ and %\verb"iopart.cls" (the IOP Publishing \LaTeXe\ preprint class file).
%\newpage

\begin{abstract}
The notion of non-trivial topological winding in condensed matter systems represents a major area of present-day theoretical and experimental research. Magnetic materials offer a versatile platform that is particularly amenable for the exploration of topological spin solitons in real space such as skyrmions. First identified in non-centrosymmetric bulk materials, the rapidly growing zoology of materials systems hosting skyrmions and related topological spin solitons includes bulk compounds, surfaces, thin films, heterostructures, nano-wires and nano-dots. This underscores an exceptional potential for major breakthroughs ranging from fundamental questions to applications as driven by an interdisciplinary exchange of ideas between areas in magnetism which traditionally have been pursued rather independently. The skyrmionics roadmap provides a review of the present state of the art and the wide range of research directions and strategies currently under way. These are, for instance, motivated by the identification of the fundamental structural properties of skyrmions and related textures, processes of nucleation and annihilation in the presence of non-trivial topological winding, an exceptionally efficient coupling to spin currents generating spin transfer torques at tiny current densities, as well as the capability to purpose-design broad-band spin dynamic and logic devices. 
\end{abstract}

\vspace{2pc}

%\noindent{\it Keywords}: topological protection, skyrmion creation, skyrmion destruction, Emergent electrodynamics, Berry phase, Skyrmions, Quantized magnetic flux, Topology, Topological Hall effect, Spin motive force, magnons, helimagnons, magnetic resonance, density functional theory, response quantities, interaction parameters, thin films, frustrated magnetism, superconductors, topological insulators, racetrack memory, non-equilibrium, melting, glass transition, metastability, skyrmions as particles, effective mass, pinning, surfaces

%\noindent{\it Keywords Back}: topological protection, skyrmion creation, skyrmion destruction
%\noindent{\it Keywords Everschor-Sitte}: Emergent electrodynamics, Berry phase, Skyrmions, Quantized magnetic flux, Topology, Topological Hall effect, Spin motive force
%\noindent{\it Keywords Fert, Cros, Reyren}: 
%\noindent{\it Keywords Garst}: magnons, helimagnons, magnetic resonance
%\noindent{\it Keywords Mankovsky, Ebert}:  density functional theory, response quantities, interaction parameters
%\noindent{\it Keywords Monchesky}: thin films, 
%\noindent{\it Keywords Mostovoy}: skyrmion, frustrated magnetism
%\noindent{\it Keywords Nagaosa}: Skyrmions, superconductors, topological insulators 
%\noindent{\it Keywords Parkin}: racetrack memory
%\noindent{\it Keywords Pfleiderer}: non-equilibrium, melting, glass transition, metastability
%\noindent{\it Keywords Rosch}: skyrmions as particles, effective mass, pinning
%\noindent{\it Keywords Takuchi, Tokura}: 
%\noindent{\it Keywords von Bergmann}: surfaces
%\noindent{\it Keywords Zang}: 

%%%%%%%%%%%%%%%%%%

\clearpage
\newpage

%%%%%%%%%%%%%%%%%%%%%%%%%%%%%%%%%%%%%%%%%%%%%%%%%
%%%%%%%%%%%%%%%%%%%%%%%%%%%%%%%%%%%%%%%%%%%%%%%%%
%%%%%%%%%%%%%%%%%%%%%%%%%%%%%%%%%%%%%%%%%%%%%%%%%

\tableofcontents

%%%%%%%%%%%%%%%%%%%%%%%%%%%%%%%%%%%%%%%%%%%%%%%%%
%%%%%%%%%%%%%%%%%%%%%%%%%%%%%%%%%%%%%%%%%%%%%%%%%
%%%%%%%%%%%%%%%%%%%%%%%%%%%%%%%%%%%%%%%%%%%%%%%%%

\ioptwocol

%%%%%%%%%%%%%%%%%%%%%%%%%%%%%%%%%%%%%%%%%%%%%%%%%

%\clearpage
%\newpage

%\input{_contributions/Taguchi_Tokura-v2}

\section[Non-centrosymmetric bulk materials hosting skyrmions\\ {\normalfont Yasujiro Taguchi, Yoshinori Tokura}]{Non-centrosymmetric bulk materials hosting skyrmions}
\label{taguchi}
{\it Yasujiro Taguchi$^1$ and Yoshinori Tokura$^{1, 2}$}

$^1$RIKEN Center for Emergent Matter Science, Wako 351-0198, Japan, 
$^2$Department of Applied Physics and Tokyo College, University of Tokyo, Tokyo 113-8656, Japan

\subsection{Status}

%This section provides a brief history and status, why the field is still important, what will be gained with further advances. (350 words max)

Magnetic skyrmions are topological spin textures that are stabilized 
in various types of magnets by different kinds of interactions. 
Non-centrosymmetric bulk magnets 
with either chiral, polar, or $D_{\rm 2d}$ symmetry provide a good arena 
to study the topological spin structures and emergent electromagnetic responses 
arising from them~\cite{Bogdanov94,Kanazawa17}.
In these magnets without inversion symmetry, Dzyaloshinskii-Moriya interaction (DMI) gradually twists 
the otherwise ferromagnetic spin arrangement, thus giving rise to helimagnetic structure in zero field 
as well as skyrmions in a certain range of magnetic field. 
Skyrmions cannot be connected to helical structure through continuous deformation, 
therefore they are topologically protected from external perturbations and hence 
appropriate for robust information carriers. Thus far, three different types of skyrmions 
as schematically shown in Fig.1
have been observed experimentally, which 
were theoretically predicted ~\cite{Bogdanov89}.

\begin{figure*}[h]
\includegraphics[width=1.0\textwidth]{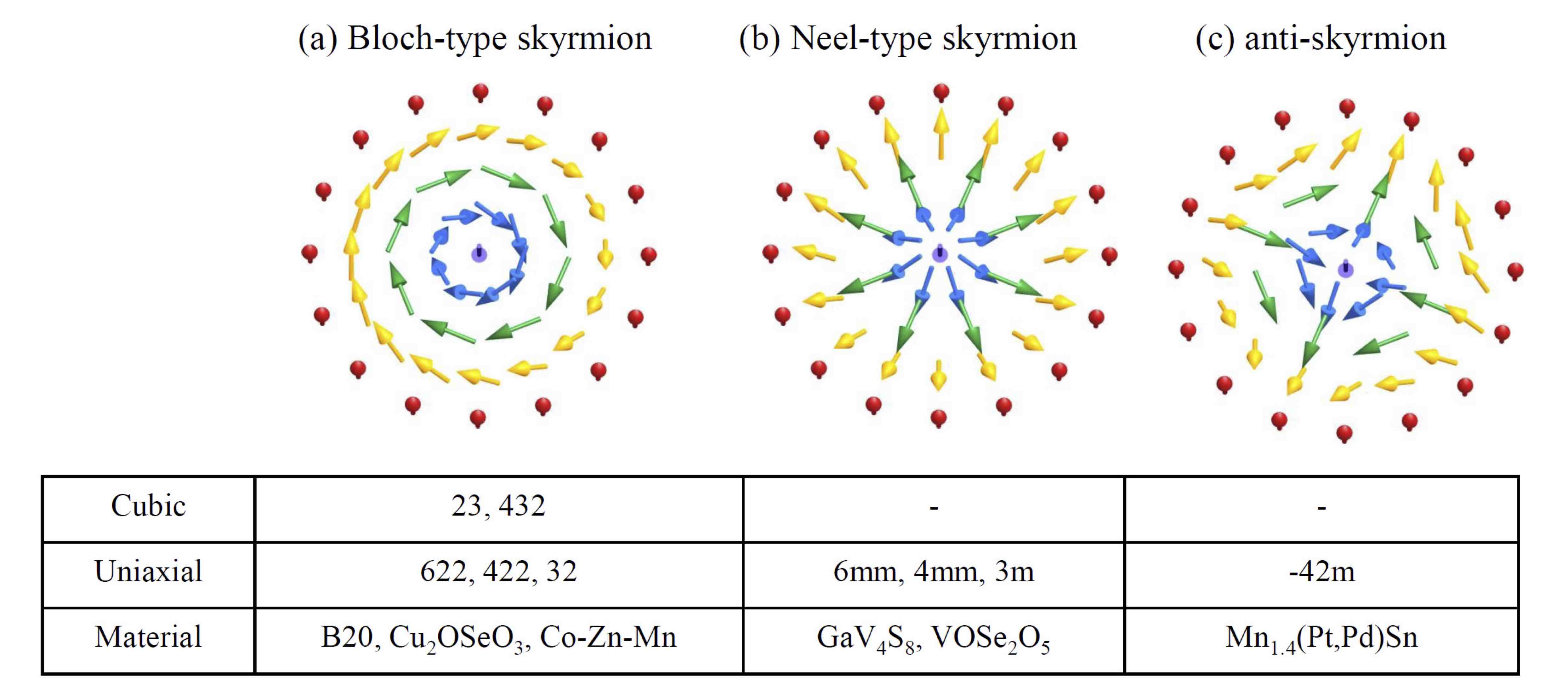}
\caption{Schematics of various skyrmions ; (a) Bloch-type skyrmion, 
(b) N\'{e}el-type skyrmion, and (c) anti-skyrmion.}
\end{figure*}

The first category is a Bloch-type skyrmion in chiral magnets, where the spins are lying 
within the tangential plane, as shown in Fig. 1(a). The $A$-phase in MnSi 
with B20-type chiral structure, which had been known for years as a mysterious phase, 
was revealed to be a skyrmion crystal, which is described as a superposition of three screw-type helices, 
by small angle neutron scattering (SANS)~\cite{Muhlbauer09}. 
Then, isolated skyrmions in addition to the skyrmion-lattice form 
were observed for B20 compounds in real space 
by Lorentz transmission electron microscopy (LTEM) technique~\cite{Yu10}. 
In 2012, an insulating and multiferroic Cu$_2$OSeO$_3$ was identified 
to host skyrmions~\cite{Seki12}. In 2015, $\beta$-Mn type 
Co-Zn-Mn alloy was discovered to exhibit skyrmions at and above room temperature, 
and the metastable skyrmion state was observed 
at zero magnetic field and room temperature~\cite{Karube17}, 
as displayed in Fig. 2.

The second class is a N\'{e}el-type skyrmion in polar magnets where the spins are lying 
within a radial plane, as shown in Fig. 1(b), and their lattice is described as a superposition
 of three cycloidal helices. The bulk material identified to host the N\'{e}el-type skyrmion is 
a polar compound GaV$_4$S$_8$ with a Lacunar spinel structure~\cite{Kezsmarki15} and related polar materials.

The third family is an antiskyrmion shown in Fig.1(c), which was discovered in 
Heusler compounds with $D_{\rm 2d}$ crystal symmetry in 2017 by using LTEM~\cite{Nayak17}. 
Application of magnetic fields stabilizes the antiskyrmion lattice over a wide range 
of temperature including room temperature. 
Cross-sections of an antiskyrmion are either screw-type helix 
or cycloidal one, depending on the azimuthal direction.

As overviewed above, some materials that exhibit (anti)skyrmions above room temperature 
have been found, but continued efforts to expand the horizon of such materials 
should be necessary to understand the fundamental physics of the skyrmions 
as well as to achieve their applications in practical devices.
Three-dimensional nature of skyrmions in non-centrosymmetric bulk materials may provide unique 
functionalities, such as directionally non-reciprocal transmission of spin excitations, 
which are difficult to observe for interfacial-DMI-based skyrmions  in magnetic multilayer systems.
In the following sections, we describe current and future callenges, followed by advances in 
science and technology to meet them, which  are commonly important for all the three types of bulk skyrmions.

\begin{figure*}[h]
\includegraphics[width=1.0\textwidth]{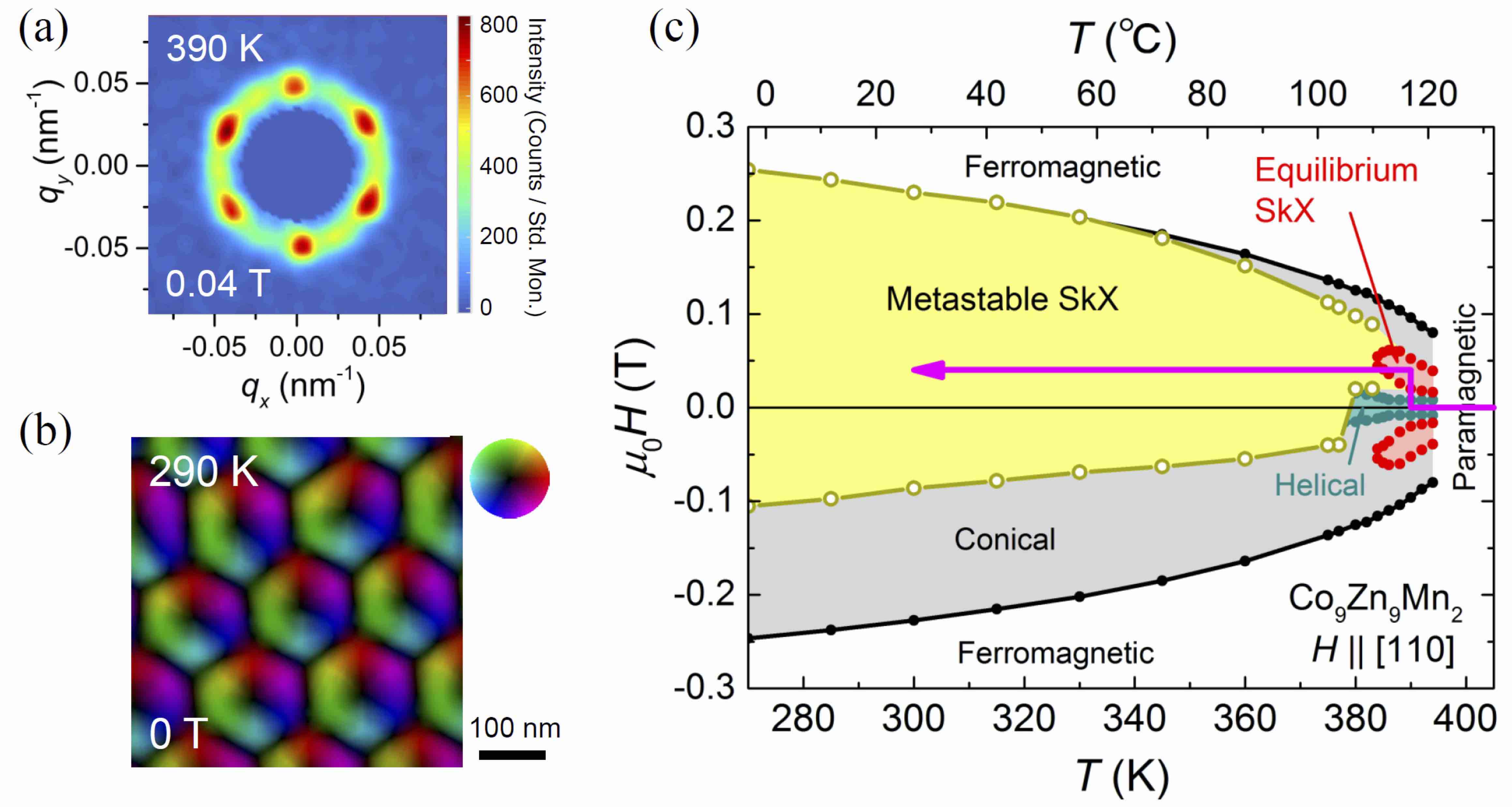}
\caption{Equilibrium and metastable skyrmions in Co$_9$Zn$_9$Mn$_2$. 
(a) Small angle neutron scattering image for equilibrium 
skyrmion crystal state at 390 K and 0.04 T. 
(b) Real-space magnetic structure deduced from Lorentz 
transmission electron microscope images via transport-of-intensity
 equation analysis for metastable state at 290 K and 0 T. 
(c) State diagram of Co$_9$Zn$_9$Mn$_2$ upon field-cooling at 0.04 T 
from 390 K. All figures are reproduced from Ref.\,\cite{Karube17} with permission.
 Copyright (2017) by the American Physical Society.}
\end{figure*}

\subsection{Current and future challenges}
%This section discusses the big research issues and challenges. (350 words max)

Toward full understanding of the intriguing physical properties of skyrmions 
as well as applying them to devices, there appears to be several challenges 
to be addressed from the materials point of view. 

1. {\it Achieving both high transition temperature and small size of a skyrmion.}

In general, transition temperature $T_c$ and the size of a skyrmion are 
proportional to $J$ and $J/D$, respectively, where $J$ and $D$ denote exchange interaction 
and DMI. To achieve high $T_c$ and small size of a skyrmion, 
$J$ should be increased, and $D$ should be even more enhanced.
Quantitative estimation of $D$ by first-principles calculations are obviously necessary.

2. {\it Increasing the (meta)stability of a skyrmion. }

As exemplified in the quintessential compound MnSi~\cite{Muhlbauer09}, 
skyrmions can  form only in a narrow phase space 
in a temperature-field plane
in many cases.
For the applications of skrymions, 
the stability (or metastability) should be enhanced so that 
the skyrmions persist in a wider phase space. The (meta)stability of the skyrmions 
has to be studied 
both experimentally and theoretically
in more detail.
	
3. {\it Three-dimensional nature of a skyrmion.}

A single skyrmion is often considered as a rigid, tube-like object. 
In reality, however, a skyrmion is better considered as a flexible string 
that can dynamically deform under the influence of pinning sites. 
Also, in the destruction process of a skyrmion, creation of monopole-antimonopole pair, 
or coalescence between two skyrmion strings are important~\cite{Milde13}. 
Another interesting issue is propagation dynamics of spin excitations along skyrmion strings and its directional  non-reciprocity.
These issues related to the three-dimensional nature of the skyrmion string
 are to be clarified
both experimentally and theoretically.

4. {\it Controlling skyrmions.}

Creation, transportation, detection, and annihilation of skyrmions 
are definitely necessary. As for the creation, current injections,
application of stress, irradiation of pulsed light have been 
reported to successfully stabilize the skyrmions.
It has been also known~\cite{Jonietz10} that skyrmions are driven by much lower current density 
than those needed for ferromagnetic domain wall motion. 
Future challenge is to control the motion of isolated skyrmions individually, 
and to keep the threshold current-density low while attaining 
the high metastability, which may show a trade-off relation.

\subsection{Advances in science and technology to meet challenges}
%This section discusses the advances in science technology needed to address the challenges. (350 words max)

Since the discovery of skyrmions in B20-type compounds, many advances 
have been achieved thus far, and also will be expected in the near future, 
to meet the challenges described in the previous section. 

1.	Recently, theoretical prescription to calculate $D$ value of metallic magnets, 
which depends sensitively on the band filling, has been established 
on the basis of first-principles calculation of band structure~\cite{Freimuth14}. 
These advances allow us to design the magnets with high $T_c$ and large value of $D$.

2.	
Another skyrmion phase disconnected from the conventional phase just below $T_c$
has recently been identified at low temperatures
in Cu$_2$OSeO$_3$ and Co$_7$Zn$_7$Mn$_6$ ~\cite{2018_Chacon_NatPhys,2018_Karube_SciAdv},
as disccused in Section 3.
Metastability of quenched skyrmion state has been investigated 
with a focus on the interplay between topological protection 
and thermal agitation. Some pump-and-probe or 
stroboscopic techniques, including the SANS, have been developed 
to detect dynamical behavior of the skyrmion-lattice formation/destruction process. 
Recent experiments have proved that metastability depends 
on the quenched randomness and the size of a skyrmion, 
but this should be investigated in more detail.

3.	Monopole-antimonopole structure as well as coalescence of skyrmions
 have been discussed from magnetic force microscopy measurement
 and theoretical consideration~\cite{Milde13}. 
These phenomena are also observed for a thin-plate sample 
by LTEM with in-plane field configuration. Emergent electric field in 
dynamically bending skyrmion strings as well as non-reciprocal transport 
along the skyrmion string have been discussed recently.

4.	Creation, transportation, and annihilation of skyrmions have recently been 
observed with various techniques, such as SANS and  LTEM
for bulk skyrmions~\cite{Jonietz10,Yu17},
and scanning transmission x-ray microscopy exploiting x-ray 
magnetic circular dichroism for interfacial skrmions~\cite{Litzius17}. 
These studies have revealed the dynamic response of skyrmions to the current. 
Skyrmions in B20-type compounds were known to be driven 
by low current density as compared with current-driven ferromagnetic 
domain wall motion~\cite{Jonietz10}, and recently current drive of skyrmions 
have been successfully demonstrated for the Co-Zn-Mn alloy 
at room temperature. Imaging experiments with further improved spatial and time resolution 
will provide further detailed information on the dynamical features of skyrmoions. 
For the detection of skyrmions, topological Hall effect due to the emergent field 
has been proved to be effective~\cite{Schulz12}.

\subsection{Concluding remarks}
%Include brief concluding remarks. This should not be longer than a short paragraph. (150 words max)

Novel non-centrosymmetric magnets hosting (anti)skyr\-mions have been found 
in the course of the intensive researches. In the future, 
further exploration of materials hosting smaller size (anti)skyrmions 
at higher temperatures should be necessary. It is also important to increase (meta)stability 
of the skyrmions while keeping the quenched randomness minimal 
to facilitate small threshold current density and high speed of skyrmion motion. 
Three-dimensional nature of the skyrmions and the associated interplay 
between the emergent fields and charge carriers should be understood 
in more detail. Creation, transportation, detection, and annihilation of individual skyrmions 
in more effective and controlled way should be further investigated
while achieving higher spatial and temporal resolutions of imaging/detecting techniques. 
Especially, highly sensitive and reliable detection of skyrmions based 
on electrical methods, such as topological Hall and planar Hall effects, should be further pursued. 

\subsection{Acknowledgements}
%Please include any acknowledgements and funding information as appropriate.

We are grateful to K. Karube and M. Ishida for their help in preparing this article.

%%%%%%%%%%%%%%%%%%%%%%%%%%%%%%%%%%%%%%%%%%%%%%%%%
%%%%%%%%%%%%%%%%%%%%%%%%%%%%%%%%%%%%%%%%%%%%%%%%%
%%%%%%%%%%%%%%%%%%%%%%%%%%%%%%%%%%%%%%%%%%%%%%%%%

\clearpage
\newpage

\section[Skyrmions in achiral magnets with competing interactions\\ {\normalfont Maxim Mostovoy}]{Skyrmions in achiral magnets with competing interactions}
\label{mostovoy}
{\it Maxim Mostovoy}

Zernike Institute for Advanced Materials, University of Groningen, Nijenborgh 4, 9747 AG Groningen, The Netherlands

\subsection{Status}
%This section provides a brief history and status, why the field is still important, what will be gained with further advances. (350 words max)

Current research on skyrmions is mostly focused on magnets with chiral crystal lattices and heterostructures with inversion symmetry broken at interfaces of magnetic layers. At the same time, non-trivial skyrmion topology that gives rise to new phenomena, such as the Topological and Skyrmion Hall Effects, is unrelated to lattice chirality. Recent theoretical studies showed that skyrmion crystals and isolated skyrmions can be stabilized by competing Heisenberg exchange interactions in Mott insulators with centrosymmetric lattices \cite{Okubo2012,Leonov2015,Batista2016} as well as by long-ranged interactions mediated by conduction electrons in itinerant magnets \cite{Ozawa2017}.   These magnetically frustrated materials with achiral lattices  are interesting because of additional collective degrees of freedom which give rise to a larger variety of topological magnetic states and more complex collective dynamics.

\begin{figure}[tbp]
\centering
\includegraphics[width=0.8\columnwidth]{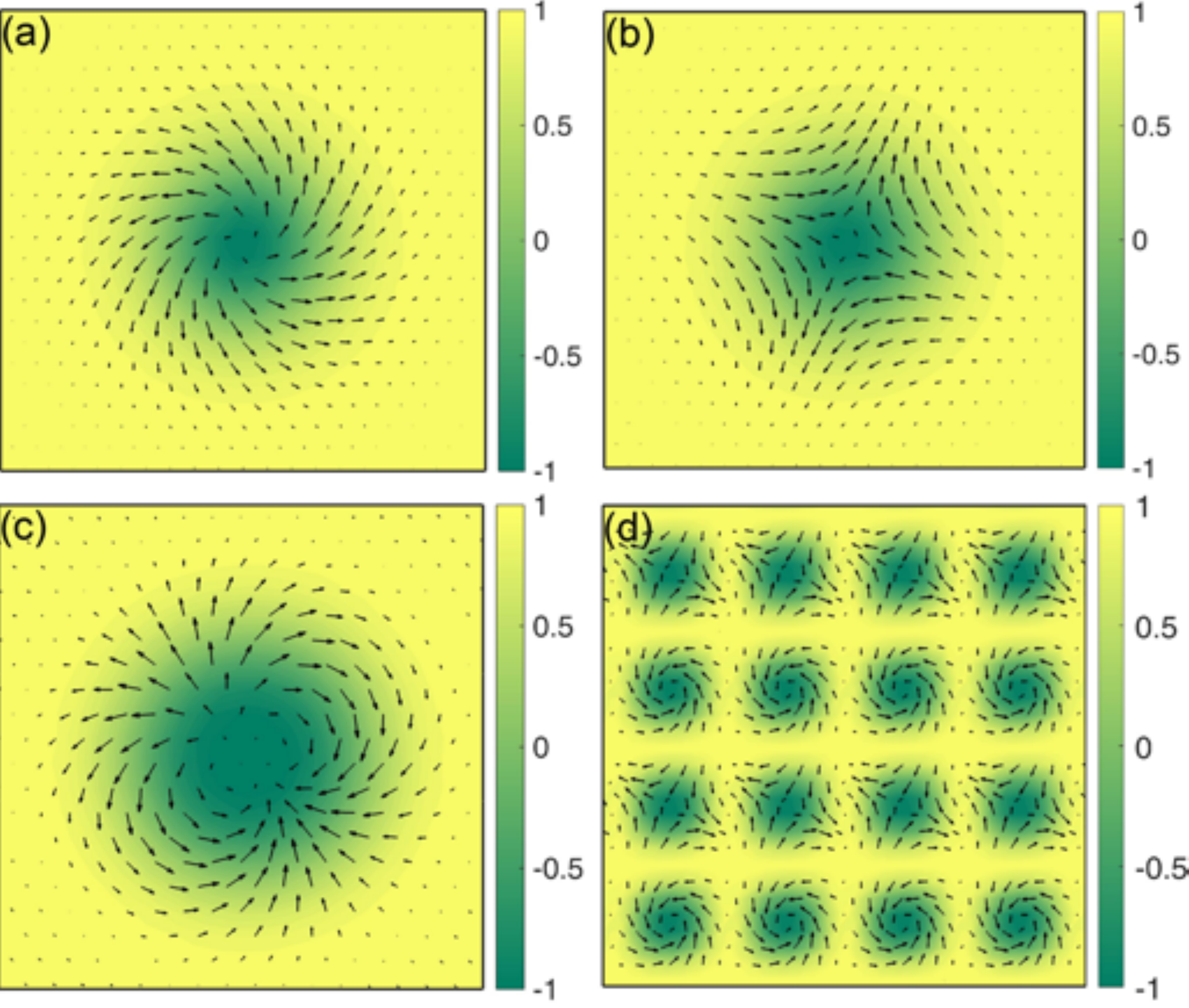}
\caption{Topological states of a frustrated magnet with a triangular lattice \cite{Leonov2015}. (a) A skyrmion with vorticity $+1$  and an arbitrary helicity. In-plane spin components are shown with arrows. The out-of-plane spin components are indicated by color.   (b) An antiskyrmion with vorticity $-1$. (c) A skyrmion with topological charge $Q = - 2$ resulting from the merger of two $Q = - 1$ skyrmions with opposite helicities. (d) A metastable array of skyrmions and antiskyrmions.
} 
\label{fig:helicityvorticity}
\end{figure}

\begin{figure}[tbp]
\centering
\includegraphics[width=0.8\columnwidth]{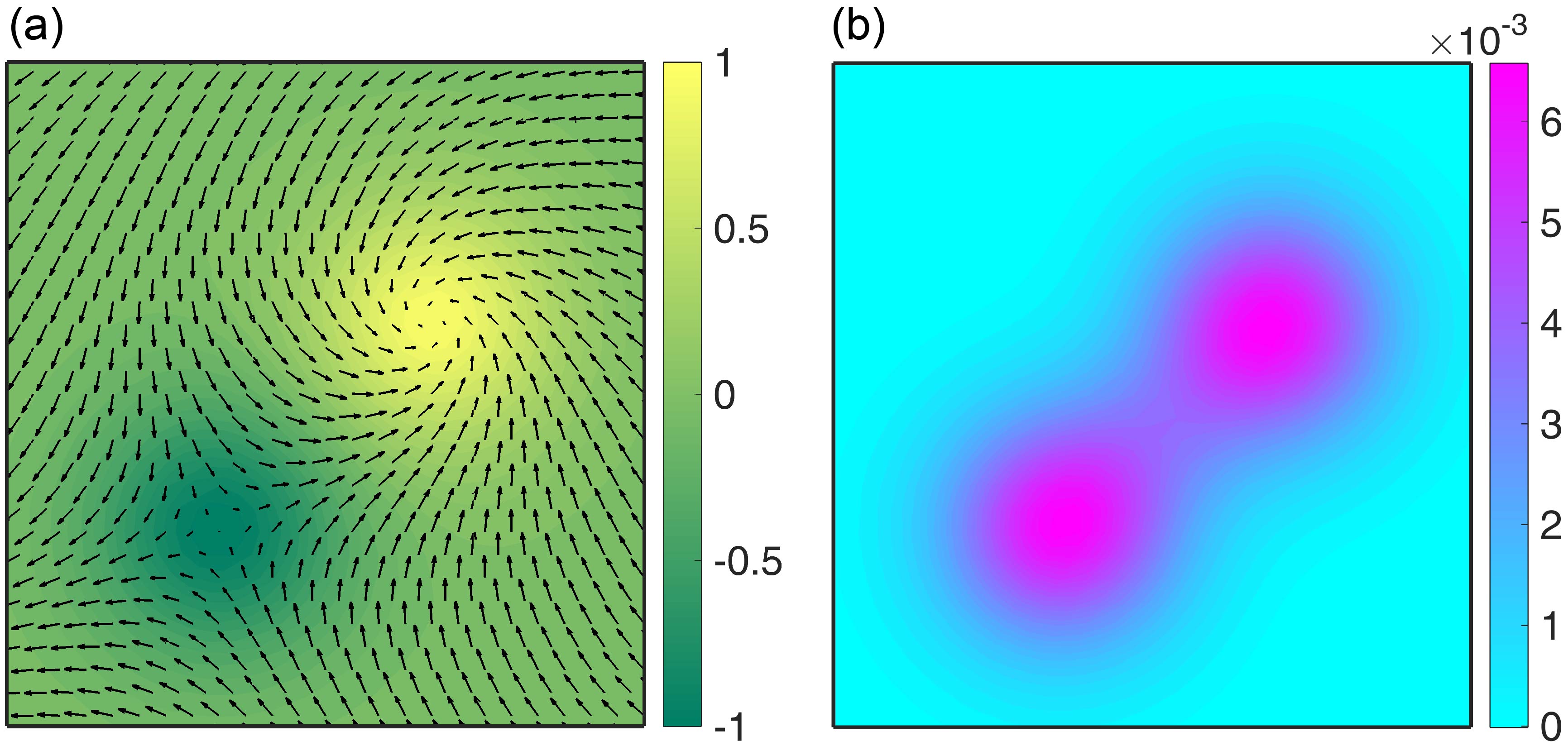}
\caption{(a) A bi-meron (vortex-antivortex pair) with topological charge $Q = +1$ in an easy-plane frustrated magnet. (b) The corresponding distribution of the topological charge density. } \label{fig:bimeron}
\end{figure}

Heisenberg exchange interactions do not select the spin rotation plane, which leads to an additional skyrmion zero mode, {\em helicity}, as there is no energy cost associated with the rotation of spins around the skyrmion symmetry axis (see Fig.~\ref{fig:helicityvorticity}a). Skyrmions with opposite helicities attract each other at short distances and form stable topological defects with higher topological charges (see Fig.~\ref{fig:helicityvorticity}c) \cite{Leonov2015,Kharkov2017}. 
The exchange energy of the skyrmion described by a classical spin model does not change under the sign reversal of an in-plane spin component, which reverses the sign of {\em vorticity},  describing the winding of spins around the skyrmion center, and the sign of the skyrmion topological charge, $Q$.   
Arbitrary vorticity allows for simultaneous presence of skyrmions and antiskyrmions in easy-axis magnets (see Fig.~\ref{fig:helicityvorticity}b,d) and vortex-antivortex pairs (bi-merons) in easy-plane magnets (see Fig.~\ref{fig:bimeron}a) \cite{Kharkov2017}. Like skyrmions, bi-merons have topological charge $Q = \pm 1$ equally divided between the vortex and antivortex carrying half-integer charges (see Fig.~\ref{fig:bimeron}b). 

Non-collinear spin orders can spontaneously break inversion symmetry and induce an electric polarization. Skyrmion with vorticity $+1$ can have a net out-of-plane electric dipole moment that depends on skyrmion helicity, while the magnetic vortex has an electric charge. The magnetoelectric coupling allows for the electric-field control of topological defects in magnetic Mott insulators. In addition, the size of skyrmions stabilized by exchange interactions can be as small as a few lattice constants, which can be of interest for high-density magnetic memory applications.
     
The search for frustrated magnets hosting skyrmions has started only recently. The intermetallic compound, Gd$_2$PdSi$_3$, shows a spiral ordering of Gd spins with 6 possible orientations of the wave vector in the hexagonal plane, which under the applied magnetic field transforms into a skyrmion array, as evidenced by the giant Topological Hall Effect (THE) \cite{Kurumaji2018}. 

\subsection{Current and future challenges}
%This section discusses the big research issues and challenges. (350 words max)

Magnetically frustrated Mott insulators hosting skyrmion crystals are yet to be found.  They have to satisfy a number of criteria, e.g., a high-symmetry axis that would allow them to accommodate multiply-periodic states. Furthermore, the mechanism for stabilization of skyrmion crystals in an applied magnetic field, which works in both chiral and achiral magnets, requires modulated  ferromagnetic orders, whereas most frustrated magnets show  antiferromagnetic (AFM) spiral states. For example, incommensurate spin spirals inducing an electric polarization have been found in dihalides, such as NiBr$_2$ and Col$_2$, with stacked triangular  lattices of magnetic ions  \cite{Kurumaji2013}. Even a weak AFM coupling between neighboring magnetic layers can suppress the skyrmion crystal state  \cite{Lin2018}.  One way to get around this problem would be to reduce thickness of these van der Waals magnets down to just one layer \cite{Klein2018}.  At the same time, stability requirements for isolated skyrmions and bi-merons in antiferromagnets are less stringent -- they can be stabilized by an easy-axis anisotropy in zero field. Non-collinear antiferromagnetic spin textures give rise to polar displacements of ions and induce a local electric polarization that can be used for electric manipulation of skyrmions.

Static and dynamics properties of a rich variety of unconventional magnetic orders and  topological spin textures in frustrated systems,  such as hopfions and magnetic hedgehogs (Bloch points), still have to be explored. Hedgehog-antihedgehog arrays have been recently observed in  SrFeO$_3$ -- a metal with the cubic perovskite structure which at $T_N  \sim 130$ K undergoes an antiferromagnetic phase transition ascribed to the onset of a helical spiral order. Later transport and neutron  measurements revealed more ordered phases with several coexisting spirals. The four-spiral state is a hedgehog-antihedgehog lattice that under an applied magnetic field shows the THE  \cite{Ishiwata2018}. It is not clear, however, whether  itinerant magnets can support stable isolated magnetic defects.

Another interesting class of materials are magnets that are both geometrically frustrated and chiral, such as  the hexagonal swedenborgite, CaBaCo$_2$Fe$_2$O$_7$ \cite{Reim2018}.  They can show complex magnetic orders resulting from the interplay between a plethora of competing spin configurations typical for geometrically frustrated systems and the tendency to form non-collinear orders due to lattice chirality.

Helicity gives rise to a new low-energy  branch of magnetic excitations in skyrmion crystals and a new collective zero mode of isolated skyrmions and antiskyrmions. Effects of this additional degree of freedom on heat transport  in frustrated magnets, such as the topological magnon Hall and spin Seebeck effects, have to be studied theoretically and experimentally. The coupling between helicity and translational modes, resulting e.g. from the dependence of skyrmion-(anti)skyrmion interactions on both positions and helicities of the topological defects  \cite{Leonov2015}, gives rise to rich and largely unexplored skyrmion dynamics, which can be excited by the spin-Hall torque, electromagnetic fields and electric currents. 

\subsection{Advances in science and technology to meet challenges}
%This section discusses the advances in science technology needed to address the challenges. (350 words max)

Magnetic frustration is very sensitive to deformations of crystal lattice, band filling and electronic configurations of magnetic ions. Therefore, the search for new skyrmion materials will require a  tuning of materials parameters, e.g. by pressure and chemical substitutions which affect  interactions between spins and magnetic anisotropy. Finding frustrated magnets with a high transition temperature and large magnetically-induced electric polarization is a challenge. The search for such materials will benefit from first-principles calculations of competing exchange interactions. {\em Ab initio}  calculation of magnetoelectric properties of non-coplanar spin textures, such as skyrmions, is an interesting problem.

It is rather challenging to understand the nature of complex multiply-periodic magnetic states, such as the hedgehog-antihedgehog array, on the basis of neutron scattering alone. This technique has to be combined with transport measurements, electric detection of skyrmions, e.g. the Spin Hall Magnetoresistance measurements, and direct imaging.  The period of magnetic modulations in frustrated magnets often does not exceed a few lattice constants, which makes the observation of small topological defects by the Lorentz Transmission Electron Microscopy and Magnetic Force Microscopy difficult and calls for development of new techniques with a higher spatial resolution. On the theory side, it will be interesting to explore  effects of quantum fluctuations on stability and physical properties of the small-sized skyrmions \cite{2019_Lohani}.

\subsection{Concluding remarks}
%Include brief concluding remarks. This should not be longer than a short paragraph. (150 words max)

The emergence of topological spin textures from magnetic frustration is a new paradigm in the search for skyrmion materials.  Experimental and theoretical study of topological magnetic states in frustrated magnets is still in its infancy. Mott insulators with competing exchange interactions as well as itinerant magnets can host tiny skyrmions with versatile physical properties and complex dynamics. The reversible electric dipole moment of skyrmions  in Mott insulators can be  controlled with an applied electric field, which provides a new route to skyrmion-based electronics with low energy losses and high density of information storage. 

\subsection{Acknowledgements}
The author acknowledges Vrije FOM-programma `Skyrmionics'.

%%%%%%%%%%%%%%%%%%%%%%%%%%%%%%%%%%%%%%%%%%%%%%%%%
%%%%%%%%%%%%%%%%%%%%%%%%%%%%%%%%%%%%%%%%%%%%%%%%%
%%%%%%%%%%%%%%%%%%%%%%%%%%%%%%%%%%%%%%%%%%%%%%%%%

\clearpage
\newpage

\section[Skyrmions far from equilibrium in bulk materials \\ {\normalfont Christian Pfleiderer}]{Skyrmions far from equilibrium in bulk materials}
\label{pfleiderer}
{\it Christian Pfleiderer}

Physik-Department, Technical University of Munich, James-Franck-Str. 1, 85748 Garching, Germany

\subsection{Status}

Condensed matter systems far from equilibrium reflect the energy landscape and statistical properties of the low-lying excitations. They may also feature novel forms of order. Systems far from equilibrium attract great current interest as potential routes towards low-dissipation information processing and long-lived non-volatile data storage.  Two non-equilibrium scenarios may be distinguished. Systems that have fallen out of thermal equilibrium under rapid cooling, and systems that are driven out of equilibrium by a non-thermal stimulus such as electromagnetic fields. In this general context, complex spin textures such as skyrmions offer premiere access to the effects of non-trivial topology on physical properties far from equilibrium.

\begin{figure*}
 \center{ \includegraphics[width=1.0\textwidth]{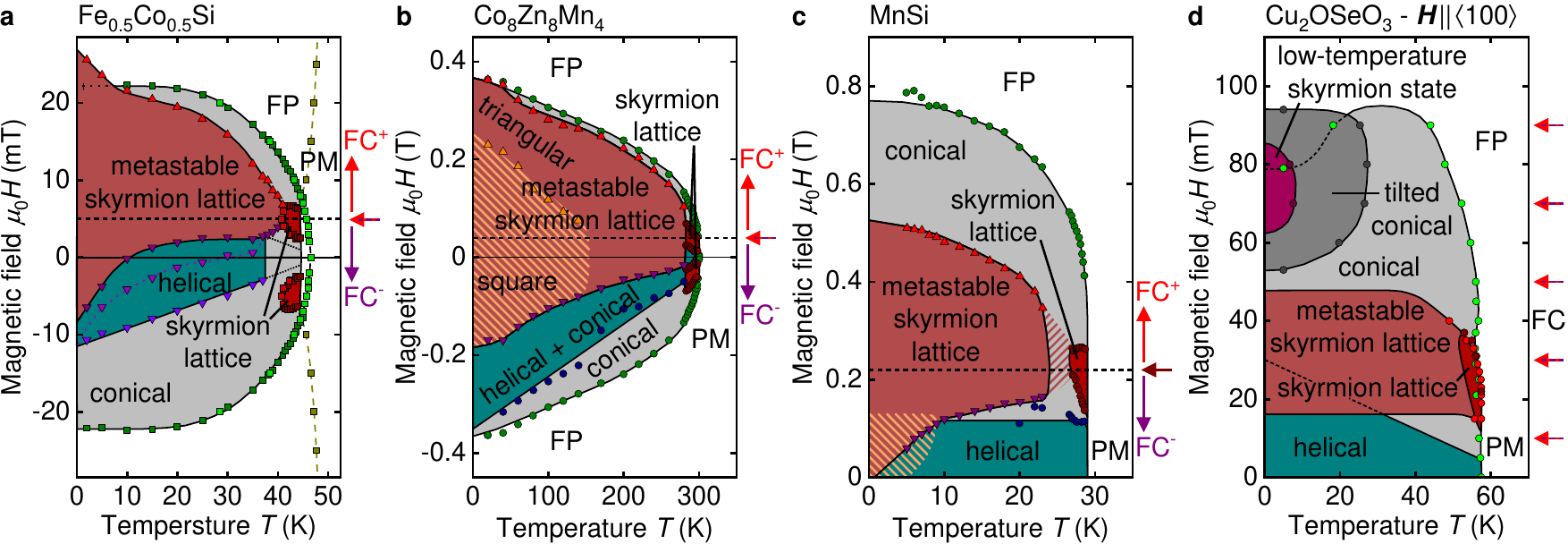}} 
  \caption{Magnetic phase diagram of selected bulk materials featuring skyrmion textures far from equilibrium under field-cooling. Diagrams shown in panels (a), (b) and (c) were recorded after field-cooling in field sweeps for increasing or decreasing field, denoted FC$^-$ and FC$^+$, respectively. The diagram shown in panel (d) was recorded under field-cooling at different field values. (a) Magnetic phase diagram of {\fcs} \cite{2016_Bauer_book}. (b) Magnetic phase diagram of {\czm} \cite{2016_Karube_NatMat}. (c) Magnetic phase diagram of MnSi after rapid thermal quenches \cite{2016_Oike_NatPhys, 2017_Nakajima_SciAdv}. (d) Magnetic phase diagram of Cu$_2$OSeO$_3$ \cite{2018_Chacon_NatPhys}.
   }
\label{Pfleiderer-Fig01}
\end{figure*}

Building on a large body of evidence that establishes skyrmion lattice order as a generic thermodynamic ground state at high temperatures, numerous studies have shown that skyrmion lattices readily fall out of thermal equilibrium under field-cooling. This was observed initially in {\fcs}, MnSi under pressure as well as {\czm} \cite{2010_Muenzer_PRB,2013_Ritz_PhysRevB,2016_Karube_NatMat}. While both, {\fcs} and {\czm}, are subject to strong disorder, extremely fast thermal quenches may be used in pure MnSi at ambient conditions \cite{2016_Oike_NatPhys}. However, despite being thermodynamically metastable, thermally quenched skyrmion lattices display exceptional stability in the low temperature limit, namely: (i) the skyrmion lattice survives deep into the field-polarized state beyond the conical phase [cf Fig.\,\ref{Pfleiderer-Fig01}\,(a)], (ii) the skyrmion lattice is stable at zero field [cf Fig.\,\ref{Pfleiderer-Fig01}\,(a), (b) and (c)], and (iii) the skyrmion lattice is stable under inversion of the applied field [cf Fig.\,\ref{Pfleiderer-Fig01}\,(a), (b) and (c)]. 

As a corollary of these observations the energy of the metastable skyrmion lattice differs only slightly from the true thermodynamic ground states, while the energy barriers are large. Using magnetic force microscopy this was exploited in {\fcs} to obtain information on the nature of the topological unwinding of skyrmions \cite{Milde13}. Further, for thin bulk samples of {\fcs}, which feature an extended parameter range of the skyrmion lattice phase in thermal equilibrium \cite{Yu10}, limitations of the topological protection due to entropy compensation were identified \cite{2017_Wild_SciAdv}. Moreover, the formation of metastable skyrmion lattice order under field-cooling has also been used, for instance, to determine the size of the intrinsic topological Hall signal \cite{2013_Ritz_PhysRevB,2014_Franz_PhysRevLett}. Further, as metastable skyrmion configurations feature long life-times their use in non-volatile memory and other applications has been advertised. Namely, skyrmions in bulk materials as well as thin films may be switched on and off by means of very rapid thermal quenches \cite{2016_Oike_NatPhys, 2013_Finazzi_PRL}.

Several studies have also addressed skyrmions and skyrmionic spin structures in systems that are strongly driven out of equilibrium by means of a non-thermal stimulus. Important examples include the exceptionally strong coupling of spin currents to skyrmion lattices, which induce skyrmion lattice flow at ultra-small current densities \cite{Jonietz10,Schulz12,2014_Mochizuki_NatMater}. In atomically thin films, it has been argued that the effects of spin currents permit to read and write skyrmions \cite{2013_Romming_Science}. Further, the creation of skyrmions with current pulses has also been employed in micro-structured films, where the nucleation process is attributed to the details of geometric constraints \cite{2015_Jiang_Science}.  On a different note, the creation of specific magnon excitations may also be used to drive skyrmion systems out of equilibrium. A remarkably complete understanding of the spectrum of magnon excitations has been achieved [cf. Sec.\,\ref{garst} and Ref.\,\cite{2017_Garst_JPhysD}]. However, in comparison to the properties of skyrmion lattices that have fallen out of thermal equilibrium, skyrmion lattices that are driven out of equilibrium by a non-thermal stimulus is much less complete both experimentally and theoretically. 

\subsection{Current and future challenges}

Major unresolved challenges concern the experimental and theoretical identification of generic aspects of metastable skyrmion lattice phases across different materials systems, as tiny details of the underlying energy landscape become important. The putative observation of metastable skyrmion textures with triangular, orthorhombic and quadratic lattice structure in {\czm} and MnSi \cite{2016_Karube_NatMat, 2017_Nakajima_SciAdv}, suggest that different morphologies may be generated far from equilibrium. This raises the question for the possible existence of liquid-crystal like skyrmion phases such as nematic or smectic configurations \cite{2018_Huang_ArXiv}. Of further interest are putative analogies with type 1 and type 2 superconductivity, namely the formation of different morphologies such as square and triangular lattice structures as well as textures akin to the intermediate mixed state of superconductors \cite{2009_Muehlbauer_PRL}. 

The identification of two independent skyrmion lattice phases in {\cso} \cite{2018_Chacon_NatPhys}, where the magnetic phase diagram under field cooling is reproduced in Fig.\,\ref{Pfleiderer-Fig01}\,(d), raises the question how different stabilization mechanisms interfere under metastable conditions. Preliminary measurements suggest, for instance, that the cubic magnetic anisotropies responsible for the low temperature skyrmion lattice phase in {\cso} allow to boost the volume fraction of the high-temperature skyrmion lattice even far from equilibrium. Further, as the low temperature skyrmion lattice phase represents a ground state in the zero temperature limit and the high temperature skyrmion lattice phase proves to be exceptionally stable under field-cooling, another open challenge concerns the nature of the nucleation process of the low temperature skyrmion phase. In addition, the effects of geometric frustration may be important, as recently explored in the Co-Zn-Mn system where they reflect the equivalence of the $\langle100\rangle$ easy axes perpendicular to an applied field along a $\langle100\rangle$ axis \cite{2018_Karube_SciAdv}. 

Unresolved questions in skyrmion lattice systems that are driven far from equilibrium by a spin current concern the observation of a rotational motion as well as an improved long-rage order (reduced mosaicity) of the skyrmion lattice under flow as predicted theoretically \cite{2012_Everschor_PhysRevB,2019_Reichhardt_PRB}. Further, the observation that the reading and writing of individual skyrmions is sensitive to specific locations in nominally homogeneous thin films suggests an important role of inhomogeneities awaiting clarification of the microscopic details \cite{2013_Romming_Science}. 

As for the spectrum of magnons of skyrmion lattices, important questions concern differences and similarities of metastable textures with particular interest on the excitation spectra of systems with different lattice morphologies. Further challenges concern the possibility to trigger the decay of metastable skyrmion lattice textures by virtue of resonant microwave pumping. Taking into account differences of the Chern numbers of different magnonic modes, it has been predicted that induced Dzyaloshinsky-Moriya interactions may be created. Finally, under very strong resonant microwave pumping an effective melting of skyrmion lattice order has been anticipated \cite{2012_Mochizuki_PhysRevLett}, raising the question for differences with a thermal melting of skyrmion lattice order. 

\subsection{Advances in science and technology to meet challenges}

Several experimental techniques have recently been implemented that allow to clarify the questions addressed above. For instance, time-resolved Lorentz transmission electron microscopy on a large field of sight has recently allowed to track system sizes exceeding several $10^4$ skyrmions \cite{2015_Rajeswari_ProcNatlAcadSciUSA}. For the investigation of ultra-slow decay mechanisms neutron scattering techniques with an exceptional dynamic range will be essential, such as resonant neutron spin echo spectroscopy \cite{2019_Franz_JPhysSocJpn}. Further examples that promise to provide new insights include neutron grating interferometry \cite{2018_Reimann_PhysRevB}, stroboscopic SANS of fast thermal quenches \cite{2018_Nakajima_PhysRevB} and time-resolved small angle scattering (TISANE) \cite{2016_Muhlbauer_NewJPhys}. For studies of periodically driven skyrmion lattice systems novel pump-probe techniques will be of interest. This concerns for instance the possibility to perform resonant elastic x-ray scattering and neutron scattering under microwave radiation.

\subsection{Concluding remarks}

In conclusion, topologically non-trivial spin textures offer a wide range of scientific insights as well as challenges for future studies that connect the notion of topology with general aspects of condensed matter systems far from equilibrium. The rather comprehensive understanding of magnetic skyrmions achieved in recent years promises to turn skyrmions far from equilibrium into a field of research in its own right. 

\subsection{Acknowledgements}
I am especially indebted to A. Bauer, A. Chacon, and M. Halder. The work reviewed in this section and related projects have received funding from the Deutsche Forschungsgemeinschaft (DFG, German Research Foundation) under Germany's Excellence Strategy EXC-2111 390814868 (MCQST), the cooperative research centre TRR80 (Projects No. E01 and F07), and the priority program SPP2137 (Skyrmionics). Support by the European Research Council [Advanced Grants TOPFIT (291 079) and ExQuiSid (788 031)] is also acknowledged. 

%\balance

%\subsection{References}
%\bibliography{Pfleiderer}

%%%%%%%%%%%%%%%%%%%%%%%%%%%%%%%%%%%%%%%%%%%%%%%%%
%%%%%%%%%%%%%%%%%%%%%%%%%%%%%%%%%%%%%%%%%%%%%%%%%
%%%%%%%%%%%%%%%%%%%%%%%%%%%%%%%%%%%%%%%%%%%%%%%%%

\clearpage
\newpage

\section[Creation, destruction and topological protection of skyrmions \\ {\normalfont Christian Back}]{Creation, destruction and topological protection of skyrmions}
\label{back}
{\it Christian Back}

Physik-Department, Technical University of Munich, James-Franck-Str. 1, 85748 Garching, Germany

\subsection{Status}
Due to their topological protection and the potentially easy control of their motion by ultra-low current densities, skyrmions - or more general chiral magnetic textures - are considered for future high density data storage or logic devices \cite{sampaio}. Skyrmion crystals (Skx) and single skyrmions have been studied in many magnetic systems ranging from bulk B20 metals such as MnSi, FeGe and Fe$_x$Co$_{1-x}$Si alloys to more complex crystalline bulk alloys and lacunar spinels and to chiral insulators such as Cu$_2$OSeO$_3$. Furthermore, recently thin film heterostructures have emerged as skyrmion hosting materials and first experiments have indicated the possibility of creating skyrmions by simple current pulses in metallic materials \cite{2015_Jiang_Science, buettner}. It has also been shown that in these materials single skyrmions and skyrmion sequences can be moved by current pulses \cite{2015_Jiang_Science,klaui}. However, in experiments it has been witnessed that in thin film materials skyrmions are easly annihilated at defects \cite{klaui,hrabec}. On the other hand it has been proposed \cite{everschor} and demonstrated \cite{buettner} that local changes in the energy landscape can be exploited to puposely create skyrmions at specific locations in the material. 

Topological protection of the skyrmion state prevents a continuous transformation of the chiral spin texture into e.g. the uniform magnetic state and vice versa. To unwind this extraordinarily stable state - a stability which in principle can also be attributed to isolated metastable skyrmions outside the stability pocket in the phase diagram - topological defects such as Bloch points \cite{schuette} or so-called hedgehogs \cite{Milde13} need to be introduced. In the continuum limit the energy of topological defects is infinite, however, this energy is reduced to physically relevant values at finite temperatures when the discrete atomic lattice is inroduced \cite{hagemeister,bessarab}. 

It should be noted that most experiments are performed at elevated temperatures. This entails that for the description of the energetics of a skyrmion hosting system the free energy needs to be taken into account. This further means that, when both, temperature induced activation and entropy are taken into account, topological protection of isolated skyrmions can be greatly reduced \cite{2017_Wild_SciAdv}.

It thus becomes clear immediately, that thorough investigations of stability on the one hand and of mechanisms to controllably create skyrmions on the other hand will become highly relevant in the near future.

\subsection{Current and future challenges}
Skyrmions in e.g. racetrack-type devices "live" in a metastable environment. In both, bulk or thin film systems this would typically be a field polarized state (ferromagnetic or ferrimagnetic). Consequently, two pressing issues arise:
\newline
i) How can metastable skyrmions be created reliably and on demand in the field polarized state.
\newline
ii) How can the stability of skyrmions be ensured.
\newline
Concerning the first point several approaches based on the interplay between a dc current or current pulses and the magnetic energy landscape exist. It has been realized early on that sykrmions may be created at positions where locally the internal stability phase diagram is altered \cite{2015_Jiang_Science,klaui,buettner,everschor} due to the combined action of the current-induced spin-transfer torque acting on the local magnetization and a magnetic energy lansdscape which varies locally on a length scale comparable to the skyrmion size \cite{everschor}. The local change of the energy landscape may be controlled by the magneto-static energy \cite{2015_Jiang_Science} or by local engeneering of the magnetic anisotropy, Dzyloshinskii-Moriya interaction or saturation magnetization \cite{everschor}. For technologically desired small skyrmion sizes below 50 nm it is thus technical challenging to realize skyrmion creation on demand. 
\newline
Concerning the second point, a larger effort must be put into the development of experimental methods enabling the detailed evaluation of skyrmion stability as proposed in \cite{bessarab,2017_Wild_SciAdv} for technologically relevant thin film materials. While it has been shown in many micromagnetic simulations - which are typically performed at zero temperature - that skyrmions in nanostructures such as racetracks avoid edges and defects \cite{sampaio} and thus escape annihilation, only few systematic studies take finite temperatures and stability under current drive at elevated temperatures into account. In a recent work by Hagemeister \textit{et. al} \cite{hagemeister}, Monte-Carlo simulations have been used to calculate the attempt frequency for annihilation and creation of isolated skyrmions while e.g. in \cite{bessarab} the geodesic nudged elastic band method has been employed to determine the energy barriers for skyrmion creation and annihilation.

\begin{figure}[h]
 \center{ \includegraphics[width=0.48\textwidth]{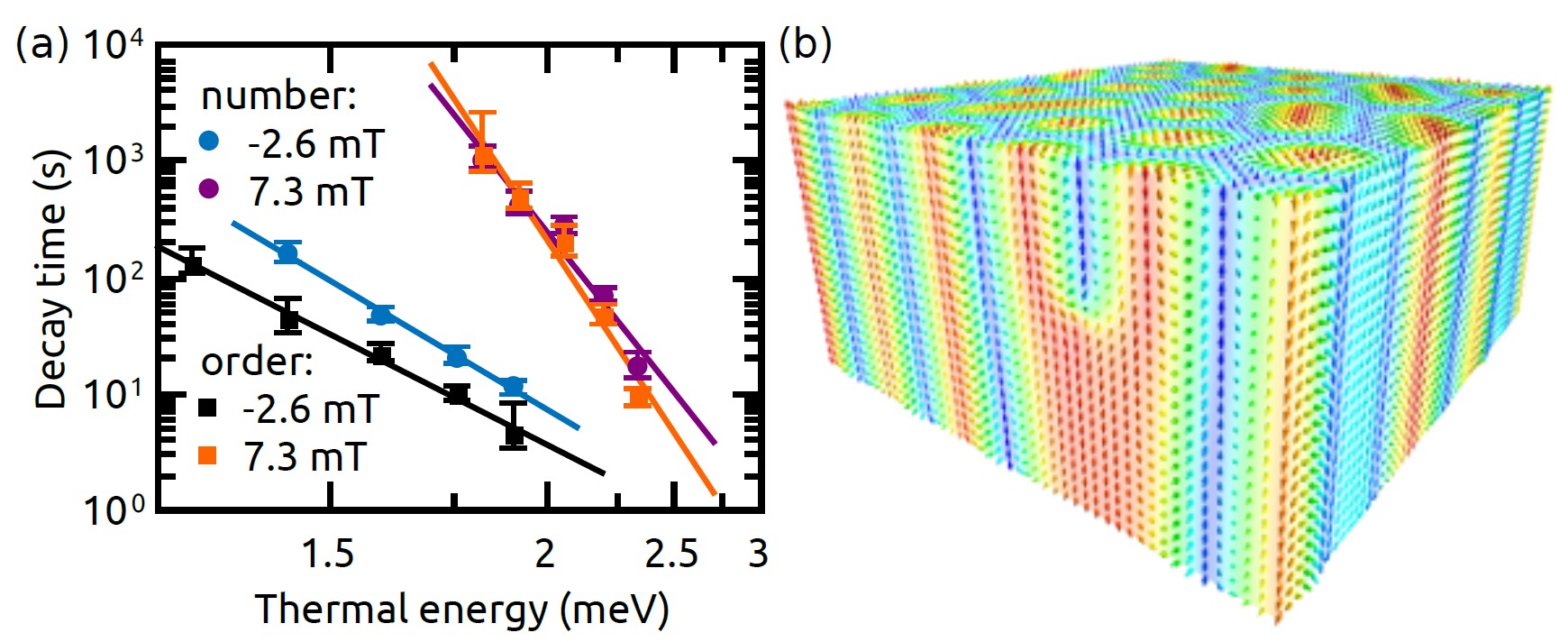}} 
  \caption{Decay rates of supercooled skyrmions in Fe$_{1-x}$Co$_x$Si ($x=0.5$). (Right) Sketch of a snapshot of skyrmion tubes zipping together to form a small section of a helical stripe corresponding to the decay from the ordered skyrmion lattice phase into the helical phase. (Left) Decay time as a function of thermal energy for increasing and decreasing magnetic fields as indicated in the figure. Shown are two different data evaluations concerning the decay of the order of the skyrmion lattice and the skyrmion number denoted order and number, respectively. Figure adapted from Ref.\,\cite{2017_Wild_SciAdv}.}
\label{Back-Fig01}
\end{figure}

\subsection{Advances in science and technology to meet challenges}
The open questions outlined above call for advancement of stroboscopic and non-stroboscopic time resolved techniques in order to be able to distinguish between repetitive events in creation/destruction of single skyrmions (or skyrmion clusters) compared to rare events. 
Possible solutions include high resolution time resolved scattering techniques which would be well suited for the observation of non-repetitive phenomena for skyrmion clusters and skyrmion crystals (in bulk systems) and non stroboscopic techniques in real space. It has been shown recently that in principle time resolved Lorentz transmission electron microscopy enables the observation of such phenomena in bulk systems \cite{2017_Wild_SciAdv}, see figure\ref{Back-Fig01} on timescales of about a millisecond, however, much effort is needed to increase the sensitivity in order  to perform similar experiments also in the technologically relevant thin film heterostructures.
High resolution techniques are also neded to observe nucleation of skyrmions and their dynamics at nucleation sites of the order of the envisoned skyrmion size below 50 nm \cite{buettner,Litzius17}.
An interesting possiblity could be the use of time resolved magneto-resistive techniques which have been employed successfully with sub-nanosecond resolution to understand subtle details in the case of domain wall nucleation/pinning and motion in recent years.

On the theory side the challenge lies in the incorporation of edge effects present in realistic racetrack devices. Boundaries might significantly alter the available paths for skyrmion creation and annihilation in particular at elevated tempertures and under current drive. A quantitative analysis of effects that arise from boundaries and the impact of edge effects on skyrmion stability at finite temperatures will be a challenge for the near future. 

\subsection{Concluding remarks}
The quantitative understanding of the stability of single skyrmions in a field polarized environment will be a challenging issues both on the experimental and the theoretical side in the years to come. Experimentally, improved thin film heterostructures and defect free racetracks will be one of the challenging issues along side with optimal recipes to prepare single skyrmions on demand. Experimental methods enabling the unambiguous observation of creation and destruction of sub 50 nm small magnetic objects must be developed. This effort has to be accompanied by theory which should deliver stability phase diagrams in realistic racetrack-type devices at elevated temperatures.

\subsection{Acknowledgements}
We acknowledge funding by the German Research Foundation via Project No. SPP2137 (Skyrmionics). This work has been funded by the Deutsche Forschungsgemeinschaft (DFG, German Research Foundation) under Germany's Excellence Strategy EXC-2111 390814868 and via TRR80 (Project No. G09).

%\balance

%%%%%%%%%%%%%%%%%%%%%%%%%%%%%%%%%%%%%%%%%%%%%%%%%
%%%%%%%%%%%%%%%%%%%%%%%%%%%%%%%%%%%%%%%%%%%%%%%%%
%%%%%%%%%%%%%%%%%%%%%%%%%%%%%%%%%%%%%%%%%%%%%%%%%

\clearpage
\newpage

\section[Collective excitations of magnetic skyrmions \\ {\normalfont Markus Garst}]{Collective excitations of magnetic skyrmions}
\label{garst}
%Collective excitations and magnonics}
{\it Markus Garst}

Institut f\"ur Theoretische Festk\"orperphysik, Karlsruhe Institute of Technology, 76131 Karlsruhe, Germany

\subsection{Status}

%This section provides a brief history and status, why the field is still important, what will be gained with further advances. (350 words max)

Magnetic skyrmions and skyrmion crystals are spatially extended, smooth topological textures of the magnetization that possess characteristic internal degrees of freedom. These give rise to a fascinating magnetization dynamics that reflects their non-trivial topology and offers interesting perspectives for applications in the fields of spintronics and magnonics; for recent reviews see Refs.~\cite{Mochizuki2015,2017_Garst_JPhysD}.

At lowest energies the skyrmion dynamics can be described in terms of a collective variable given by the first-moment, $\vec R = \int dx dy\, \vec r \rho_{\rm top}$, of the topological charge density $\rho_{\rm top} = \frac{1}{4\pi} \hat M (\partial_x \hat M \times \partial_y \hat M)$ within the $(x,y)$-plane where $\hat M$ is the local orientation of the magnetization. The quantity $\vec R$ can be identified with the linear momentum of the texture that obeys the so-called Thiele equation. This effective equation of motion captures the translational motion of skyrmions induced by spin-transfer or spin-orbit torques in spintronic applications, and it describes the skyrmion Hall effect, i.e., the deflection of skyrmions away from the direction of the spin current. 
%This low-energy dynamics is covered in more detail in section {\it (Spintronics with skyrmions -- S. Parkin)}.

\begin{figure*}[t]
\begin{center}
\includegraphics[width=12cm]{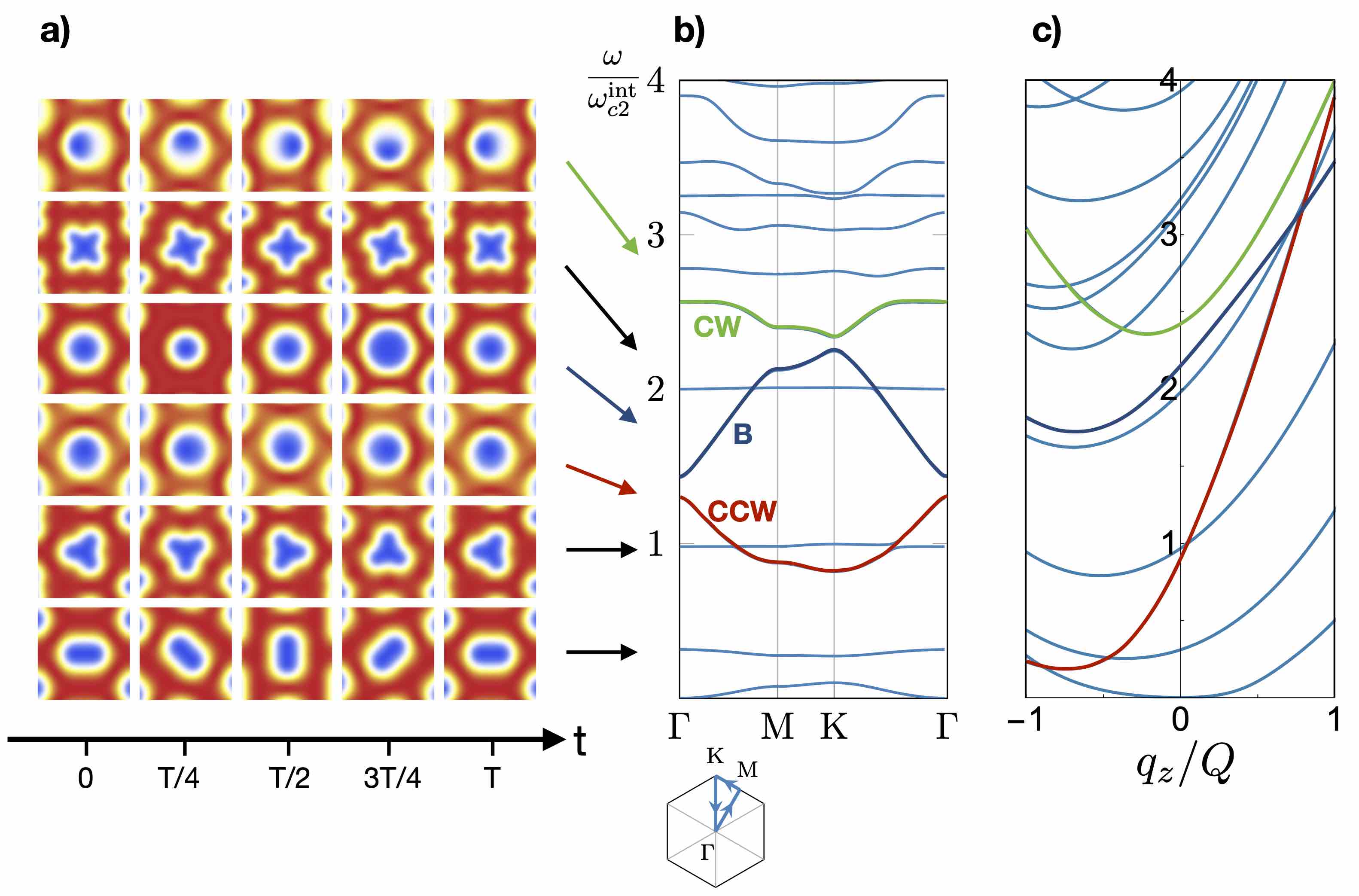}
\caption{Magnon band structure of a skyrmion crystal in a chiral magnet. a) Illustration of excitation modes with zero wavevector, $\vec q=0$, as a function of time $t$; $T$ is the corresponding period. The color represents the $z$-component of the magnetization. Only the counterclockwise (CCW), breathing (B) and clockwise (CW) mode are associated with a homogeneous ac magnetic moment. Magnon band structure for wavevectors b) within the first Brillouin zone of the hexagonal skyrmion crystal and c) for wavevectors along the applied magnetic field showing a non-reciprocal spectrum $\omega(q_z) \neq \omega(-q_z)$. Note that the limit $\vec q \to 0$ for the CCW, B, and CW modes is non-analytic due to dipolar interactions. Parameters for Cu$_2$OSeO$_3$ with $a_{\rm SkX} = 2\pi/Q \approx 60$ nm, $\omega^{\rm int}_{c2}/(2\pi) \approx 2.3$ GHz and $H/H_{c2} = 0.5$, see Ref.~\cite{2017_Garst_JPhysD}.}
\label{Fig_Garst}
\end{center}
\end{figure*}

The rigidity of skyrmions results in additional dynamical excitations of the magnetization at higher energies that, for ferromagnetic materials, are located in the range of microwave frequencies. The associated magnetic resonances of two-dimensional skyrmion crystals were first theoretically identified by M. Mochizuki \cite{2012_Mochizuki_PhysRevLett}. They comprise two modes where the global magnetization oscillates within the plane of skyrmions either in a counterclockwise (CCW) or clockwise (CW) fashion and, in addition, a breathing mode (B) where the magnetization oscillates out-of-plane. These resonances were experimentally detected with the help of coplanar waveguides (CPW) in the cubic chiral magnets Cu$_2$OSeO$_3$, FeGe, Fe$_{0.8}$Co$_{0.2}$Si and MnSi \cite{Onose2012,Schwarze2015}, that host Bloch-skyrmions, and in rhombohedral GaV$_4$S$_8$ with N\'eel-skyrmions \cite{Ehlers2016}.

The full magnon spectrum is, however, much richer. When spin waves propagate across the periodic texture of magnetic skyrmion crystals, they experience Bragg scattering that leads to a backfolding of their dispersion resulting in a magnon band structure. The extension of the corresponding Brillouin zone, $2\pi/a_{\rm SkX}$, is determined by the lattice constant of the skyrmion crystal $a_{\rm SkX}$, that is typically on the order of several tens of nanometers.
%, for the materials listed above, is typically on the order of several tens of nanometers. 
This backfolding leads to various spin-wave modes at the $\Gamma$-point of the Brillouin zone but, due to selection rules, only three of them can be excited by magnetic resonance. The other modes with finite frequencies are associated with higher magnetic multipole excitations, see Fig.~\ref{Fig_Garst}.

Isolated single skyrmions also possess resonances. Whereas the skyrmions found in chiral magnets are mainly stabilized by the Dzyaloshinskii-Moriya interaction and are relatively stiff with only a few internal modes \cite{2017_Garst_JPhysD}, the properties of so-called skyrmion bubbles, often found in magnetic multilayers, are dominated by dipolar interactions leading to larger skyrmion sizes with more internal modes \cite{Makhfudz2012,Montoya2017}. 
%
%The internal excitation modes of skyrmions are also important for 
%the determination of their lifetime \cite{}.
%
At high energies, the spin wave scattering  off single skyrmions is governed by an emergent electrodynamics that is directly linked to the non-trivial topology of skyrmions, see section \ref{everschor}. It results in a characteristic skew scattering that can give rise to a topological magnon Hall effect \cite{2014_Mochizuki_NatMater}. For skyrmion crystals, the emergent electrodynamics leads to magnon Landau levels leading to finite Chern numbers of certain magnon bands \cite{2017_Garst_JPhysD,RoldanMolina2016}. As a result, skyrmion crystals with sufficiently low Gilbert damping should be accompanied with topologically protected magnon edge states. 

For magnonics, that aims to control spin waves for microwave applications and information processing, the magnetic insulator Cu$_2$OSeO$_3$ is especially interesting as it possesses a small Gilbert damping $\alpha \sim 10^{-4}$ \cite{Stasinopoulos2017}. In addition, its magnetoelectric coupling allows to access the magnetization dynamics with the help of electric fields with various intriguing magnetoelectric phenomena \cite{Mochizuki2015}.

\subsection{Current and future challenges}
%This section discusses the big research issues and challenges. (350 words max)

{\it Magnon band structure of skyrmion crystals --- }
Up to now, the experimental investigation of the magnetization dynamics of skyrmion crystals is limited to its three magnetic resonances. 
A future challenge is the exploration of the magnon band structure and its dispersing spin wave modes with various complementary experimental techniques like neutron scattering, spin-wave spectroscopy and Brillouin light scattering.
A particularly interesting aspect is the non-reciprocal dispersion, $\omega(q_z) \neq \omega(-q_z)$, for wavevectors $q_z$ along the magnetic field, i.e., perpendicular to the skyrmion crystal plane, see Fig.~\ref{Fig_Garst}c). In bulk chiral magnets, skyrmions are extended into strings that are aligned with the applied field, and these strings transfer magnons in a non-reciprocal fashion \cite{Seki2020}.  
%
%Inelastic neutron scattering is, in principle, a powerful method to map out this magnon band structure over large regions in energy and reciprocal space. However, the typical wavevector $2\pi/a_{\rm Sky}$ for most of the skyrmion materials is too small 
%to resolve the magnon dispersion within the first magnetic Brillouin zone with current neutron detectors. The largest known $2\pi/a_{\rm Sky}$ is found in MnSi so that this material is most promising for a neutron scattering study. 
%%
%In the opposite limit of small wavevectors, spin wave spectroscopy and Brillouin light scattering can be employed to access the dispersing character of the magnetic resonances close to the $\Gamma$-point of the Brillouin zone.

{\it Standing spin wave modes and surface twist -- }
The repeated reflections of magnons at surfaces leads to standing spin wave modes in small samples with low Gilbert damping \cite{Stasinopoulos2017}. Their theoretical description is challenging for 
%The theoretical description of such standing spin waves in chiral magnets is not yet available and challenging for 
two reasons. First, the Dzyaloshinskii-Moriya interaction leads to a surface reconstruction of the magnetization, i.e., a so-called surface twist that in turn results in non-trivial boundary conditions for the reflected spin waves even for free surfaces. The penetration depth of this surface twist itself is, however, not yet fully understood \cite{Zhang2018}. Second, the regime of wavevectors for the standing spin waves is dominated by dipolar interactions giving rise to a non-local interaction between surface twist and spin waves that is challenging to treat theoretically. 

{\it Magnetization dynamics of single skyrmions and skyrmion strings -- }
Another challenge is the experimental identification of resonances, like the breathing or the gyrotropic counterclockwise mode, attributed to single skyrmions either in chiral magnets or magnetic multilayers. This requires materials with sufficiently low damping in order to resolve resonances with a small weight proportional to the skyrmion density. 
An isolated skyrmion or a finite cluster of skyrmions would also provide a platform to investigate experimentally the emergent electrodynamics for spin waves, their skew scattering and the topologically protected magnon edge states preferably with microscopic methods like magneto-optic Kerr effect (MOKE). On the theoretical side, the magnetization dynamics of isolated skyrmion strings needs to be further investigated, that can act as non-reciprocal transmission lines for spin waves. The analogy with vortex filaments in superfluids suggests that skyrmion strings possess an interesting non-linear dynamics that remains to be explored. In general, the non-linear magnetization dynamics of textured magnets both under equilibrium and non-equilibrium conditions is poorly developed, see also section \ref{pfleiderer}.

{\it Skyrmion dynamics in frustrated ferromagnets and chiral antiferromagnets -- } 
Skyrmions stabilized by exchange-frustration will gain more attention with the increasing availability of materials. The magnetization dynamics of such skyrmions is richer due to the additional helicity degree of freedom that influences the translational motion relevant for spintronics \cite{Leonov2015}. 
Similarly, the interest in antiferromagnetic skyrmions \cite{Barker2016} is expected to increase fuelled by the growing general interest in antiferromagnetic spintronics and, in particular, by their vanishing skyrmion-Hall effect, that is advantageous for spintronics. Antiferromagnetic skyrmions are anticipated to possess internal eigenmodes with THz frequencies with interesting applications in electronics. The excitation spectrum of skyrmions in both frustrated ferromagnets and chiral antiferromagnets \cite{Kravchuk2019} remains to be  studied further. 

\subsection{Advances in science and technology to meet challenges}
%This section discusses the advances in science technology needed to address the challenges. (350 words max)

For the comprehensive  study of the magnon spectrum of skyrmion crystals different experimental techniques need to be combined. Inelastic neutron scattering is, in principle, a powerful method to map out excitation spectra over large regions in energy and reciprocal space. However, the typical wavevector $2\pi/a_{\rm SkX}$ for most of the skyrmion materials is too small to resolve the magnon dispersion within the first magnetic Brillouin zone with current neutron detectors. The largest known $2\pi/a_{\rm SkX}$ is found in MnSi so that this material is most promising for a neutron scattering study. 
In the opposite limit of small wavevectors, spin wave spectroscopy and Brillouin light scattering can be employed to access the dispersing character of the magnetic resonances close to the $\Gamma$-point of the Brillouin zone.
Experiments with MOKE, NV-centers or spin-resolved STM might be able  to resolve spatially  the  dynamics of skyrmion crystals or even single skyrmions. 

The study of skyrmions in ferromagnetic multilayers is often hampered by disorder that results in pinning of skyrmions, strongly hysteretic effects and large damping, see section \ref{fert}. Better samples will be beneficial for spintronic applications and will also facilitate the study of skyrmion resonances and magnon-skyrmion interactions in these systems. 
For antiferromagnetic skyrmions, multilayers realizing synthetic antiferromagnets or ferrimagnets close to their compensation point need to be realized. In order to further advance the study of skyrmions in  frustrated ferromagnets and  antiferromagnets, additional bulk materials also need to be identified. 

On the theoretical side, numerical simulations will be instrumental to guide the analysis of the dynamics of skyrmion string textures in bulk systems and magnetic multilayers.

\subsection{Concluding remarks}
%Include brief concluding remarks. This should not be longer than a short paragraph. (150 words max)

The non-trivial topology of magnetic skyrmions has a strong impact on its magnetization dynamics giving rise to an emergent electrodynamics for spin waves, that leads to skew scattering and topological magnon bands. Whereas the uniform resonances of skyrmion crystals have been successfully studied in the past, many interesting properties of propagating spin waves in skyrmion textures remain to be investigated. Technologically, these might be exploited to control and manipulate magnetic microwave excitations with potential applications in magnonics. 

\subsection{Acknowledgements}
This research was funded by DFG via SFB 1143 'Correlated Magnetism: From Frustration to Topology', grant GA 1072/5-1 and grant GA 1072/6-1.

%Please include any acknowledgements and funding information as appropriate.

%%%%%%%%%%%%%%%%%%%%%%%%%%%%%%%%%%%%%%%%%%%%%%%%%
%%%%%%%%%%%%%%%%%%%%%%%%%%%%%%%%%%%%%%%%%%%%%%%%%
%%%%%%%%%%%%%%%%%%%%%%%%%%%%%%%%%%%%%%%%%%%%%%%%%

\clearpage
\newpage

\clearpage
\newpage

\section[Emergent electrodynamics of skyrmions \\ {\normalfont Karin~Everschor-Sitte}]{Emergent electrodynamics of skyrmions}
\label{everschor}
{\it Karin~Everschor-Sitte}

Institute of Physics, Johannes Gutenberg-University, 55128 Mainz, Germany

\subsection{Status}
\begin{figure*}
 \center{ \includegraphics[width= 0.9 \textwidth]{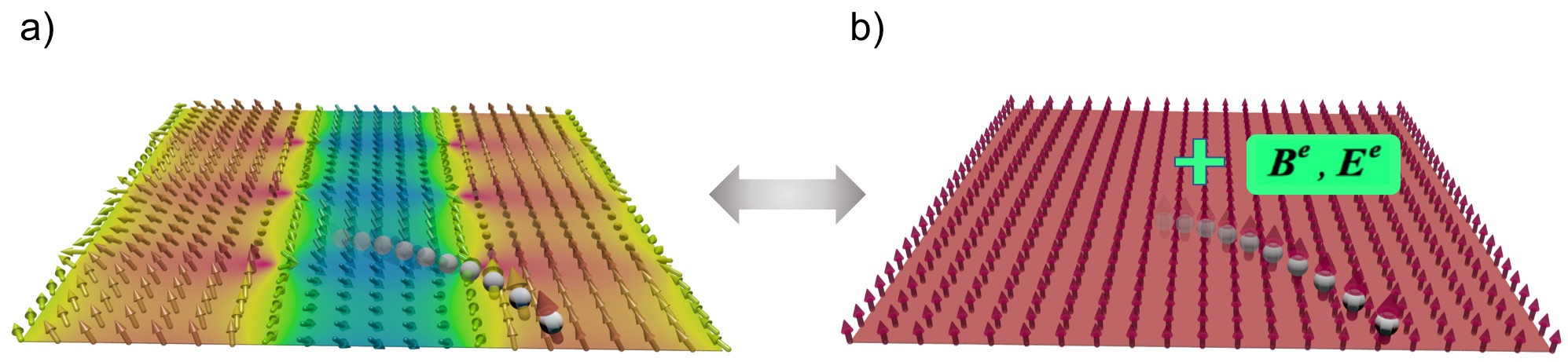}} 
  \caption{ An electron traverses adiabatically a spatially inhomogeneous smooth magnetic texture. This problem can be mapped to one where the electron moves in a ferromagnetic background but instead ``feels'' additional emergent magnetic and electric fields.
   }
\label{fig:mapping}
\end{figure*}

A central theme in spintronics is the complex interplay between electric currents and magnetization dynamics.
An electric current exerts forces on non-collinear magnetic structures and vice versa. 
In the adiabatic limit where the magnetic structure varies smoothly compared to the atomic lattice spacing and band-structure effects are negligible, the current-carrying electron picks up a Berry phase while aligning its magnetic moment along the direction of the local magnetization direction. 
The forces induced by the magnetic structure on the electrons are elegantly accounted for via the description of an emergent electrodynamics, mapping the complex interplay of an electron traversing a magnetic texture to that of a charged particle subject to electric and magnetic fields as in classical electrodynamics, see Fig.~\ref{fig:mapping}. 

\begin{figure*}
\center{ \includegraphics[width= 0.9 \textwidth]{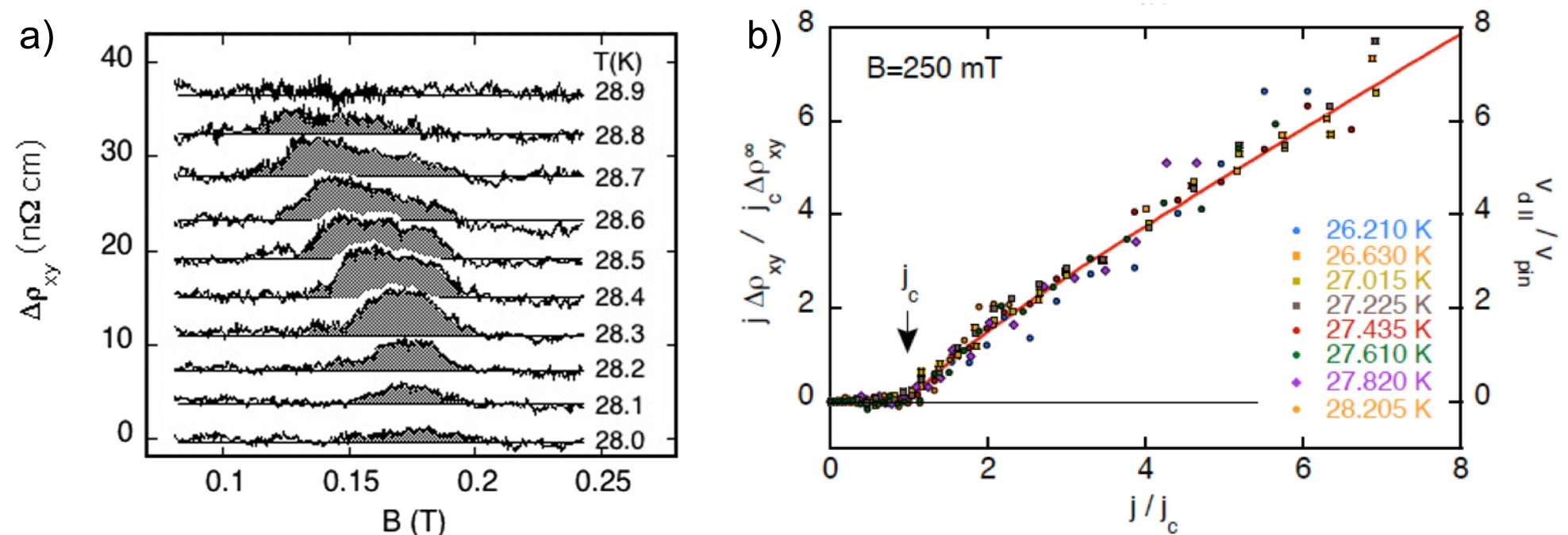}}
  \caption{Experimental observation of a) the emergent magnetic field b) the emergent electric field for skyrmions through a corresponding change in the Hall signal.  Figures reprinted with permission from Refs.~\cite{Neubauer2009, Schulz12}.} 
\label{fig:exp}
\end{figure*}

The key idea of emergent electrodynamics is already encoded in a one-band free-electron Stoner model.\cite{Volovik1987, Bazaliy1998, Everschor-Sitte2014} Here the electron ``feels'' emergent magnetic and electric fields
\begin{eqnarray}
\label{eq:Be}
\vect B^e_i &=& \frac{\hbar}{2 e} \epsilon_{ijk} \hat{\vect M} \cdot (\partial_j
\hat{\vect M} \times \partial_k \hat{\vect M}), \ \mathrm{and}\\ 
\vect E^e_i &=& \frac{\hbar}{e} \hat{\vect M} \cdot (\partial_i \hat{\vect M} \times
\partial_t \hat{\vect M}),
\label{eq:Ee}
\end{eqnarray}
induced by the smooth magnetic texture 
$\vect{M}(\vect{r}, t) = M_s \hat{\vect M}(\vect{r}, t) $ with saturation magnetization $M_s$. 
Smooth magnetic skyrmions are tailor-made to study the emergent electrodynamics, as their quantized winding number 
\begin{equation}
\label{eq:windingnumber}
\mathcal{W} = \frac{1}{4 \pi} \int \hat{\vect M} \cdot
(\partial_x \hat{\vect M} \times \partial_y \hat{\vect M})\, dx\, dy
\end{equation}
ensures the ``emergent magnetic flux'' per skyrmion to be
quantized $ \int  e \vect B^e \cdot d \vect \sigma = 4 \pi \hbar \, \mathcal{W}$.
As for topological trivial magnetic phases the winding number vanishes, no emergent magnetic field arises. This allows to detect the non-trivial skyrmion topology via the emergent
magnetic field using Hall measurements.\cite{Neubauer2009, Lee2009} The additional contribution to the Hall signal based on the non-zero winding number is denoted as the topological Hall effect (THE), see Fig.~\ref{fig:exp}a). 

For a time-dependent magnetic texture an emergent electric field arises. The force induced by a moving non-collinear magnetic texture on the conduction electrons is denoted as spin motive force. It has been observed for domain walls\cite{Yang2009} and for skyrmions.\cite{Schulz12} For skyrmions, the quantization of the emergent magnetic field carries over to the emergent electric field. Thus, a Hall measurement with a current density applied above the depinning threshold, allows for a detection of both the emergent electric field and the drift velocity, see Fig.~\ref{fig:exp}b). 

\subsection{Current and future challenges}
While qualitative features in experiments with smooth magnetic skyrmions can be described quite well by the simple model and the emergent fields in Eqs.~(\ref{eq:Be}) and (\ref{eq:Ee}) originating in real-space Berry phase effects only, the real situation is however far more complex.
For quantitative agreements with experiments in particular with smaller skyrmions, several effects need to be taken into account, including non-adiabatic processes, modifications due to the band structure and multiband-effects, fluctuations of the magnetization amplitude and spin-orbit coupling. 
In skyrmion systems with weak spin-orbit coupling the emergent fields obtain a chiral contribution\cite{Kim2013}
 and the effects of momentum space and mixed Berry phases become relevant.\cite{Freimuth2013}
Furthermore, even though Hall measurements are a simple technique to establish the topological character of a magnetic structure, it is necessary to carefully disentangle the different contributions to the Hall signal, like normal, anomalous, and topological contributions. 
Complementary studies in combination with theory predictions are needed to obtain a systematic and thorough interpretation of the results for extracting individual contributions such as the THE. 

 Recently, a direct relation between the emergent magnetic field in real-space and the emergence of chiral and topological orbital magnetism independent of spin-orbit coupling has been established.\cite{Dias2016} This result allows for new experimental fingerprints of topological magnetic textures and paves the way for chiral orbitronics. Another important aspect to uncover the full potential of emergent electromagnetism is going beyond the adiabatic limit. In a recent article\cite{Nakazawa2018} the authors have developed a unified theory for the THE to treat adiabatic and non-adiabatic components of the spin gauge field, thereby being able to cover the range from the strong to the weak coupling regime of the electron spin following the magnetic texture. 

Besides the exciting developments in ferromagnetic materials, the emergent electromagnetism for antiferromagnetic skyrmions has been predicted.\cite{Cheng2012, Buhl2017} The main essence can be understood within a simple model for a collinear antiferromagnet in which the emergent magnetic fields on individual sub-lattices have opposite signs. This means that i) the net THE is zero and ii) electrons of opposite spins are deflected in opposite directions, thereby inducing a topological spin Hall effect. With the recent effort of making antiferromagnets active elements in spintronics, emergent electromagnetism for antiferromagnets promises to become an interesting direction.

Another perspective is to go into the direction of quantum skyrmionics. Once the skyrmion size becomes comparable to the size of the underlying lattice, quantum effects become important.\cite{Ochoa2018} Here one can speculate to observe a quantum version of emergent electrodynamics in analogy to the emergent quantum electrodynamics, that has recently been discovered for a quantum spin ice system.\cite{Sibille2018} 

\subsection{Advances in science and technology to meet challenges}
Progress in science towards understanding the complex interplay between electric currents and magnetization dynamics is made by tackling the physics from various directions. 

Experimentally there has been a tremendous progress in engineering materials to design their topological properties. For example via constructing  multilayer systems, typically out of magnetic and heavy element-based layers, or doping of bulk materials, it is possible to not only fabricate devices with desired real-space skyrmion properties, but also create different (topological) band structures.  
While initially the emergent electrodynamics has been almost an exclusive way to address the topological properties of a magnetic texture and its dynamics, several experimental magnetic imaging techniques have become available, such as Lorentz transmission electron microscopy, spin-polarized tunneling microscopy, magnetic force microscopy, Kerr microscopy, scanning transmission X-ray microscopy and electron holography, which are complemented by momentum space techniques like small angle neutron scattering. 

The theoretical description, of the emergent electrodynamics has gone beyond the simplest formulation given in Eqs.~(\ref{eq:Be}) and (\ref{eq:Ee}). 
Yet, so far electronic (magnetic) degrees of freedom are treated mostly separately, and the magnetic (electronic) effects are only taken into account effectively. While analytic theories will only be capable to treat certain limits of the complicated interplay of electric currents and magnetization dynamics, and ab-initio techniques so far can model only small system sizes, new powerful simulation tools will need to be developed to capture the physics of emergent electrodynamics. 

\subsection{Concluding remarks}

The concept of emergent electrodynamics provides an intuitive picture for the physics of charge currents traversing a spatially inhomogeneous magnetic texture in terms of a classical electrodynamics description. Exploiting the physics of emergent electrodynamics allows to explore topological properties of magnetic textures by all-electrical means. While the basic theory has been developed already in 1987,\cite{Volovik1987} there are several challenging directions to explore in the future including mixed topologies such as in real and momentum space and going beyond (smooth) ferromagnetic textures.

\subsection{Acknowledgements}
Fruitful discussion with past and present collaborators is gratefully acknowledged with special thanks to Kyoung-Whan Kim.
Support by the German Research Foundation (DFG) under the Project No. EV 196/2-1, the priority program on skyrmionics (SPP2137), and the Transregional Collaborative Research Center (SFB/TRR) 173 SPIN+X is gratefully acknowledged.

%\subsection{References}
%\bibliography{Everschor-Sitte}

%%%%%%%%%%%%%%%%%%%%%%%%%%%%%%%%%%%%%%%%%%%%%%%%%
%%%%%%%%%%%%%%%%%%%%%%%%%%%%%%%%%%%%%%%%%%%%%%%%%
%%%%%%%%%%%%%%%%%%%%%%%%%%%%%%%%%%%%%%%%%%%%%%%%%

%\clearpage
%\newpage

%\input{_contributions/Bluegel}

%%%%%%%%%%%%%%%%%%%%%%%%%%%%%%%%%%%%%%%%%%%%%%%%%
%%%%%%%%%%%%%%%%%%%%%%%%%%%%%%%%%%%%%%%%%%%%%%%%%
%%%%%%%%%%%%%%%%%%%%%%%%%%%%%%%%%%%%%%%%%%%%%%%%%

\clearpage
\newpage

\section[Skyrmions as particles \\ {\normalfont Achim Rosch}]{Skyrmions as particles}
{\it Achim Rosch}$^{1,2}$\\
\label{rosch}
$^{1}$Institute for Theoretical Physics, University of Cologne, Cologne, Germany,  $^{2}$Department of Physics, Harvard University, Cambridge MA 02138, USA

\subsection{Status}
A single magnetic skyrmion in a ferromagnetic film is a magnetic texture characterized by its topology. The shape of the skyrmion is determined by the interplay of frustrating magnetic interactions. Here relativistic spin-orbit interactions are key to obtain small skyrmions. When a skyrmion gets smaller, typically its rigidity increases. While large $\mu$m-sized topological textures may best be viewed as fluctuating circular domain walls, small skyrmions with sizes of $100\,$nm and below are for many practical purposes a rigid object, described by 
\begin{equation}
\vec M(\vec r_i,t) = \vec M_0(\vec r_i - \vec R(t)) + \delta \vec M(\vec r_i,t) 
\end{equation}
where  $\vec M(\vec r_i,t)$ is the local magnetization on a microscopic level, $\vec R(t)$ is the coordinate of the skyrmion and $\delta \vec M(\vec r_i,t)$ parametrizes small corrections to the rigid texture  $\vec M_0(\vec r_i - \vec R(t))$. The latter can be obtained by minimizing the (free) energy of the system.

Under the condition that fluctuations $\delta \vec M(\vec r_i,t) $ can be neglected, it was pointed out a long time ago by Thiele \cite{1973_Thiele_PRL}, that one can obtain a simple equation of motion for magnetic textures by projecting the equations of motions of the spin-system (often approximated by the Landau-Lifshitz-Gilbert (LLG) equation) onto the coordinate $\vec R(t)$ only.
 \begin{equation}
 \vec G \times \dot{\vec R}(t)+\Gamma \, \dot{\vec R}(t)=\vec F(\vec R(t),t)
 \label{dynamics}
 \end{equation}
 Here $\vec G$ is the gyrocoupling arising from the Berry-phases of the spins. It is given by the product of the topological winding number of the skyrmion and the magnetization density $m$, $\vec G=m \int d^2 \vec r \, \hat\vec M \cdot(\partial_x \hat \vec M \times \partial_y \hat\vec M)$. The damping term $\Gamma$ is within the Landau-Lifshitz-Gilbert equation proportional to the Gilbert damping $\alpha$, but the validity of this description is unclear, see below.
 
There are many different sources for external forces $\vec F(\vec R(t),t)$ including electric, spin and thermal currents, field gradients, nano-structure, sample boundaries and defects.
 
\subsection{Current and future challenges}
Understanding the dynamics of skyrmions as particles is essential for the interpretation of all experiments manipulating single skyrmion. This poses both theoretical and experimental challenges. Detailed experiments testing theoretical models of forces are especially important as often the precise microscopic physics underlying the forces is unclear.
%In the following, we will, first, discuss questions related to the forces described by Eq. (\ref{dynamics}). Second, we address the question to what extent Eq. (\ref{dynamics}) is valid, for example, to what extent the concept of a skyrmion mass is needed to complement Eq.~(\ref{dynamics}). Finally, we consider the role of internal degrees of the skyrmion.

While the value of the first term, the gyrocoupling in Eq.~(\ref{dynamics}) is largely fixed by topology and the average magnetization, it is much less clear, which physical processes can be responsible for the damping $\Gamma$. Within the LLG equation  the Gilbert damping $\alpha$ is used to parametrize phenomenologically microscopic processes which lead to a decay of the uniform magnetization in the system which may arise from the coupling to electrons, phonons, or magnons. Within the Thiele approach, the same parameter determines the damping of skyrmions. It is far from obvious under which conditions and for which materials this approach is correct. For example, if one considers a model system where the total magnetization $S_z$ is exactly conserved, such that $\alpha=0$, there will still be skyrmion friction arising, e.g.,  from the spin-conserving scattering of thermally excited magnons or electrons. For example, a moving skyrmion induces an emergent electric field \cite{Schulz12} which accelerates thermal magnons or electrons and transfers energy from the skyrmion to its environment. Note that the majority of all skyrmion experiments are performed in a regime where temperature is much larger than the magnon gap. Especially in insulators with small Gilbert damping $\alpha$ one can expect that skyrmion friction is not dominated by Gilbert damping. Comparing experiments measuring spin precession in the ferromagnetic state (and thereby $\alpha$) and skyrmion damping (e.g., by tracking skyrmion trajectories) separately will allow to investigate quantitatively this question.

An issue which is, however, more important are the forces on the right-hand side of Eq.~(\ref{dynamics}). One the one hand, these are forces used to manipulate the skyrmion, arising from currents, field gradients or from nanostructuring of the sample, e.g., in a wire geometry. On the other hand, this also includes forces which are less under experimental control as they arise from defects or skyrmion-skyrmion interactions. Defects result in skyrmion pinning \cite{Schulz12,2015_Mueller_PRB,2018_Fernandes_NatComm}, strongly affect skyrmion motion and also skyrmion creation and destruction. As skyrmions are very smooth topological textures, pinning forces from local defects are often suppressed but such a mechanism does not exist for defects which have a length scale similar to the skyrmion size, arising, e.g., from surface roughness of magnetic layers.

\begin{figure}
\center 
\includegraphics[width=0.48 \textwidth]{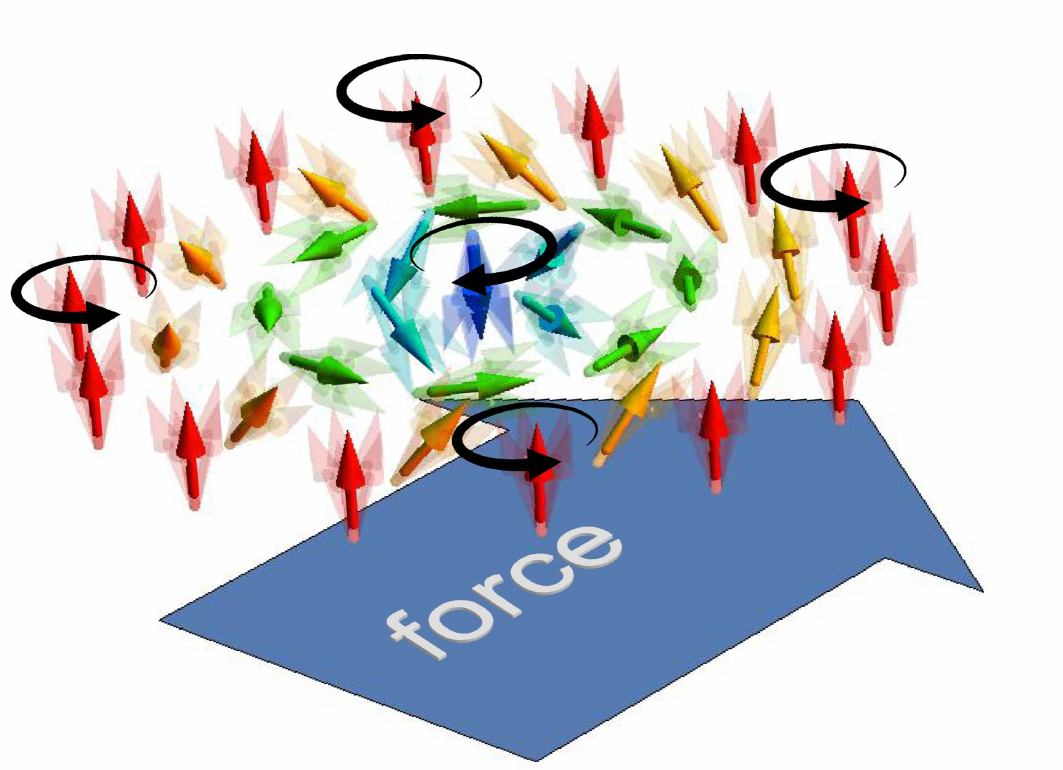}
\caption{When a force is applied to a skyrmion, internal degrees of freedom are excited. These internal excitations lead to retardation effects which can be described as an effective mass of the skyrmion. Different forces lead to different types of deformations and therefore to different values of the effective mass. The figure is taken from Ref.~\cite{2015_Mueller_PRB}. \label{fig1}}
\end{figure}

A widely discussed question concerns the effective mass of the skyrmion arising when the corrections $\delta \vec M$ to the rigid skyrmion picture are taken into account. The effective skyrmion mass $M_s$ leads to an inertial force of the form $M_s \frac{d^2 \vec R}{d t^2}$ and can be used to parametrize corrections to Eq.~(\ref{dynamics}). There is a large disagreement in the current literature on the value of the effective skyrmion mass. Some recent publications claim that the mass is exactly zero \cite{2018_Kravchuk_PRB}, others find large intertial effects both in experiments and in simulations of skyrmion dynamics \cite{2012_Makhfudz_PRL}. As has been shown in Ref.~\cite{2014_Schuette_PRB}, there is not a single effective mass, but different apparent effective masses should be considered depending on the type of force which is applied. As the mass ultimately arises from deformations of the moving skyrmion, see Fig.~\ref{fig1}, one obtains different values whether one considers skyrmions in nanostructures \cite{2012_Makhfudz_PRL,2017_Loss_PRX}, driven by field gradients, or driven by spin-torques. For the response to some  hypothetical force which does not deform the skyrmion, the effective mass vanishes, while large values scaling with the skyrmion size are obtained when the response to thermal fluctuations \cite{2014_Schuette_PRB} is considered, arguably the most generic way to define an effective mass. From a more general point of view, one should not only consider the effective mass but the full frequency dependent response \cite{2014_Schuette_PRB} to external forces, especially as considering forces proportional to $\frac{d \vec R}{d t}$ (gyrocoupling and friction) and $\frac{d^2 \vec R}{d t^2}$ (mass and 'gyrodamping') only turns out to violates basic causality relations \cite{2014_Schuette_PRB}.

An effective description of a skyrmion in terms of its coordinate ignores the role of internal degrees of freedom of the skyrmion. These internal degrees of freedom can, however, be used to manipulate the skyrmion and often become important under 'extreme conditions', when the skyrmion is subject to strong forces, or is close to an instability. In these cases, one should generalize the Thiele approach to include extra degrees of freedom, see, e.g., Refs.~\cite{2018_Ritzmann_NatElec}. In frustrated magnets with weak dipolar or spin-orbit interaction, for example, the motion of skyrmions can excite the internal precession of spins which in turn affects the skyrmion coordinate \cite{2018_Ritzmann_NatElec}. Here a promising direction is to actively use the internal excitations to control the skymion motion. An oscillating magnetic field can, e.g., induce via a ratchet-like mechanism a form which leads to a motion of skyrmions \cite{2016_Moon_SciRep,2018_Psadouraki}. Another opportunity is provided by the recent observation of skyrmions in inversion symmetric systems \cite{2019_Kurumaji_Sience}. In the presence of inversion symmetry, skyrmions with opposite helicity, i.e., left-handed and right-handed skyrmions are exactly degenerate and one can use this internal degree to store information \cite{Okubo2012,Leonov2015}. For ultrasmall skyrmions in magnetic insulators also quantum effects may become important and it is possible to entangle the motional degree of freedom with internal degrees of freedom \cite{2019_Lohani}. Especially interesting in this context is the quantum tunneling of skyrmions to antiskyrmions in frustrated magnets which can delocalize quantum skyrmions \cite{2019_Lohani}.

Another very interesting and largely unresolved question (beyond the scope of the discussion of this section) is to what extent skyrmion lattices and, e.g., their melting transition can be described by particle models and (two-body) skyrmion-skyrmion interactions.

%\subsection{Advances in science and technology to meet challenges}
%his section discusses the advances in science technology needed to address the challenges. (350 words max)

\subsection{Advances in science and technology to meet
challenges}
To unravel the forces governing skyrmion dynamics is a challenge which can only be met by careful experiments and their comparison with theory. Here the main difficulty will be to distinguish between intrinsic forces (e.g., due to scattering from thermal magnons) and extrinsic forces (arising, e.g., from defects). Ideally, the question can be studied in ultraclean systems with single skyrmions whose positions can be measured, e.g., by electron microscopy or other techniques to image skyrmions. Equally important are systematic experimental and theoretical studies on the pinning of skyrmions both by single defects and a finite density of defects. In both cases the effects of thermal fluctuations will be important. On the theoretical side, it will be interesting to use field-theoretical techniques to investigate which thermal effects can and which cannot be described by numerical simulations of stochastic Landau-Lifshitz Gilbert equations. A classical description may fail in situations where classical physics predicts that modes with frequencies larger than $k_B T/\hbar$ contribute (which, for example, is the case for the thermal blackbody radiation).

\subsection{Concluding remarks}
Treating a skyrmion as a point particle is an extremely useful concept and the basis for most theories discussing the dynamics of skyrmions. Obtaining a quantitative understanding of the forces arising from damping, deformations of the skyrmion, from currents and, especially, short and long ranged defects has to be a central goal of the field of skyrmionics. Internal degrees of freedom provide further opportunities to manipulate and control skyrmions and to realize new types of functionalities.

\subsection{Acknowledgements}
Support of the DFG within the priority program on skyrmionics (SPP2137) and the CRC 1238 (project C04) is gratefully acknowledged. A. R. would also like to thank the Department of Physics at Harvard for hospitality and M. Garst, J. Iwasaki, V. Lohani, J. Masell, C. Sch\"utte, and Naoto Nagaosa for useful discussions.

%%%%%%%%%%%%%%%%%%%%%%%%%%%%%%%%%%%%%%%%%%%%%%%%%
%%%%%%%%%%%%%%%%%%%%%%%%%%%%%%%%%%%%%%%%%%%%%%%%%
%%%%%%%%%%%%%%%%%%%%%%%%%%%%%%%%%%%%%%%%%%%%%%%%%

\clearpage
\newpage

\section[Investigations of skyrmion systems using density functional theory \\ {\normalfont Sergiy Mankovsky and Hubert Ebert}]{Investigations of skyrmion systems using density functional theory}
\label{mankovsky}
{\it Sergiy Mankovsky and Hubert Ebert}

LMU Munich, Department of Chemistry,
Butenandtstrasse 11, D-81377 Munich, Germany

% %%%%%%%%%%%%%%%%%%%%%%%%%%%%%%%%%%%%%%%%%%%%%%%%%%%%%%%%%%%%%%%%%%%%%%%%%%%%%%
\subsection{Status}
% %%%%%%%%%%%%%%%%%%%%%%%%%%%%%%%%%%%%%%%%%%%%%%%%%%%%%%%%%%%%%%%%%%%%%%%%%%%%%%

Theoretical investigations on skyrmionic systems using first-principles
electronic structure methods on the basis of density functional theory
(DFT) already contributed in an indispensable way to the insight
concerning their formation, stability and properties. In particular,
dealing with realistic materials, DFT-based theories provide a direct
parameter-free connection to experiment, giving access to the
material-dependent physics behind the observed effects, and support this
way the development of new materials possessing desired properties.

On the one hand side, reliable computational schemes are now available to calculate the necessary input parameters for subsequent simulations using a variety of statistical methods. Inclusion of spin-orbit coupling (SOC) within the underlying electronic structure calculations allow not only to calculate the isotropic exchange coupling parameters entering the classical atomistic Heisenberg Hamiltonian but also the magnetic
anisotropy and Dzyaloshinskii-Moriya interaction (DMI) parameters of its
extended version. The same holds for the corresponding coarse grained
micro-magnetic parameters entering the Landau-Ginsburg free energy used
as an effective Hamiltonian for the continuous magnetization field
$\vec{m}(\vec{r})$. This way a firm basis e.g.\ for subsequent Monte
Carlo simulations is provided that lead in particular to realistic phase
diagrams of two- and three-dimensional
systems \cite{RSP+16}. Spin-dynamics simulations 
on the basis of the Landau-Lifshitz-Gilbert (LLG) equation in addition
have to account for dissipation processes represented by the Gilbert
damping parameter. Again numerical schemes have been developed during
the last decade for its reliable calculation \cite{EMKK11}. 

Due to various developments made in related fields during the last two
decades first-principles methods not only provide indirectly to the
understanding of the formation and stability of skyrmionic systems but
also of their physical properties that are of central importance for any
type of application. The current driven Skyrmion dynamics for example
depends on the current induced spin-orbit torque (SOT) and spin transfer
torque (STT) as well as the spin Hall effect (SHE). Efficient and
reliable schemes to calculate the linear response quantities describing
these and related phenomena as for example the anomalous Hall effect
(AHE) for suitable reference systems are available
nowadays. This holds for any intrinsic as well as extrinsic 
contribution to these  response quantities \cite{LKE10}. Corresponding calculations with and without SOC
allowed in 
particular to decompose the response quantity into its SOC- and topology
related contributions with the later one connected to a non-collinear
spin configuration \cite{HWD+15}. 
In this context, one has to stress that
with the Berry phase and corresponding curvature a powerful concept 
is available that links the  intrinsic contribution of the 
various response quantities 
among each other as well as to the
underlying electronic structure \cite{Freimuth14}.   

Because of the large size of a single skyrmion most first-principles
investigations had to be restricted so far to suitable reference
systems. Only in few cases very small two-dimensional skyrmions have
been investigated up to now using a periodic super-cell or embedding
technique, respectively.

% %%%%%%%%%%%%%%%%%%%%%%%%%%%%%%%%%%%%%%%%%%%%%%%%%%%%%%%%%%%%%%%%%%%%%%%%%%%%%%
\subsection{Current and future challenges}
% %%%%%%%%%%%%%%%%%%%%%%%%%%%%%%%%%%%%%%%%%%%%%%%%%%%%%%%%%%%%%%%%%%%%%%%%%%%%%%

In spite of the progress made concerning the first-principles calculation of 
the various parameters needed for subsequent statistical simulations there are
still a number of limitations to be removed. Concerning the exchange
coupling including the DMI it was already demonstrated by various authors
 that it is in
general important to go beyond the standard bilinear pair interaction expansion
that is often restricted to nearest neighbors to obtain accurate results for
skyrmionic systems.
Moreover, the extension of  the Heisenberg Hamiltonian by adding higher-order
multi-spin interaction terms can lead to new physics,
e.g.\ to the stabilization of a skyrmion lattice without an external magnetic
field \cite{refHEINZE}. 
As these parameters
are always calculated for a  reference system -- often in a
ferromagnetic configuration -- it is important to have the reference
system as close as possible to the situation of the real skyrmionic
system of interest. This implies in particular that the non-collinear
spin structure as well as the impact of a finite temperature should be
accounted for when calculating interaction parameters. This applies the 
same way for the Gilbert damping parameter \cite{EMC+15}.

Another great challenge is the realistic calculation of response properties.
Concerning chirality-driven transport properties for example, one should stress
the topological contribution; e.g.\ the topological Hall effect
(THE) in case of the AHE. Contributions of that type  may be rather pronounced 
as it was known from experiment and could be shown by first-principles
calculations for artificial nanoskyrmion lattice \cite{HWD+15}. 
This study reflects the size gap as a central problem
when applying first-principles methods to skyrmionic
systems; i.e.\ the difference between the 
large size of a real single skyrmion and the number of atomic sites
that can be handled on a computer using super-cell or embedding techniques. 
This applies in particular
when calculating response quantities as the torquance or spin
conductivity related to SOT and SHE, respectively.
Again inclusion of
temperature and non-collinearity within such calculations is an
important issue.

When dealing with spatially very extended skyrmions the rather slowly varying spin modulation justifies certain simplifications. In particular the 
magnetic texture may be described in terms of emergent electrodynamics 
(see section 'Emergent Electrodynamics' by K. Everschor-Sitte) allowing to calculate the linear response quantities for a collinear magnetic state in the presence of an effective 'emergent' magnetic field. Obviously, the applicability of this approach should be further explored.

% %%%%%%%%%%%%%%%%%%%%%%%%%%%%%%%%%%%%%%%%%%%%%%%%%%%%%%%%%%%%%%%%%%%%%%%%%%%%%%
\subsection{Advances in science and technology to meet challenges}
% %%%%%%%%%%%%%%%%%%%%%%%%%%%%%%%%%%%%%%%%%%%%%%%%%%%%%%%%%%%%%%%%%%%%%%%%%%%%%%

Concerning the necessary extension of the Heisenberg Hamiltonian by higher order exchange interaction parameters various  first-principles 
schemes have been suggested, focusing in particular on the chiral interaction
terms. These are the so-called chiral biquadratic \cite{BDL19} as well
as chiral three-spin and four-spin \cite{LRP+19,MPE19b} interactions
that can be accounted for in future simulations.

\begin{figure}[h]
\includegraphics[width=1.0 \linewidth]{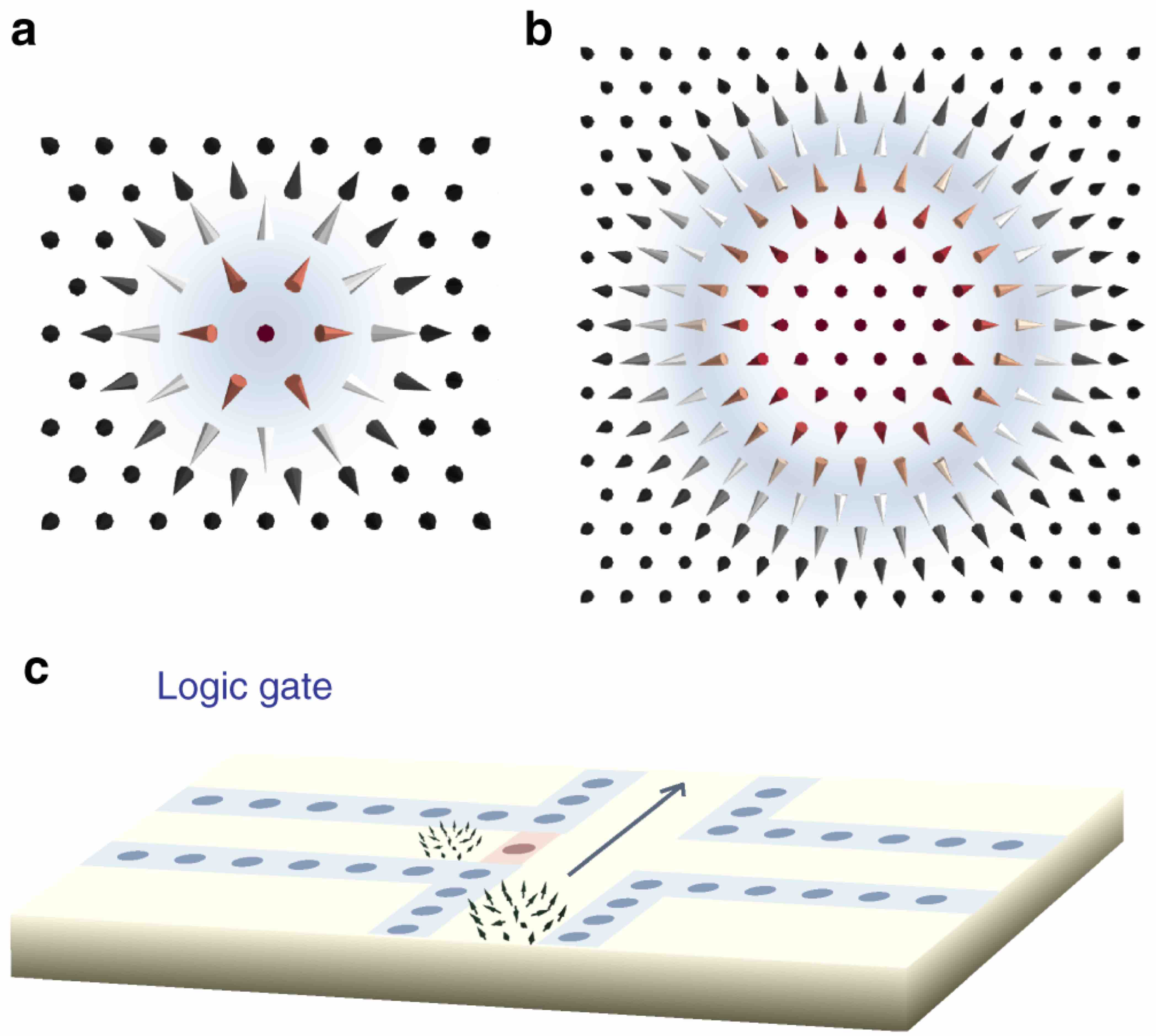}
\caption{\label{fig:DFT} Spintronic devices with defects guiding the
 motion of a skyrmion. (a, b) Non-collinear regions (blue area) defining
  pinning or repulsion depending on the chemical nature of the
  impurities. A pinning impurity can lock a skyrmion within the blue
  region while a repulsive impurity can pin a large skyrmion on the core
  region, which is a local minimum energy. (c) Pinning (red) usually
  hinders skyrmion mobility and can be used to decelerate them. 
  Figures reproduced from Ref.\,\cite{2018_Fernandes_NatComm}.}
\end{figure}

To achieve at detailed description of skyrmionic systems of realistic size
that may include in addition certain aspects of a device on a first-principles
level the embedding technique seems to be most appropriate. 
In fact the first example to be found in the literature that investigates the
influence of neighboring impurities acting as pinning or repulsive
centers seems to be very promising \cite{2018_Fernandes_NatComm} (see Fig.\ \ref{fig:DFT}).
%
% ffffffffffffffffffffffffffffffffffffffffffffffffffffffffffffffffffffffffffffff
% ffffffffffffffffffffffffffffffffffffffffffffffffffffffffffffffffffffffffffffff
%
Nevertheless, the underlying techniques have to be 
further optimized to close the mentioned size gap.

To include the impact of thermal phonons and spin fluctuations
explicitly e.g.\ in terms of a corresponding realistic self-energy seems
to be too demanding.  
The use of an adequate effective medium determined on the basis of the 
alloy analogy model that accounts in particular for the 
effective finite life time of an electronic state proved to be
sufficiently accurate for bulk systems  \cite{EMKK11}.  
Combination of this technique with the embedding technique will allow to
account for the impact of finite temperature effects 
when calculating exchange coupling parameters as well as response
quantities in real space.

To calculate for example the THE for skyrmion systems via the emergent field approach, the  diagonal and ordinary Hall conductivities have to be calculated  for different spin channels in the presence of the effective magnetic field associated with the skyrmion magnetic texture. This gives access to the topological Hall resistivity via the expression
$\rho^T_{yx} = \frac{\sigma^{\uparrow}_{xy} - \sigma^{\downarrow}_{xy}}{
  (\sigma_{xx} + \sigma_{yy})^2}  $ \cite{GFS+15}. 
This approach that gives very valuable insights into the mechanism behind the topological contributions should be complemented by direct calculations of the response quantities.   

To account for the impact of a non-collinear magnetic structure on the Gilbert damping, it may be expressed in terms of a wave vector dependent tensor $\underline{\alpha} (\vec{q})$ expanded in powers of the wave-vector $\vec{q}$;  assuming  again a slow modulation of the magnetization. The calculation of the  expansion coefficients done within the framework of linear response theory \cite{MWE18} should again be complemented be studies on realistic skyrmion structures eventually done in real space.

% %%%%%%%%%%%%%%%%%%%%%%%%%%%%%%%%%%%%%%%%%%%%%%%%%%%%%%%%%%%%%%%%%%%%%%%%%%%%%%
\subsection{Concluding remarks}
% %%%%%%%%%%%%%%%%%%%%%%%%%%%%%%%%%%%%%%%%%%%%%%%%%%%%%%%%%%%%%%%%%%%%%%%%%%%%%%

Due to many fundamental as well as technical developments made during the last two decades DFT based investigations on skyrmionic systems will surely provide a more and more detailed insight into the properties of these fascinating systems. In particular the handling of large system sizes should allow in the future to investigate realistic structures including device aspects relevant for applications.  The full dynamical behavior of such systems, on the other hand, will still have to be described using complementary coarse grained and statistical methods.  Corresponding work, however, can be based now on system parameters calculated from first-principles accounting for all relevant aspects including spin non-collinearity as well as finite temperature. 

\subsection{Acknowledgements}

Financial support by the DFG via SFB 1277 (Emergent Relativistic Effects in Condensed Matter - From Fundamental Aspects to Electronic Functionality) is gratefully acknowledged.

%%%%%%%%%%%%%%%%%%%%%%%%%%%%%%%%%%%%%%%%%%%%%%%%%
%%%%%%%%%%%%%%%%%%%%%%%%%%%%%%%%%%%%%%%%%%%%%%%%%
%%%%%%%%%%%%%%%%%%%%%%%%%%%%%%%%%%%%%%%%%%%%%%%%%

\clearpage
\newpage

\section[Spintronics with skyrmions - towards devices \\ {\normalfont Tianping Ma and Stuart S.P. Parkin}]{Spintronics with skyrmions - towards devices}
\label{ma}
{\it Tianping Ma and Stuart S.P. Parkin}

Max Planck Institute for Microstructure Physics, Halle (Saale), Germany

% %%%%%%%%%%%%%%%%%%%%%%%%%%%%%%%%%%%%%%%%%%%%%%%%%%%%%%%%%%%%%%%%%%%%%%%%%%%%%%
\subsection{Status}
% %%%%%%%%%%%%%%%%%%%%%%%%%%%%%%%%%%%%%%%%%%%%%%%%%%%%%%%%%%%%%%%%%%%%%%%%%%%%%%

Today spintronic devices play a major role in two distinct technologies for digital data storage \cite{2003_Parkin}.  On the one hand spin-valve magnetic recording read heads have allowed for massive increases in the storage capacity of magnetic hard disk drives (HDDs)\cite{parkin1995giant}.  On the other hand, spintronics offers the possibility of solid-state memories that have no moving mechanical parts. One of these, magnetic random access memory (MRAM), is a highly interesting memory technology that stores data as the direction of magnetization of a magnetic element\cite{1999_Parkin}. 

The magnetic bits typically form one magnetic electrode of a magnetic tunnel junction (MTJ) that is composed of two magnetic electrodes separated by an ultra-thin tunneling barrier.  The resistance of such an MTJ depends on the relative magnetic orientations of the electrodes, whether the magnetizations are parallel or anti-parallel. The development of MRAM based on magnetic tunnel junctions was made possible by the discovery of very large room temperature tunneling magnetoresistance in devices that have body centered cubic CoFe or CoFeB electrodes separated by magnesium oxide (MgO) tunnel barriers \cite{2004_Parkin} and the concept of spin transfer torque that enables switching of the state of the magnetic elements by using spin angular momentum derived torques from current passed through the elements \cite{1996_Slonczewski}.  Although the original concept of MRAM using MTJs dates from 1995 when DARPA funded a program to prove the possibility of such a concept and that the fundamental principle was realized in 1999 \cite{1999_Parkin}, it has taken more than twenty years to realize MRAM with necessary properties for widespread applications beyond niche applications for which small scale MRAM chips with tiny capacities were earlier developed.  

This history of the development of MRAM is typical of novel technological concepts, especially those that involve the use of new materials and physics, that can take several decades for their fruition. Turning to skyrmions, the subject of this article, the use of skyrmions for data storage clearly has its antecedents in Magnetic Racetrack Memory, a novel memory technology proposed by one of the authors in 2002  \cite{2004_Parkin_patent,2008_Parkin,2015_Parkin}. Unlike any other solid-state memory device under consideration today, this memory shifts magnetic data that is encoded in chiral magnetic domain walls, back and forth along magnetic nanowires, the magnetic racetracks.  By storing many domain walls -- perhaps a 100 or more -- in a single racetrack a very dense memory is possible.  The potential is for a solid-state non-volatile memory device that could have storage capacities comparable to magnetic disk drives but that would have much greater performance, use much less energy and that would be much more compact.

The essential physics that underlies Racetrack Memory is that a series of magnetic domain walls can be moved together in the same direction along the racetrack when current is applied. The first demonstration of this principle in 2008 \cite{2007_Hayashi} used current that was spin-polarized from spin-dependent scattering within the magnetic material of the racetrack itself. The maximum velocity of the domain walls was ~100 m/s and was limited by the maximum current density that did not result in significant heating of the nanowire.  Similar velocities from such a volume spin transfer torque (STT) mechanism were later found in nanowires formed from magnetic multilayers of ultra-thin Co and Ni layers that exhibit perpendicular magnetic anisotropy (PMA) so that the domain wall (DW) width can be reduced to just a few nanometers, thereby achieving high storage density.

Over the past 5 years or so there have been some extraordinary advances in the physics of the current induced motion of PMA domain walls (DWs) via spin-orbit derived torques.  Two essential ingredients are the formation of chiral N{\'e}el domain walls via an interface Dzyaloshinskii-Moriya Exchange Interaction (DMI) and spin currents generated by the spin Hall effect in metallic layers adjacent to the magnetic layers that form the racetrack.  The spin-orbit torques are much more efficient than volume STT torques and allow for chiral DWs to be moved at several hundred m/s \cite{2013_Ryu}, as shown in Fig. 12(a).  The final major development was the discovery of a giant exchange torque in synthetic antiferromagnetic (SAF) racetracks \cite{2015_Yang}, as shown in Fig. 12(b).  Such SAF structures (or, also termed artificial antiferromagnets,) solve a major problem that plagues nearly all magnetic nano-devices, namely the magnetostatic fields that they generate\cite{parkin1995giant}. In addition to overcoming interactions between DWs within and between racetracks, SAF racetracks, the antiferromagnetic coupling (AF) exchange field within the SAF also gives rise to a new mechanism to drive the domain walls so that DW velocities of more than ~1 km/s are possible \cite{2015_Yang}. Research into the current induced motion of domain walls, especially with regard to its use in racetrack memory, has a history dating back more than a decade and is now well established. Thus, Racetrack Momory has now entered into the industrial application phase, where important matters especially relating to scaling to tiny dimensions, reliability and reproducibility of DW motion, and thermal stability of DWs will be explored. The nearly identical comcept of using skyrmions or anti-skyrmions as the memory-storage element in racetracks means that progress on such concept can be accelerated.

% %%%%%%%%%%%%%%%%%%%%%%%%%%%%%%%%%%%%%%%%%%%%%%%%%%%%%%%%%%%%%%%%%%%%%%%%%%%%%%
\subsection{Current and future challenges}
% %%%%%%%%%%%%%%%%%%%%%%%%%%%%%%%%%%%%%%%%%%%%%%%%%%%%%%%%%%%%%%%%%%%%%%%%%%%%%%

There have been several proposals to replace the DWs in a racetrack with skyrmions or other related objects, some of which are discussed elsewhere in this review. For example, the current induced motion of individual skyrmions has been demonstrated in racetracks that are nearly identical in structure to those discussed above with regard to chiral domain walls.  A typical structure is an ultra-thin CoFeB ferromagnetic layer coupled to a heavy metal such as Pt or W: the interface generates a DMI as well as providing PMA, as shown in Fig. 12(c).  The interfacial nature means that the magnetic layer must be very thin. Simplistically, the skyrmions that are observed in such systems can be considered as a pair of chiral domain walls and the current driving mechanisms are very similar to those of the individual DWs except that the topological charge of the skyrmion causes them to move partly sideways perpendicular to the current flow direction.  This makes their use in racetracks challenging.

Interest in skyrmions for racetrack memory data bits was originally based on the observation that bulk skyrmions in the single crystalline B20 compound MnSi could be moved with very low current densities \cite{Jonietz10}, orders of magnitude below the critical current density above which chiral DWs can be moved with spin-orbit torques.  It is an inevitable consequence that these very low current densities also correspond to very slow motion of the skyrmions.  For some applications of racetrack memory, for example, to replace vertical NAND flash memories, small current densities may be advantageous and the consequent slow motion of the magnetic bits might be acceptable.  In any case the ability to be able to vary performance and to trade off energy with speed in racetrack devices is very attractive.

As discussed in Tokura et al several types of bulk skyrmions, both Bloch and Néel, have been discovered in various materials as well as, more recently, anti-skyrmions in an inverse tetragonal Heusler compound \cite{Nayak17}.  Each of these objects has distinct properties that could make them of interest for various device applications.  Anti-skyrmions are of special interest in that they are innately stable spin textures \cite{2019_Saha}.   To date no current induced motion of anti-skyrmions has been demonstrated but, theoretically, the motion under spin-orbit torques is more complex than for skrymions. The latter move sideways when current is applied, as mentioned above, but anti-skyrmions are predicted to move in straight lines or to the left or to the right when excited by spin-orbit torques, depending on the current direction relative to the complex wall structure of the anti-skyrmion.  Whilst this is potentially of use this means that if the racetrack is formed from polycrystalline material the motion of an anti-skyrmion in each grain could be in different directions. Moreover, it is not clear that there wouldn't be a considerable barrier to the passage of the anti-skyrmion from one grain to the next since the anti-skyrmion's structure is directly tied to the underlying crystal lattice. This is a very interesting question and a possible challenge to the use of anti-skrymions for racetrack memory devices. The recent discovery that anti-skyrmion co-exist or can be transformed into a bloch like skyrmions\cite{elliptical_parkin,peng2020controlled} makes them even more interesting for device applications.

A major challenge is to scale racetracks to dimensions that would make them technologically relevant (this is also a challenge for DW based racetrack devices). This means exploring the motion of skyrmions in racetracks that are perhaps 20-50 nm wide and where the skyrmions themselves are just 20 nm in diameter and spaced similarly. There are some recent experiments which demonstrated the current induced motion of skyrmions are, however, comparatively large (>100 nm)\cite{Litzius17,woo2018current,klaui,hrabec}. To make applications of skyrmions possible, much more progress is needed in decreasing their size. This is clearly not easy and their detection is much more difficult than chiral DWs for several reasons.  Firstly, DWs are straight and secondly, the domains with opposite magnetizations between the DWs can be detected, whereas for skyrmions, their boundaries are curved and the difference in magnetization between their interior and their exterior is much smaller. The detection of skyrmions is likely possible via integrated magnetic tunnel junctions but it is not easy to fabricate such magnetic tunnel junctions nor to create them with sufficient signal to differentiate individual skrymions.

It is clear that zero moment skyrmions akin to SAF chiral domain walls are needed in order to avoid interactions between the skyrmions via dipole fields.  For skyrmion bubbles it is, of course, obvious that the concept of the SAF racetrack is directly transferable by simply replacing an SAF DW with an SAF skyrmion, as shown in Fig. 12(d).  For bulk skyrmions, similar concepts are possible. Another possibility is to tune the magnetization in the host compound to zero: this would be possible, for example, in ferrimagnetic host materials, such as inverse tetragonal Heusler compounds that host anti-skyrmions \cite{2010_Nayak}. It was shown possible some time ago to switch the magnetization of a ferrimagnetic layer whose magnetization is zero by current by relying on sub-lattices with distinct interactions with the current \cite{2006_Jiang}.  Recently it has similarly been shown that domain walls can be moved by current even when the magnetization of a ferrimagnetic racetrack formed from a Co-Gd amorphous alloy is zero \cite{2018_Blaesing}: indeed the domain wall velocity was shown to be largest not when the magnetic moments of the Co and Gd sub-lattices are equal (and opposite) but rather when their angular momenta are exactly compensated. Much progress has been made on this topic recently\cite{woo2018current,legrand2020room}.

Other major challenges include understanding how to overcome pinning of skyrmions at defects or at edges of racetracks, and to ensure sufficient thermal stability of skyrmions at dimensions that are useful over long periods of time (10 years for useful devices).  Low current densities at which skyrmions move means that they may be unstable to thermal fluctuations so that mechanisms to trap the individual skyrmions are needed. Whilst it is straightforward to envisage trapping of skyrmions in large devices, for example, by forming notches or anti-notches, this is very difficult to achieve in ~20 nm sized objects.

%
% ffffffffffffffffffffffffffffffffffffffffffffffffffffffffffffffffffffffffffffff
\begin{figure}[h]
\includegraphics[width=1.0\linewidth]{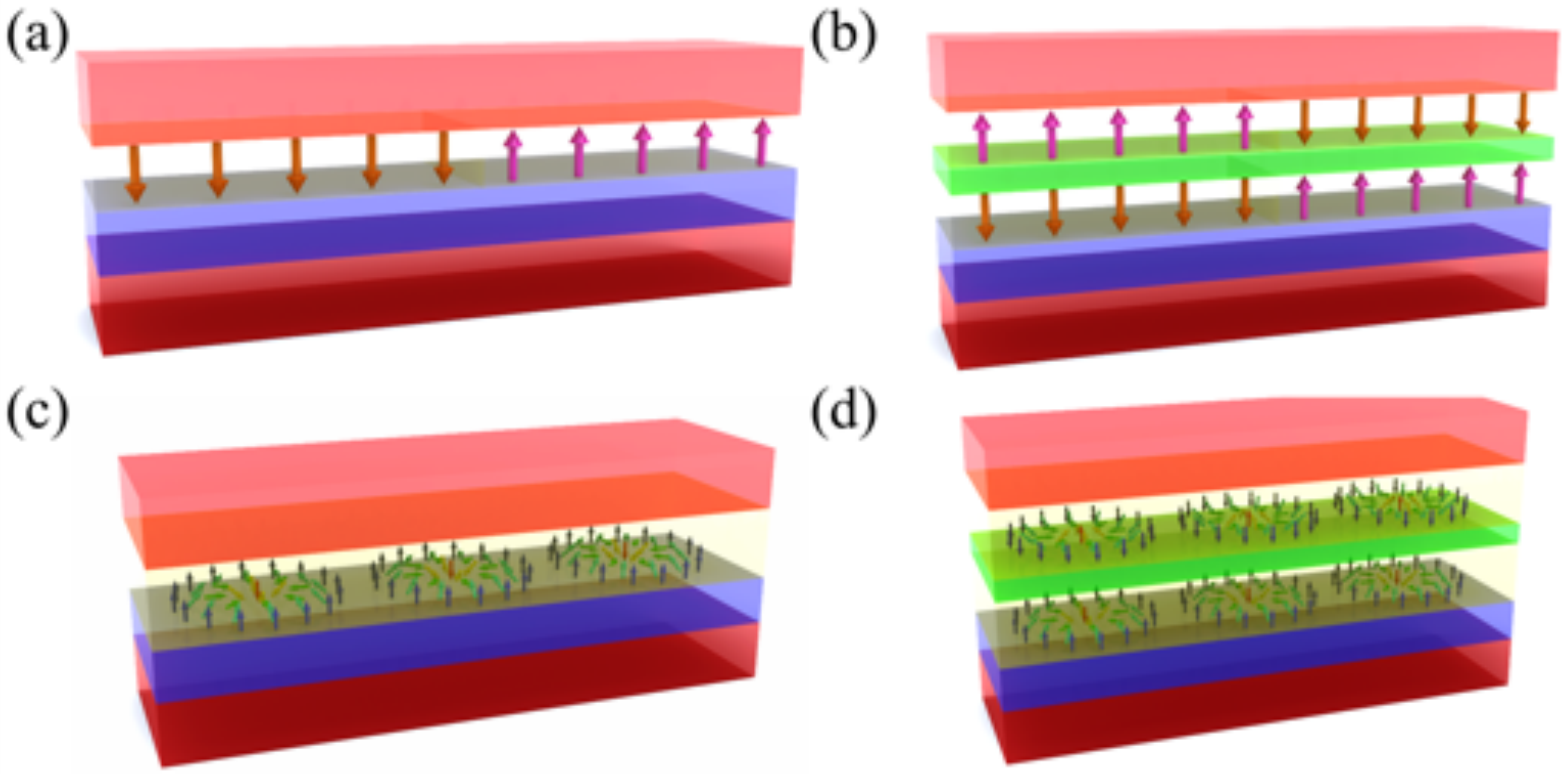}\;
\caption{\label{fig:Parkin1} Schematic diagram of (a) domain Wall in a single PMA ferromagnetic layer (SL) based racetrack device, (b) domain wall in a synthetic antiferromagnetic (SAF) PMA racetrack, (c) skyrmion in a SL layer and (d) skyrmion  in a SAF coupled system. Red indicates substrate and capping layer. Blue indicates heavy metal layer. Yellow indicates a ferromagnetic layer with PMA. Green layer indicates Ru spacer layer which gives rise to a RKKY interaction: its thickness is chosen to correspond to AF coupling \cite{1990_Parkin,1991_Parkin_PRB}.}
\end{figure}
% ffffffffffffffffffffffffffffffffffffffffffffffffffffffffffffffffffffffffffffff
%

%\includegraphics [width=0.8\linewidth]{_figures/Nagaosa-Fig2.eps}

% %%%%%%%%%%%%%%%%%%%%%%%%%%%%%%%%%%%%%%%%%%%%%%%%%%%%%%%%%%%%%%%%%%%%%%%%%%%%%%
\subsection{Advances in science and technology to meet challenges}
% %%%%%%%%%%%%%%%%%%%%%%%%%%%%%%%%%%%%%%%%%%%%%%%%%%%%%%%%%%%%%%%%%%%%%%%%%%%%%%

To date most studies on bulk skyrmions have focused on single crystals or on lamellae cut from these crystals by focused ion milling techniques.  Major advances in preparing thin films of these host compounds are needed.  It is very likely that films prepared on amorphous or polycrystalline layers will be needed: thus, the development of novel techniques, such as the recent development of chemical templating layers \cite{2018_Filippou} for preparing thin layers of Heusler compounds as thin as only one unit cell, are needed. 

Another technological area where advances are needed is in suitable computer architectures that can take full advantage of, especially, racetrack memories. Today's computer architectures are not used to the data itself being movable from one location to another.  This leads to very interesting questions for racetrack memories that would allow for the dynamic trade-off between density and performance by restricting or limiting the number of chiral domain walls or skyrmions in a single racetrack.

The use of accelerated materials discovery techniques to identify new materials that may host various types of skyrmions is very interesting.  Advances in calculating equilibrium and metastable magnetic ground states in complex compounds in which spin-orbit coupling effects are key are needed. 

Other areas of interest include 2D materials that can be exfoliated and 2D heterostructures that combine 2D magnetic layers with, for example, ferroelectric or anti-ferroelectric layers.  Engineered multi-ferroic structures could be important for creating and manipulating skyrmions by the use of electric fields rather than by currents.  In this regard the possibility of creating and manipulating skyrmions using optical means is very interesting but will require major advances in photonic engineering to allow on-chip light sources.

Skyrmion have proposed for several other applications including neuromorphic, magnonic, and logic devices\cite{zazvorka2019thermal,zhang2015magnetic}. These proposals are very much in their ''infant'' stages but provide fertile ground for invention.

% %%%%%%%%%%%%%%%%%%%%%%%%%%%%%%%%%%%%%%%%%%%%%%%%%%%%%%%%%%%%%%%%%%%%%%%%%%%%%%
\subsection{Concluding remarks}
% %%%%%%%%%%%%%%%%%%%%%%%%%%%%%%%%%%%%%%%%%%%%%%%%%%%%%%%%%%%%%%%%%%%%%%%%%%%%%%

The family of non collinear chiral nano-objects continues to expand with an increasing number of skyrmion and anti-skyrmion objects with distinct properties depending on the symmetry of the host material. These objects have extraordinary properties that make them of great scientific interest. The roadmap to technological applications is very challenging with the need to develop devices that are perhaps just 10-20 nm in size, in which the skyrmions can be both written and read, in which the skyrmions do not significantly interact with one another, and in which they are sufficiently thermally stable.  When these challenges are solved chiral non-collinear spin textures could herald the dawn of new all solid-state spintronic memory-storage devices that would have storage capacities, performance and reliability that go well beyond today's technologies.

\subsection{Acknowledgements}

We thank funding support from the European Research Council (ERC) under the European Union's Horizon 2020 research and innovation program (grant agreement No 670166) and the Deutsche Forschungsgemeinschaft (DFG, German Research Foundation) -- Project number 403505322.

%%%%%%%%%%%%%%%%%%%%%%%%%%%%%%%%%%%%%%%%%%%%%%%%%
%%%%%%%%%%%%%%%%%%%%%%%%%%%%%%%%%%%%%%%%%%%%%%%%%
%%%%%%%%%%%%%%%%%%%%%%%%%%%%%%%%%%%%%%%%%%%%%%%%%

\clearpage
\newpage

\section[Epitaxial thin films derived from bulk materials hosting skyrmions \\ {\normalfont Theodore L. Monchesky}]{Epitaxial thin films derived from bulk materials hosting skyrmions}
\label{monchesky}
{\it  Theodore L. Monchesky}

Dalhousie University

\subsection{Status}
%This section provides a brief history and status, why the field is still important, what will be gained with further advances. (350 words max)

In thin films, the Dzyaloshinskii-Moriya Interaction (DMI) that gives rise to skyrmions is produced either by broken inversion symmetry from the interface or by inversion asymmetry in the bulk of the film, or both.    Interfacial DMI enables the use of common centrosymmetric ferromagnets such as Co to be employed, see Sect.~\ref{fert}.
%({\it Skyrmions in multilayers and tailored heterostructures -- A. Fert et al.}). 
Such systems require relatively complex precious-metal multilayers, and their magnetic properties are averaged across the film stack.  On the other hand, chiral magnets that derive their DMI from bulk inversion asymmetry require only one single layer and may prove to be more resilient to skyrmion annihilation through sample edges. The major challenge with this second class of material is to find a way to prevent crystallographic grains with both left and right-handed chirality from forming when deposited on centrosymmetric substrates -- an essential requirement for applications such as racetrack memories, see Sect.~\ref{ma}.
%, see section ({\it Spintronics with skyrmions -- S. Parkin}).

Of all the currently available chiral magnetic films fabricated from materials that host skyrmions in bulk crystal, the vast majority of films are from the archetypal B20 family. The B20 silicides and germanides are well lattice matched to Si(111) and grow well on these technologically relevant substrates, but can also be grown epitaxially on MgO and SiC(0001).  Numerous examples exist in the literature, including MnSi, Fe$_x$Co$_{1-x}$Si, FeGe, Fe$_{1-x}$Co$_x$Ge, Fe$_{1-x}$Mn$_x$Ge and MnGe (see references in\cite{Meynell:2017prb} and \cite{Spencer:2018prb}).  High quality films on large area substrates create the possibility to explore the influence of anisotropy and finite size effects on chiral magnets.  Anisotropy provides an important mechanism for increasing the skyrmion stability, which is enhanced along the easy direction, or in an easy-plane, but is reduced along a hard-axis or in a hard-plane relative to the competing cone phase \cite{Wilson:2014prb}.

Finite size effects arise from the presence of DMI in the zero-torque boundary conditions, which creates surface states that decay into the bulk of the film on a length scale set by the helical wavelength, $L_D$ and the applied magnetic field \cite{Meynell:2014prb1}.  One-dimensional surface twists first were first observed in polarized neutron reflectometry measurements in the field induced ferromagnetic state of MnSi/Si(111) \cite{Wilson:2014prb}. These twists produce a confining potential for skyrmions that is crucial for devices, since film edges remove the skyrmions' topological protection \cite{Meynell:2014prb1}. Micromagnetic calculations show that the surface states produce a cross-over in the magnetic behaviour for film thicknesses below approximately $8 L_D$, where skyrmions are stabilized over a large portion of the phase diagram \cite{Rybakov:2016njp, Leonov:2016prl}.  This provides a natural definition for what is considered a thin film in the case of a chiral magnet.

\subsection{Current and future challenges}
%This section discusses the big research issues and challenges. (350 words max)

\begin{figure}
	\centering	
	\includegraphics[width=0.8
	\linewidth]{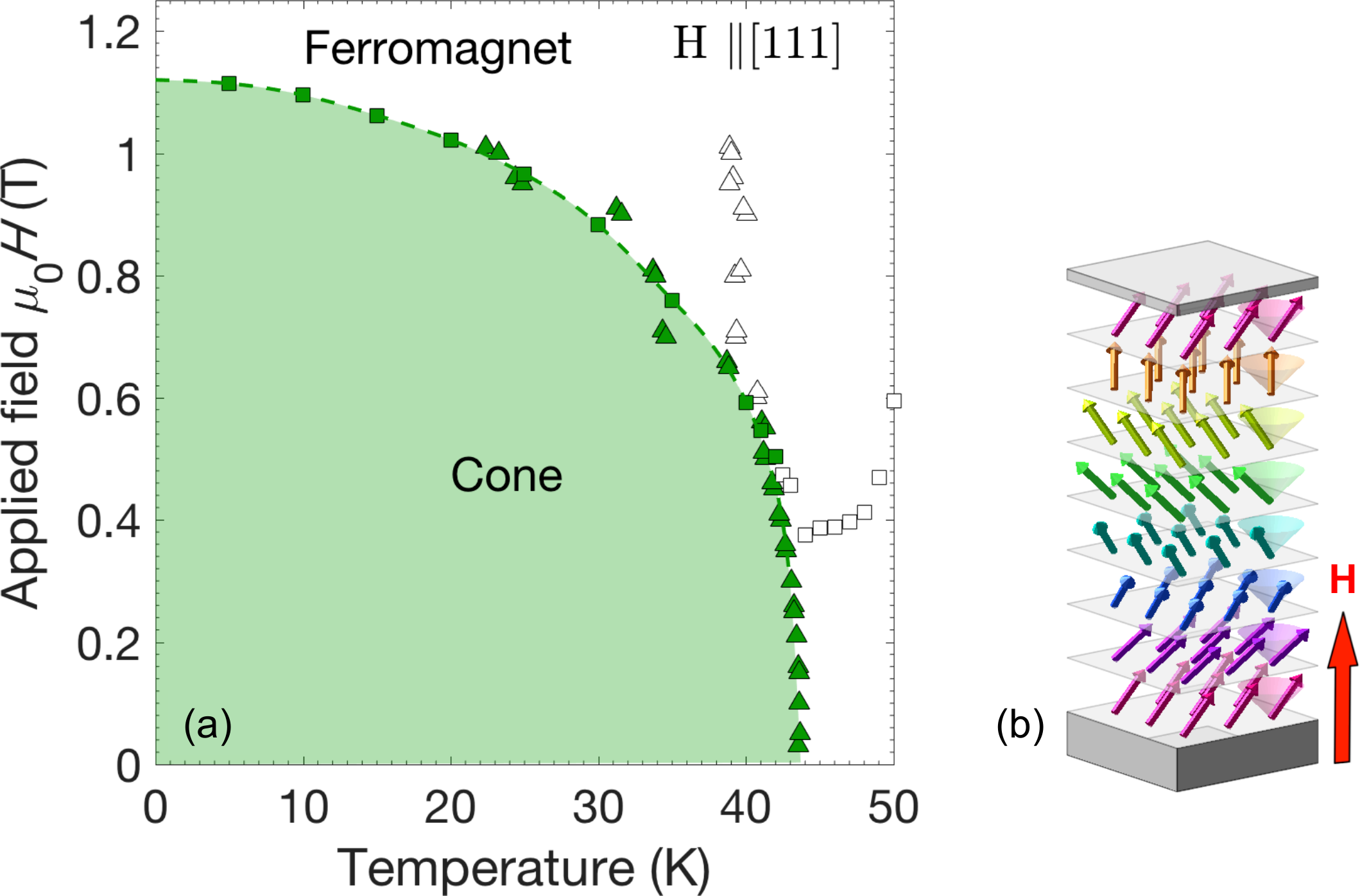}
	\caption{(a) Magnetic phase diagram of a 25.4 nm = $1.83 L_D$  thick MnSi(111) film grown on Si(111) derived from magnetometry measurements.  The magnetic field $H$ is applied along the film normal.  An illustration of the cone phase is shown in (b).  Adapted from Ref.~\cite{Wilson:2014prb}.}
	\label{fig:PD1}
\end{figure}

Despite numerous examples of high-quality films, there exists disagreement over the magnetic structure of these materials.  In the case of MnSi, Si(111) and SiC(0001) substrates create a hard-axis anisotropy that reduces the skyrmion stability relative to the cone phase along the axis.  Whereas the magnetometry measurements shown in Fig.~\ref{fig:PD1}, and Lorentz transmission electron microscopy measurements find the cone phase as the sole thermodynamically stable state along this axis, anomalous features appear in the Hall effect measurements in this region of the phase diagram.  Non-adiabatic spin transport through the cone phase provides an interpretation that reconciles these observations that is also consistent with bulk measurements (see references in \cite{Meynell:2014prb2}).  The observer Hall signal may be an example of the chiral Hall effect\cite{Lux:2019arXiv}. This proposal needs further theoretical and experimental studies that will be crucial for using the Hall effect to detect topological textures.  

Muon-spin rotation experiments add to the debate over the phase diagram  by demonstrating that the spin-textures are more complex than a simple cone phase for MnSi films in out-of-plane magnetic fields \cite{Lancaster:2016prb}.  This calls for detailed micromagnetic calculations of the magnetic structure at chiral grain boundaries.

\begin{figure}
	\centering
	\includegraphics[width=1.0\linewidth]{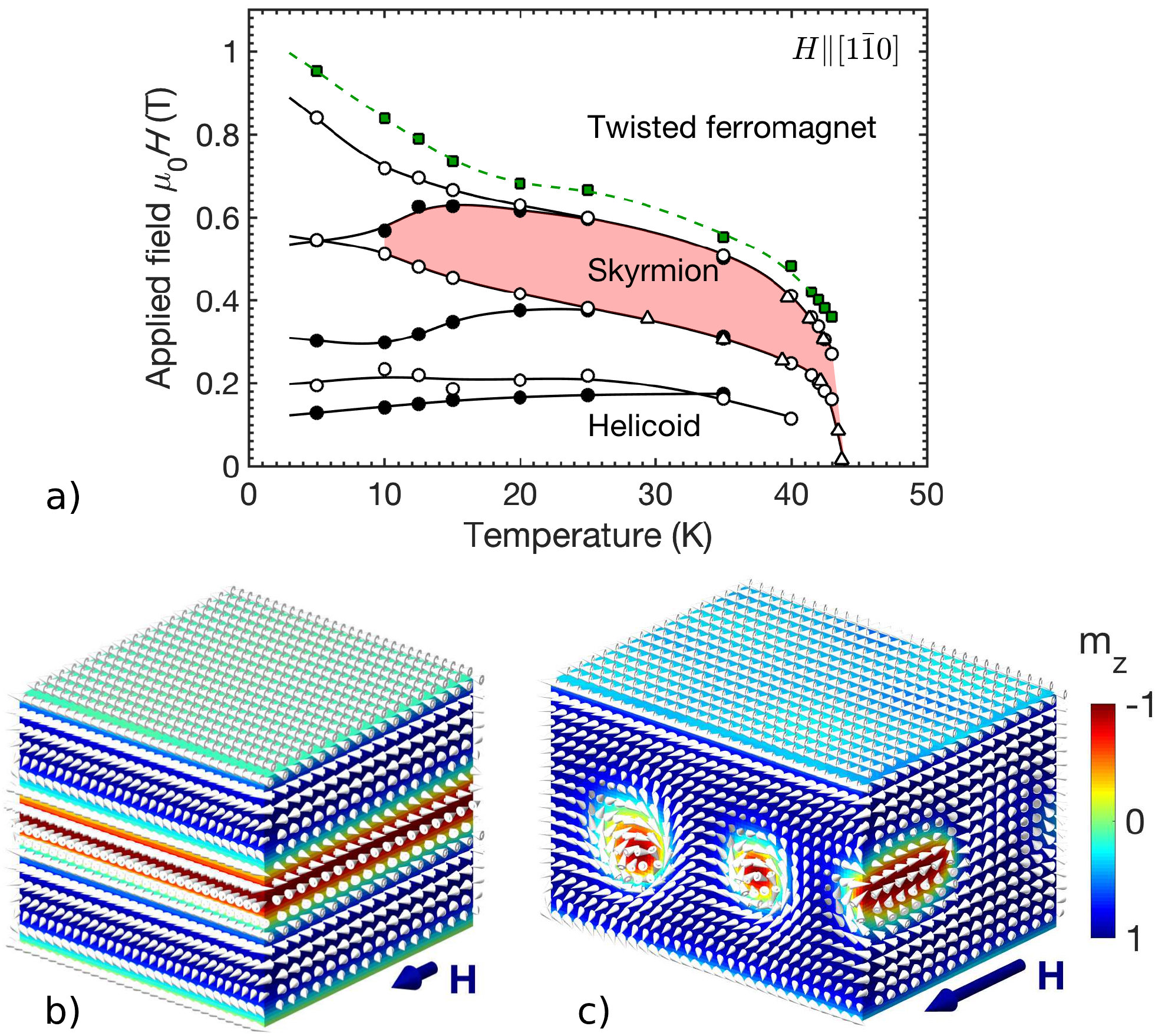}	
	\caption{(a) Magnetic phase diagram for a 26.7 nm = 1.92 $L_D$ thick MnSi(111) film grown on Si(111).  The magnetic field is applied in the plane of the film.  The phase boundaries shown by the circles are determined from peaks in the static magnetic susceptibility.  White (black) circles are for increasing (decreasing) magnetic fields.  The squares correspond to transitions determined from minima in $dM^2/dH^2$ for increasing fields. The magnetic phases are identified by PNR and SANS.  Micromagnetic calculations of the film show a helicoidal phase (b) and a disordered arrangement of torons (c).  The color bar shows the component of the magnetization along the field directions, $z$. Adapted from Ref.~\cite{Meynell:2017prb}.}
	\label{fig:PD2}
\end{figure}

For in-plane magnetic fields, polarized neutron-reflectometry finds low temperature discrete helicoidal states, and a highly disordered in-plane skyrmion state at higher temperatures, as shown in Fig.~\ref{fig:PD2}.  This is supported by small-angle neutron scattering (SANS) \cite{Meynell:2017prb} from a stack of 8 films.  The SANS data is explained by magnetic states consisting of short skyrmion tubes terminated by magnetic Bloch points\cite{Meynell:2017prb}, identified as torons\cite{Leonov:2018prb1}. However, evidence of skyrmions is not observed in a glancing-incidence SANS measurement of a single film \cite{Wiedemann:2017arxiv}.  Interpretation of the electron transport measurements are at odds with the phase diagram derived from neutron scattering \cite{Yokouchi:2015jpsj}, which calls for further theoretical investigations of the planar-Hall effect. 

Discrepancy also exists for FeGe. Although sputtered FeGe/Si(111) thin films show an easy axis anisotropy and thin films grown by molecular beam epitaxy show a hard axis anisotropy, both give a large anomalous Hall effect signal attributed to the presence of skyrmions (see reference in \cite{Spencer:2018prb}).  More surprising is the signal for Fe$_x$Co$_{1-x}$Ge that is much larger than expected theoretically \cite{Spencer:2018prb}. However, there are to date no reports of direct observations of skyrmions in either FeGe/Si(111) or FeGe/MgO(001): neutron scattering \cite{Kanazawa:2016prb} and resonant x-ray magnetic scattering \cite{Zhang:2017sr} find no evidence of a skyrmion lattice phase, and muon-spin rotation does not show signatures of complex magnetic textures in FeGe/MgO(001)\cite{Zhang:2017sr}.

To understand the conflicting reports of on B20 films, it will be important to systematically investigate the influence grain size on the magnetic phase diagram.  An additional theoretical challenge is to provide an account of the influence of chiral grain boundaries on the nucleation and stability of skyrmions.  The competing nucleation of torons from the film interfaces and the influence of surface anisotropies also needs further exploration.
 
\subsection{Advances in science and technology to meet challenges}
%This section discusses the advances in science technology needed to address the challenges. (350 words max)
Epitaxial strain is a parameter that could be valuable in resolving the controversy over the MnSi thin film system.  Although the SiC(0001) substrate was expected to reverse the sign of the uniaxial anisotropy in MnSi, this is not the case.   A suitable seed layer or substrate needs to be found that would produce an in-plane compressive strain in order to create out-of-plane skrymions that could be imaged with Lorentz microscopy.  

Further microscopy measurements are also needed in order to sort out the magnetic structure of the interfaces and the complex grain boundaries.  The recent advances in 3D magnetic tomography may be helpful to shed light on this difficult issue and reveal what influence surface states such as chiral bobbers  \cite{Rybakov:2016njp}  may play in the observed electron transport phenomena.

One of the immediate challenges for chiral epitaxial films is the stabilization of samples with a single chirality.  To address the influence of strain and misfit dislocations, B20 films could be grown on bulk B20 single crystals. A large range of lattice mismatch is in principle possible in the large B20 family.  For applications, however, the skyrmion materials need to be grown on common semiconducting substrates.  The fact that the most commonly available substrates are centrosymmetric poses a significant challenge: an enantiomorphic grain structure in the films is unavoidable without some additional symmetry breaking at the substrate interface, such as with a chiral surfactant or perhaps carefully prepared vicinal surfaces.

The discovery of skyrmions in other crystallographic points groups creates new opportunities to address the challenges in epitaxial systems.  Antiskyrmions are observed in bulk crystals of Mn$_{1.4}$PtSn ($D_{2d}$ point group), and recently have been grown in thin film form on MgO substrates \cite{Swekis:2019prm}. This inverse Heusler alloy may be only the first of a large class of new skyrmion materials that have yet to be stabilized in epitaxial form.  Unlike the B20 family, these materials are not enantiomorphic.  The challenge to grow epitaxial inverse Heuslers, however, will be to avoid twinning from a 90$^{\circ}$ relative rotation of (001) grains about the  [001] direction.  In polar crystals with $C_{nv}$ point group symmetry, such as GaV$_4$S$_8$, the broken inversion symmetry that produces the DMI derives from the stacking of the atomic layers along the [111] direction, analogous to Pt/Co/Ir multilayers.  This non-enantiomorphic class of crystals may prove to be the most resilient to the formation of chiral defects and may have distinct advantages for application.
   
\subsection{Concluding remarks}
%Include brief concluding remarks. This should not be longer than a short paragraph. (150 words max)

The impact of misfit dislocations and chiral grain boundaries on spin textures remain important questions to resolve for the B20 epitaxial films.  An understanding of these issues is important in order to obtain a better understanding of how defects influence skyrmion nucleation and stability.  Important for the development of future skyrmion devices will be the discovery of single crystal epitaxial films with a high-Curie temperature, $T_C$.  It will be necessary to find new epitaxial systems outside of the B20 family with higher $T_C$ and where the chirality of the film structure can be more easily controlled.

\subsection{Acknowledgements}
I would like to thank M. L. Plumer, D. Kalliecharan and J. McCoombs for their critical reading of the manuscript.

%%%%%%%%%%%%%%%%%%%%%%%%%%%%%%%%%%%%%%%%%%%%%%%%%
%%%%%%%%%%%%%%%%%%%%%%%%%%%%%%%%%%%%%%%%%%%%%%%%%
%%%%%%%%%%%%%%%%%%%%%%%%%%%%%%%%%%%%%%%%%%%%%%%%%

%%%%%%%%%%%%%%%%%%%%%%%%%%%%%%%%%%%%%%%%%%%%%%%%%
%%%%%%%%%%%%%%%%%%%%%%%%%%%%%%%%%%%%%%%%%%%%%%%%%
%%%%%%%%%%%%%%%%%%%%%%%%%%%%%%%%%%%%%%%%%%%%%%%%%

\clearpage
\newpage

\section[Skyrmions in multilayers and tailored heterostructures \\ {\normalfont Albert Fert, Vincent Cros, and Nicolas Reyren}]{Skyrmions in multilayers and tailored heterostructures}
\label{fert}
{\it Albert Fert, Vincent Cros, and Nicolas Reyren}

Unit\'{e} Mixte de Physique, CNRS, Thales, Univ. Paris-Sud, Universit\'{e} Paris-Saclay, 91767, Palaiseau, France

\subsection{Status}
Studies of the magnetic properties in spin glasses revealed that chiral magnetic interactions between neighboring spins can play a key role in metallic based systems. Extending the 3-sites RKKY model to bilayer systems, it has been predicted  \cite{refFERT1990} that a large Dzyaloshinskii Moriya interaction (DMI) should be present at the interface between a magnetic film and a heavy material with large spin-orbit coupling. The pioneer works on surface magnetism in various model thin films realized in the 2000s by SP-STM confirmed this prediction with the observation of spin spiral states induced by a large DMI and ultimately of a skyrmion lattice phase in one monolayer of Fe on Ir (111) \cite{refHEINZE}. The investigation of magnetic skyrmions in such ultrathin epitaxial films represents a fundamental interest but are hardly relevant for the observation and then the use of skyrmions at room temperature (RT). 

\begin{figure}
  \includegraphics[width=1.0\columnwidth]{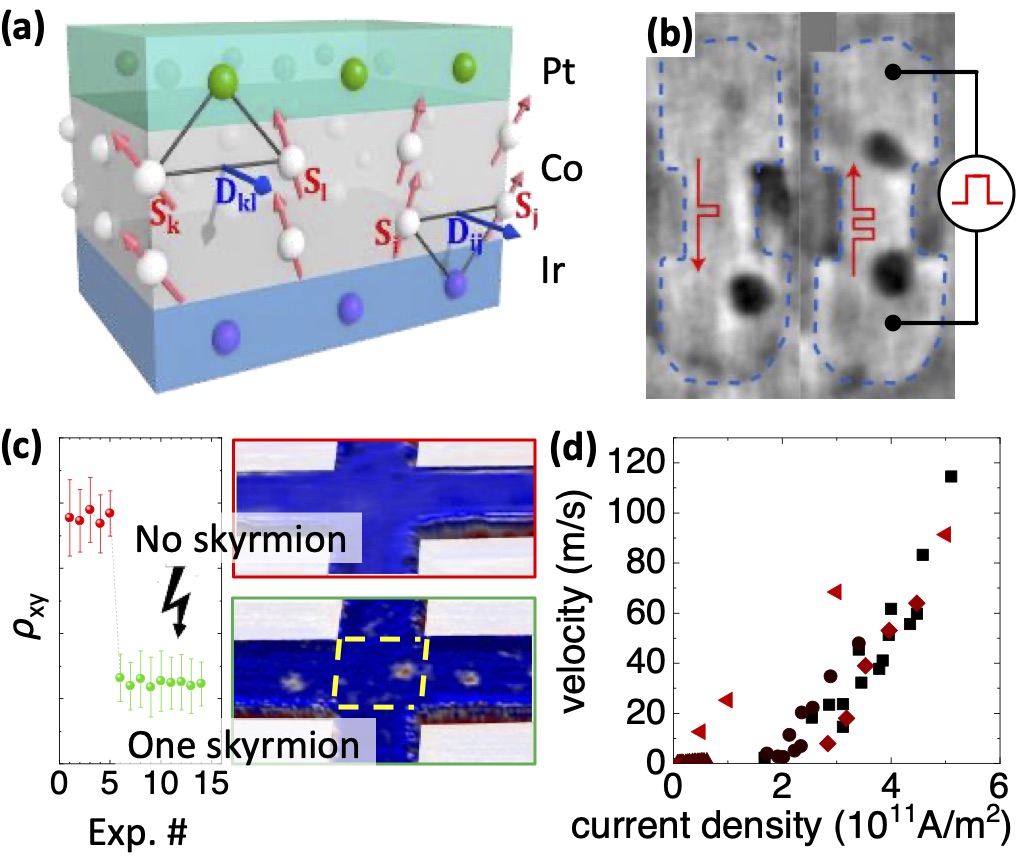}
    \caption{\label{Fig. 1}(a) Unit cell of a (Ir/Co/Pt)$\times 20$ multilayer with additive DMI at the interfaces of a 0.6-nm-thick Co layers with Ir and Pt. The thin Ir and Pt layers couple the skyrmions in the different layers in a single column of skyrmions \cite{ref31NMR}.  (b) Creation of skyrmions at the edge of a constriction in a (Pt/CoFeB/MgO)$\times$15 multilayer by successive current pulses in opposite directions (track width is 400 nm) \cite{refBUTTNERetal}. (c) Detection of the presence of a single skyrmion by the change in AHE (25 n$\Omega$\,cm in this specific example using the multilayer of panel (a)) \cite{refDM}. (d) Skyrmion velocity as a function of the current density in different types of multilayers (from recent literature). }
\end{figure}

For this purpose, the best solution up to now seems to increase the effective magnetic volume by using multilayer stacks composed of multiple repetitions of thin magnetic metal layers separated by heavy metal nonmagnetic layers grown by sputtering deposition  (figure~\ref{Fig. 1}(a)). This approach enables the increase of the thermal stability of columnar skyrmions, that are coupled in the successive layers, leading to the recent observation of sub-100 nm skyrmions stable at RT in different multilayer systems \cite{ref31NMR}\cite{refBoulle2016}\cite{ref32NMR}. Another advantage of the multilayer approach is that it offers multiple opportunities to finely tailor the skyrmion characteristics, e.g. by changing the material combination (magnetic metals, heavy metals, ferroelectrics), by tuning the different interactions or by varying the number of repetitions in the multilayers. Initially randomly created using sweeping of a perpendicular field, often at a defect position, skyrmions in multilayers can be created by spin transfer torques and thermal effects induced by current pulses near constrictions, as in figure~\ref{Fig. 1}(b) \cite{refBUTTNERetal} or local current injector \cite{hrabec}\cite{Finizio2019}. Recently, Hall resistivity measurements (mostly Anomalous Hall Effect, AHE) was used to electrically detect single sub-100 nm skyrmions \cite{refDM}\cite{Zeissler2018} (figure~\ref{Fig. 1}(c)). Finally, these skyrmions are moved using spin-orbit torque associated to spin Hall effect \cite{klaui} (figure~\ref{Fig. 1}(d)), with velocities that are for the moment around 100 m/s for current densities of a few 10\textsuperscript{11} A/m\textsuperscript{2}, i.e. in the range of current densities used today in spintronic devices. 

\subsection{Current and future challenges}
Up to now, a large dispersion of skyrmion sizes has been found from less than 100 nm \cite{ref31NMR}\cite{ref32NMR} to a few  $\mu$m \cite{2015_Jiang_Science} even if the multilayer structures only slightly differ. Actually, the skyrmion size depends on the balance between the magnetic interactions: DMI, exchange, anisotropy, interlayer dipole fields (related to magnetization with additional effects of the number of repetitions in case of multilayers \cite{reflegrand}). A future challenge is to refine the material and interfacial properties to reach the best compromise for achieving the desired skyrmion properties in terms of size, 3D shape, stability and mobility at low current density. Further efforts on material science are also necessary to reduce the influence of grains and/or local material inhomogeneities on the actual skyrmion shape \ref{Zeissler2017}. This perturbation is found to be particularly important when grains and skyrmions are of similar sizes, which is often the case in today?s multilayer \cite{Legrand2017}. Concerning the nucleation of skyrmions, the next challenge is a better control leading to a deterministic nucleation both in space and time. In pursuing this aim, the exploration of other mechanisms such as vertical injection of pure spin current \cite{ref42NMR} or vertical electric field \cite{ref67NMR}\cite{refTitiksha} are promising techniques. Also it is predicted that the topology of the magnetization distribution should give rise to an additional effect, the Topological Hall Effect (THE). However, in metallic heterostructures, a clear physical picture did not emerge yet and it is difficult to conclude whether this THE effect exists or not \cite{refDM}\cite{Zeissler2018} as it is extremely difficult to single out from the predominant more standard AHE \cite{MaccarielloNatNano2018}. Moreover, the scaling of THE for skyrmions down to 10 nm is unknown. Finally, the improvement of the spin torque induced motion of skyrmions probably remains the largest challenge as the interaction of skyrmions with the defects and inhomogeneities in today structures largely impedes linear motion and large disparities between skyrmions velocities are observed. Another challenge, related to skyrmion topology is the so-called Skyrmion Hall Effect (SkHE) leading to a generally undesired oblique motion particularly in 1D devices such as racetracks.

\subsection{Advances in science and technology to meet challenges}
\begin{figure}
 \includegraphics[width=1.0\columnwidth]{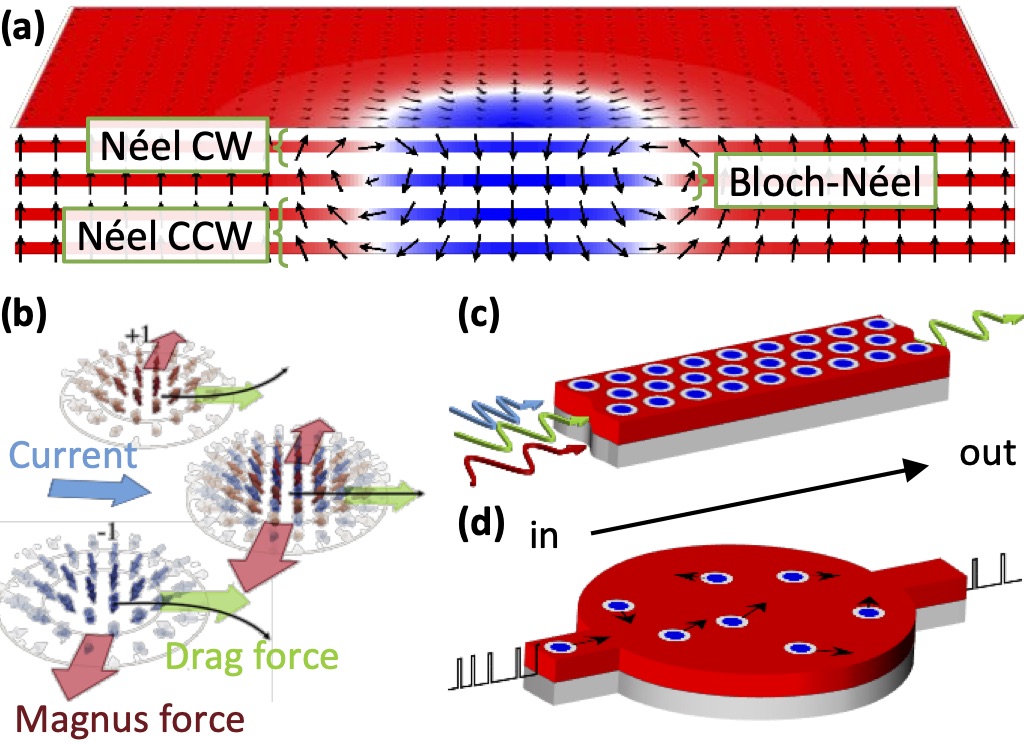}
   \caption{\label{Fig. 2}(a) Multilayer exhibiting a typical 3D spin texture with opposite chiralities in the top and bottom layers. The 3D texture comes from the competition between DMI and dipole interactions \cite{reflegrand}. (b) On the left, SkHE of the current-induced motion of skyrmions of opposite polarities (opposite transverse components of the motion for opposite polarities). On the right, purely longitudinal motion for coupled skyrmions with opposite polarities \cite{Barker2016}. (c)  Illustration of a skyrmion magnetic crystal filtering magnons of specific frequencies. (d) Illustration of a skyrmion reshuffler, a basic bloc for some neuro-inspired devices in which the signal is a sequence of pulses (transposed in skyrmions)}
\end{figure}

Reducing the skyrmion diameter down to 10 nm or less requires further improvements of the engineering of the multilayer properties, and particularly of the control of the interlayer dipolar field effects that can even lead to layer-dependent size and chirality of the skyrmions in terms of CW, CCW, N{\'e}el, or Bloch-like rotation of the magnetization through the skyrmion diameter  \cite{reflegrand} (figure~\ref{Fig. 2}(a)). Key advances for size reduction and improved mobility are anticipated from analytical modelling considering all the magnetic interactions in heterostructures, including interlayer RKKY interaction allowing the development, beyond ferromagnetic heterostructures, of synthetic ferrimagnetic and synthetic antiferromagnetic (SAF) multilayers. The observation of such ultra-small skyrmions and their 3D shaping will require sophisticated X-ray based techniques (i.e. magnetic holography, tomography, etc.), advanced electron microscopy and NV center microscopy to characterize the 3D magnetization textures, possibly with compensated magnetization. On the material side, targeting systems with reduced magnetization to decrease dipolar fields, ferrimagnetic transition metal-rare earth alloys have recently been used allowing the RT stabilization of skyrmion in the 20 nm range  \cite{refCaretta}. Pushing further this strategy, heterostructures including a Ru layer to induce a RKKY antiferromagnetic coupling between the magnetic layers allow cancelling the detrimental action (increase of diameter, hybrid texture, ...) of the interlayer dipolar fields \cite{refCaretta}\cite{2019_Legrand_NatMat}. In such systems in which antiferromagnetic skyrmions have been observed \cite{refdenisov} the interfacial DMI becomes again the key parameter for the control of size and stability. Interestingly, it has been recently discovered that large DMI are not only found at interfaces with heavy metals, but also with graphene or oxides, which further opens the field of possible heterostructures. Another great interest of SAF multilayers is that the SkHE should be opposed in layers of opposite magnetization, thus resulting in a purely longitudinal motion, probably desirable in future devices (figure~\ref{Fig. 2}(b)). The improvement of the materials will involve the use of multilayers having much less defects, such as amorphous layers or crystalline systems, as well as materials with larger efficiency than heavy metals for the production of spin current, e.g. Rashba interfaces and/or topological insulators. Reproducible creation of skyrmions in a precise location by current near local sources of inhomogeneity requires the development of technologies (FIB, ion implantation, etc.) allowing the introduction of local defects or nanoconstrictions. This approach could be seen as complementary to the existing demonstration using nano-patterned electrodes \cite{hrabec}\cite{Finizio2019}. The detection of skyrmions should be extended to smaller sizes by improvement of AHE techniques or detection by vertical structures (TMR or GMR). Concerning the THE, a better understanding should be obtained in models going beyond the adiabatic limit and also taking into account the finite mean free path of the electrons.

\subsection{Concluding remarks}
The strategic choice made a few years ago of exploiting the interfacial DMI in multilayers to stabilize isolated skyrmions at room temperature has paid off. Moreover, in only a couple of years, a concerted effort from several communities allowed to demonstrate the basic functions required for any types of skyrmion devices: nucleation, manipulation, detection and deletion. Furthermore, a specificity of skyrmions in multilayers is the possible modulation of their properties, size or chirality, using the heterostructure composition. Beyond such 3D skyrmions, using novel 3D topological spin textures such as bobbers or hopfions could provide some promises, unexplored yet, towards 3D devices \cite{Rybakov:2016njp}. Also interesting, the skyrmions can be used as interacting particles in 2D systems. Leveraging on this unique behavior, several conceptual skyrmion devices have been proposed, such as reconfigurable magnonic crystals (figure~\ref{Fig. 2}(c)) and probabilistic devices (figure~\ref{Fig. 2}(d)). Finally, all these recent advances change the status of skyrmions from being an intriguing topological spin texture to one of the most promising nanoscale micromagnetic particles ready to shake up many technologies.

\subsection{Acknowledgements}

Financial support from the Agence Nationale de la Recherche, France, under grant agreement No. ANR-17-CE24-0025 (TOPSKY), the Horizon2020 Framework Programme of the European Commission, under FET-Proactive Grant agreement No. 824123 (SKYTOP) and FET-Open grant agreement No. 665095 (MAGicSky), and the DARPA TEE program, through grant MIPR \# HR0011831554 is acknowledged.

%\balance

%%%%%%%%%%%%%%%%%%%%%%%%%%%%%%%%%%%%%%%%%%%%%%%%%
%%%%%%%%%%%%%%%%%%%%%%%%%%%%%%%%%%%%%%%%%%%%%%%%%
%%%%%%%%%%%%%%%%%%%%%%%%%%%%%%%%%%%%%%%%%%%%%%%%%

\clearpage
\newpage

\section[Skyrmions in atomically thin layers and at interfaces \\ {\normalfont Kirsten von Bergmann}]{Skyrmions in atomically thin layers and at interfaces}
\label{vonbergmann}
{\it Kirsten von Bergmann}

Department of Physics, University of Hamburg, 20355 Hamburg, Germany.

\subsection{Status}
% This section provides a brief history and status, why the field is still important, what will be gained with further advances. (350 words max)

Magnetic skyrmions are stabilized by the Dzyaloshinskii-Moriya interaction (DMI), which imposes a unique rotational sense on the spin texture. The requirement for the occurrence of the DMI of broken inversion symmetry is fundamental to all surfaces and interfaces. The first examples of interface-induced magnetic skyrmions were in model-type systems of highly ordered epitaxially grown atomic layers on single crystal surfaces~\cite{refHEINZE,2013_Romming_Science}. Skyrmions in such systems range from less than $3$\,nm~\cite{refHEINZE,2013_Romming_Science,Romming_PhysRevLett.114.177203} over about $50$\,nm~\cite{HerveNC2018} to approx.\ $200$\,nm~\cite{Chen2015} in diameter. Real-space spin-polarized scanning tunneling microscopy (STM) experiments together with \emph{ab-initio} calculations have provided a very good understanding of the relevant energies and ground state formation mechanisms~\cite{refHEINZE,2013_Romming_Science,HerveNC2018,Dupe2014}. Whereas ultrathin films have mostly been studied at low temperature and skyrmions are induced by magnetic fields on the order of 1\,T, some systems can be described by micromagnetic theory~\cite{Romming_PhysRevLett.114.177203,HerveNC2018}, like most bulk-DMI and magnetic multilayer samples. In other systems exchange interactions between more distant atoms have been found to play a role~\cite{refHEINZE,Dupe2014,Malottki_SR2017}, which can act together with the DMI to favor non-collinear states over the ferromagnetic state (see also Sec.\,\ref{mostovoy}).

\begin{figure*}[h]
\includegraphics[width=1.0\textwidth]{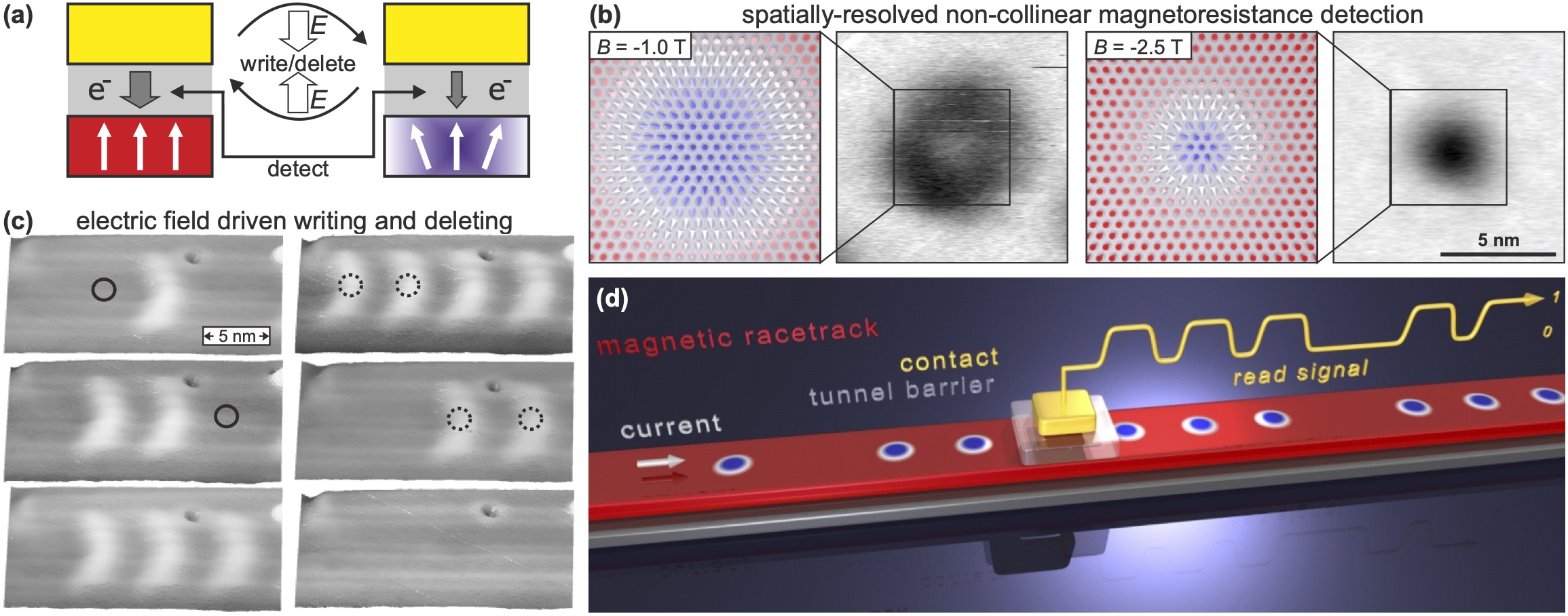}
\caption{
\textbf{(a)}~Schematics of electric field ($E$) induced switching and tunnel current (e$^-$) detection of the magnetic state (red: ferromagnetic, blue: skyrmion interior) using a planar tunnel junction with a non-magnetic (yellow) electrode; the $E$ writes or deletes a skyrmion in the magnetic film depending on the direction~\cite{ref67NMR}; due to the non-collinear magnetoresistance, which does not require a magnetic electrode or spin-orbit interaciton, the tunnel current is different for a collinear and a non-collinear magnetic configuration~\cite{Hanneken_NatNano2015}.
\textbf{(b)}~Sketches of magnetic skyrmions in a PdFe bilayer on Ir(111)~\cite{2013_Romming_Science,Hanneken_NatNano2015} at two different magnetic field values ($B$) as indicated (colored according to the out-of-plane magnetization component, each cone represents an atom and points in the direction of its magnetic moment); corresponding differential conductance maps (gray scale) obtained with a non-magnetic tip, where the signal is a measure for the local non-collinearity of the skyrmion spin texture; note that the position of largest non-collinearity (darker) shifts from the in-plane region for larger skyrmions (smaller $|B|$) to the center for smaller skyrmions (larger $|B|$).
\textbf{(c)}~Consecutive spin-polarized scanning tunneling microscopy measurements showing the same sample area of three atomic layers of Fe on Ir(111) with a different number of skyrmions (in this strained film the skyrmions are not axially symmetric but distorted, in agreement with the structural arrangement of the atoms); circles indicate positions of locally applied electric field between tip and sample, solid circles for negative $E$ (left) and dashed circles for positive $E$ (right).
\textbf{(d)}~Sketch of a magnetic racetrack memory device with a magnetic stripe carrying the information in the form of presence or absence of magnetic skyrmions (blue dots), uniting the non-collinear magnetoresistance detection and the electric field switching in one read/write-head made out of a non-magnetic contact (yellow) on a planar tunnel junction. Panel (b) adapted from~\cite{Hanneken_NatNano2015}, (c) adapted from~\cite{ref67NMR}, (d) adapted from K.\ von Bergmann and A.\ Kubetzka, Phys.\ Unserer Zeit 3, 118 (2017).
\label{KvB}
}
\end{figure*}

The advantage of atomic layer model systems is that the fundamental properties of skyrmions, e.g.\ magnetic-field dependent size and shape, can be extracted and directly compared to results from theory~\cite{Romming_PhysRevLett.114.177203}. In addition, exploiting the high spatial resolution of STM together with an energy-resolved density of states measurement, it is possible to disentangle different magnetoresistive effects: it has been shown that in addition to a contribution to the Hall effect in lateral transport, a localized non-collinear magnetic texture shows a different signature also in vertical tunnel magnetoresistance measurements, i.e.\ the non-collinear magnetoresistance effect~\cite{HerveNC2018,Hanneken_NatNano2015}, see Fig.\,\ref{KvB}(a),(b).

An important aspect regarding applications is a controlled switching between the topologically distinct states of a skyrmion and a local ferromagnet, i.e.\ to write and delete skyrmions, and different dominating mechanisms have been found experimentally in atomic layer model systems: a tunnel electron induced mechanism has been realized for skyrmions pinned to atomic defects~\cite{2013_Romming_Science}, whereas for a different sample system a highly deterministic electric field driven switching has been demonstrated~\cite{ref67NMR}, cf.\ Fig.\,\ref{KvB}(a),(c).

\subsection{Current and future challenges}
% This section discusses the big research issues and challenges. (350 words max)

Various proposals have been put forward how to exploit skyrmions in different types of applications ranging from data storage in the racetrack, via logic circuits, to concepts utilizing skyrmion assemblies in two-dimensions. Challenges on the fundamental level include the profound understanding of different mechanisms to manipulate and control individual magnetic skyrmions in terms of switching (i.e.\ writing and deleting) and lateral motion. Furthermore the details of skyrmion interactions with neighboring skyrmions, with boundaries of a magnetic structure, or with defects needs to be understood to facilitate different functionalities by either avoiding effects like pinning, or utilizing them for designated features of a device. Along these lines also a reproducible experimental demonstration how to merge or duplicate skyrmions in a controlled fashion is lacking.

Model-type systems such as atomically thin magnetic layers provide a platform to gain a fundamental understanding of the physics of magnetic skyrmions. However, it is not \emph{a priori} clear that it is directly transferrable to more applied systems, where the interfaces may be subject to roughness and intermixing. In addition magnetic properties also depend on the structural order within the magnetic film, i.e.\ the degree of strain or disorder. Again, such effects may be either detrimental to a proper functioning of a device or can be exploited to fine-tune magnetic properties when there is sufficient knowledge how the spin texture is affected.

Finally, antiferromagnetic skyrmions have been identified as promising building blocks for spintronic applications due to the absence of deflection under lateral currents, a vanishing magnetic strayfield, and the expected fast dynamics. In principle antiferromagnets constitute a large material class that might host antiferromagnetic skyrmions. Whereas a reduced transverse motion has been observed recently in a ferrimagnet~\cite{woo2018current}, up to now no experimental observation of skyrmions in antiferromagnets has been reported.

\subsection{Advances in science and technology to meet challenges}
% This section discusses the advances in science technology needed to address the challenges. (350 words max)

The experimental toolbox to characterize nanometer-scale skyrmions or antiferromagnetic skyrmions in atomically thin layers is still limited: techniques that achieve spin-sensitive measurements with a resolution down to the atomic scale are spin-polarized STM and exchange force microscopy. For larger skyrmions also other techniques with a resolution on the order of 10\,nm are applicable, like spin-polarized low energy electron microscopy~\cite{Chen2015}, magnetic force microscopy~\cite{refDM}, or scanning transmission X-ray microscopy~\cite{klaui}; the first one usually requires ultra-high vacuum conditions and single crystals, and the last one the growth of the magnetic film on a membrane.

A fundamental understanding of skyrmioninc systems can be achieved by the synergy of experiments and calculations for the same system. \emph{Ab initio} calculations can tackle model systems of atomically thin magnetic layers on single crystals and derive the relevant magnetic interactions and energies~\cite{refHEINZE,HerveNC2018,Dupe2014}. This can be used as input for simulations to directly compare results for, e.g., skyrmion emergence, switching, stability, and current-induced motion to experimental findings. In this way the parameter space for different physical mechanisms can be identified. In an attempt to make predictions for more applied sample systems theory is trying to mimic disorder and intermixing to understand the impact on the magnetic properties.

There have been advances in the electrical detection of single magnetic skyrmions without the need for high spatial resolution via Hall measurements, however the origin of the signal is still under debate~\cite{refDM,Zeissler2018}. The non-collinear magnetoresistance, as found experimentally with STM for a model system~\cite{Hanneken_NatNano2015}, see~Fig.\,\ref{KvB}(a),(b), has not been singled out in spatially averaging magnetoresistance measurements. On the contrary, the concept of electric field induced switching of magnetic skyrmions demonstrated for a model system using the tip of an STM~\cite{ref67NMR}, see Fig.\,\ref{KvB}(a),(c), has been transferred to a planar insulator in contact with a magnetic film and an influence of the electric field on the magnetic state has been shown~\cite{SchottNL2017}.

\subsection{Concluding remarks}
% Include brief concluding remarks. This should not be longer than a short paragraph. (150 words max)

Regarding possible future skyrmionic devices the study of atomically thin magnetic films is important for an understanding of the fundamental properties of skyrmions on a microscopic scale. A fine-tuning of the magnetic properties of a material can be guided by such investigations and a realization of lateral movement of skyrmions in an STM setup could resolve many open questions regarding interactions of skyrmions with each other and structural defects. The study of model-type systems facilitates the discovery and detailed understanding of novel physical mechanisms which are expected to advance the field. For instance the combination of all electrical detection of skyrmions with vertical tunnel currents and an electric field driven writing and deleting of skyrmions could be expoited in future skyrmion-based devices using a single planar tunnel junction as read/write head, see Fig.\,\ref{KvB}(a),(d).

\subsection{Acknowledgements}
% Please include any acknowledgements and funding information as appropriate.

Support by Andr\'{e} Kubetzka is gratefully acknowledged. This work has received funding from the Deutsche Forschungsgemeinschaft (DFG, German Research Foundation) - 402843438; - SFB668-A8; and from the European Unions Horizon 2020 research and innovation program under Grant Agreement No. 665095 (FET-Open project MAGicSky).

%\subsection{References}
%Limit of ten references. Please provide the full author list, and article title, for each reference to maintain style consistency in the combined roadmap article. Style should be consistent with all other contributions.

%%%%%%%%%%%%%%%%%%%%%%%%%%%%%%%%%%%%%%%%%%%%%%%%%
%%%%%%%%%%%%%%%%%%%%%%%%%%%%%%%%%%%%%%%%%%%%%%%%%
%%%%%%%%%%%%%%%%%%%%%%%%%%%%%%%%%%%%%%%%%%%%%%%%%

\clearpage
\newpage

\section[Skyrmions in Confined Geometries \\ {\normalfont Jiadong Zang}]{Skyrmions in Confined Geometries}
\label{zang}
{\it Jiadong Zang}

Department of Physics and Astronomy, University of New Hampshire, Durham, New Hampshire 03824, USA

\subsection{Status}
%This section provides a brief history and status, why the field is still important, what will be gained with further advances. (350 words max)

To boost the application of magnetic skyrmions in information technology, it is essential to thoroughly understand both static and dynamical behaviors of skyrmion in various nanostructures with sizes comparable to skyrmions. Nanoribbons and nanowires can be used as a prototype of skyrmion-based racetrack memory \cite{Fert2013}, and nanodisks can be potentially used as skyrmion-based nano-oscillators \cite{Garcia-SancheZ2016}. The device perspectives are extensively discussed in Chapter 9. In these geometries, extra magnetic anisotropy at the surface and skyrmion-edge interactions become important due to large surface-bulk ratio, and long-range dipolar interaction between magnetic moments also plays a significant role. As a result, skyrmion textures therein can exist even in the absence of external magnetic field, in contrast to bulk samples. Other exotic topological spin textures, such as bobbers and hopfions \cite{Rybakov2015, Liu2018}, will also emerge. 

\begin{figure}
\includegraphics[width=1.0\columnwidth]{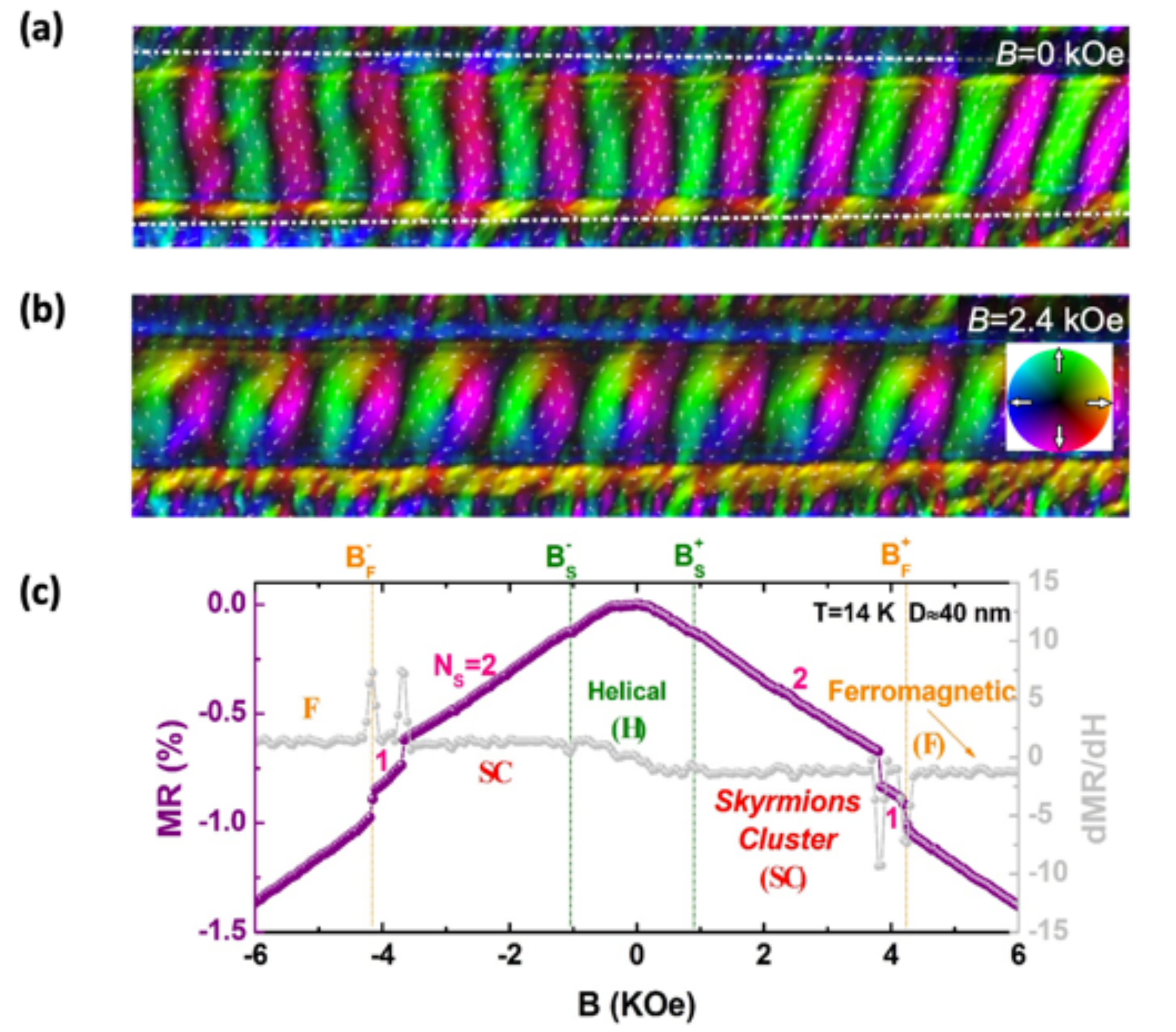} 
\caption{Skyrmions in nanoribbons and nanowires. Real space image of (a) helical state and (b) a single skyrmion chain in the ribbon. Color in both figures represent in-plane spin directions dictated by the color wheel in (b). (d) Magnetoresistance along a MnSi nanowire shows two jumps at the skyrmion phase, corresponding to change of skyrmioin numbers therein. Reproduced from Refs.\,\cite{Du2015, Du2015b}}
\label{figure_confine_fig1} 
\end{figure}

There have been extensive investigations of skyrmions dynamics driven by external stimuli, such as electric current, in bulk samples. However, nontrivial topology of skyrmion induces a transverse motion, dubbed as the skyrmion Hall effect \cite{Zang2013}, and a collision between skyrmions and the edge is thus inevitable in confined geometries. Due to the missing spins across the sample boundary, magnetic moments form a chiral edge mode with opposite rotation sense as the skyrmions therein, so the interaction between skyrmions and the boundary is highly nontrivial. On the other hand, the constricted geometry provides possibility to create individual skyrmions. As an example, sending a current through a notch in a nanoribbon, a single skyrmion can be generated \cite{Iwasaki2013}. That is fundamentally attributed to the chiral edge state. Topological charge around the notch is nonzero even at zero field, so that skyrmions can be easily nucleated. The same scenario applies to ultra-narrow nanoribbons \cite{Du2015}. At zero field, helices are formed with their wave vector pointing along the ribbon. On the upper and bottom edges, helical stripes connect to the edge spin state and form local distribution of the topological charge (figure \ref{figure_confine_fig1}(a)). Therefore, confined geometry provides a platform for investigating nontrivial topological textures and opens many opportunities to manipulate skyrmions.

\subsection{Current and future challenges}
%This section discusses the big research issues and challenges. (350 words max)

Studying skyrmions in confined geometries faces both experimental and theoretical challenges. Most skyrmions of interests have a size of sub-100nm, so that confined geometry reveals its unique feature only when its size is on the same scale. An advanced nanostructure fabrication is thus required, which can be achieved by lithography and etching technique based on high quality skyrmion thin films, or directly use focused ion beam for bulk crystals although an intricate receipt is needed. Imaging skyrmions and resolving fine spin structures in these nanostructures are also extremely difficult. Lorentz transmission electron microscope (TEM) is broadly used for nanometer magnetic imaging. However, at the sample boundary, the sudden drop of electrostatic potential leads to strong Fresnel fringes and give rise to artificial magnetic contrast across hundreds of nanometers. It is thus difficult to differentiate the true spin textures from these artifacts. Other than the spatial resolution, reaching high temporal resolution and tracking the motion of skyrmion in nanoribbons will be a milestone of the field. A skyrmion velocity of 100m/s requires a temporal resolution of nanoseconds in order to monitor the motion of a 100nm-size skyrmion. Recent advances of time-resolved pump-probe X-ray microscopy has been used to study skyrmion in thin films \cite{Litzius17,Finizio2019}. Extra effort will be paid to push its spatial resolutions.

Theoretically modeling the stability, dynamics and interactions of skyrmions in confined geometry require innovations in analytical approaches. On the numerical side, high performance micro- or atomistic magnetic simulation with dipolar interaction and possibly extra surface interaction taken into account. It was recently suggested that the interaction between the skyrmion and sample edge is not always repulsive, but can be attractive due to the background spin configurations\cite{Du2018}. Therefore, integration between the dynamic simulation and the study of phase diagram in confined geometry is important. The experimental discovery of target skyrmion in a FeGe nanodisk of finite thickness indicates the importance of three-dimensional (3D) confinement \cite{Zheng2018}. Further exploration of 3D topological spin textures becomes an important topic in the emerging stream of 3D spintronics, thus detailed simulations need to be done. However, 3D magnetic simulation becomes unexpectedly heavy due to rapid increase of sampling size and the existence of many competing configurations. Highly parallelized magnetic simulation capable of automatic searching of lowest energy state is still lacking.

All-electric detection of skyrmions is another direction worth pursuing. The topological Hall effect discussed in Chapter 6 and Chapter 11 is a widely used method, but it is usually intertwined with other physics and its signal is weak in finite-sized samples. Other signals thus need to be explored. In confined geometries, finite number of skyrmions can lead to quantized magnetoresistance, as recently discovered in MnSi nanowires  and Pt/Co/Ir nanodisks\cite{Du2015b, Zeissler2018}. Other than that, attempts of using tunneling magnetoresistance to characterize skyrmions will be high rewarding from the device application perspective.

%%%%%%%%%%%%%%%%%%%%%
%%%%%%%%%%%%%%%%%

\subsection{Advances in science and technology to meet challenges}
%This section discusses the advances in science technology needed to address the challenges. (350 words max)

Imaging of skyrmions in nanoribbons and nanodisks with significant reduction of artifical boundary fringes have been enabled already by Lorentz TEM. An amorphous Pt layer coated at the edge was employed to successfully reduce the boundary Fresnel fringes in these nanosized samples \cite{Du2015}. In the FeGe nanoribbon with width down to 130nm, the chiral edge spin state was clearly identified. At low fields, helical stripes with ${\bf k}$ vector parallel with the ribbon is the ground state, and a single skyrmion chain is nucleated once the field is elevated (figure \ref{figure_confine_fig1}(b)). Interestingly, even in wider ribbons at low temperatures, only one skyrmion chain is generated, which appears on the ribbon edge first and gradually shifts to the ribbon center when field increases. This suggests an exotic field-dependent interaction between skyrmions and the edge. Recent micromagnetic simulation has revealed a Lennard-Jones nature of this interaction \cite{Du2018}. Further local magnetic imaging is desired and NV magnetometry might be a promising method at this point\cite{Dovzhenko2018}. 

Discrete number of skyrmions in confined geometries leads to quantized signal of magnetoresistance. Recent study of skyrmions in MnSi nanowire has shown cascading resistance changes when the field is applied along the wire \cite{Du2015b}. At the field increasing process, jumps of magnetoresistance were identified when the skyrmion number changes (figure \ref{figure_confine_fig1}(c-d)). This shows promising possibility of using simple electric measurement to monitor skyrmion numbers in confined geometries.

\begin{figure}
\includegraphics[width=1.0\columnwidth]{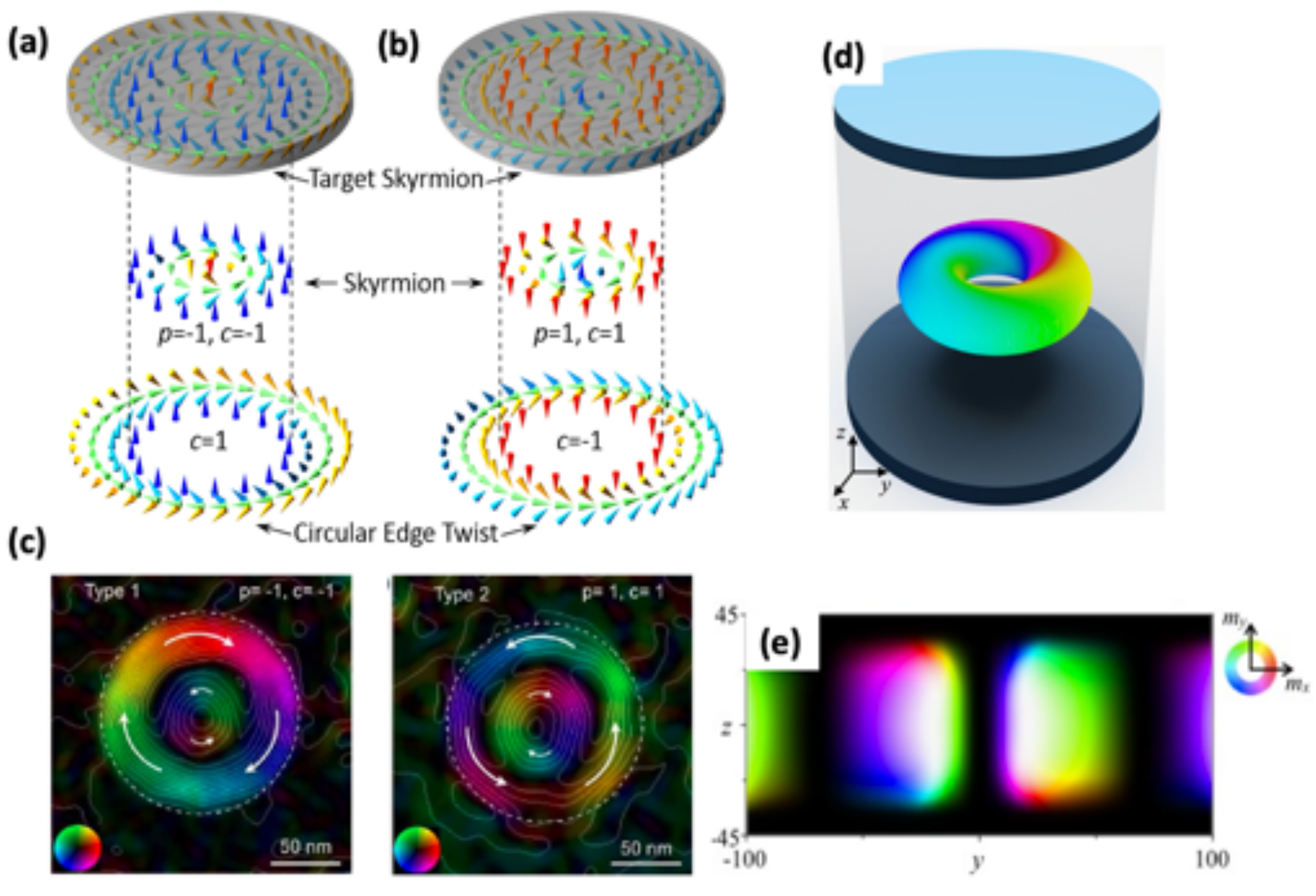} 
\caption{Skyrmions in 3D confined geometries. (a-b) Sketch of two species of target skyrmions stabilized at zero field, and (c) are their corresponding real space image under electron holography. (d) a sketch of a hopfion confined in a nanodisk sandwiched between two spin-polarized layers. (e) Lateral cross-section of a hopfion. It consists of a pair of skyrmion and anti-skyrmion. Reproduced from Refs.\,\cite{Liu2018, Zheng2018}}
\label{figure_confine_fig2} 
\end{figure}

Interests have emerged on studying skyrmions and other relevant topological spin textures in 3D confined geometries. In bulk samples, skyrmions stack quite uniformly along the field direction and form arrays of tubes. However, at the sample boundary, skyrmion tube twists due to imbalanced DM interaction, and the bobber state can appear where the skyrmion tube terminates at a Bloch point \cite{Rybakov2015}. Furthermore, target skyrmion, a state consists of multiple concentric spin helices\cite{Leonov2014}, was observed in nanodisks with thickness comparable to the skyrmion size \cite{Zheng2018} (figure \ref{figure_confine_fig2}(a-c)). Importantly, this state is stabilized by the dipolar interaction and persists in the absence of magnetic field. As a consequence, two species of target skyrmions are degenerate, and can be switched from one to the other by an external magnetic field. Once this nanodisk is sandwiched in between two spin-polarized layers, another spin texture called hopfion has been theoretically predicted \cite{Liu2018}, whose lateral cross-section consists of a pair of skyrmion and anti-skyrmion (figure\ref{figure_confine_fig2}(d-e)). Its dynamical behavior and the influence of geometrical confinement are worth exploring in the future. Material realization of the hopfion and other 3D spin textures is worth extensive study\cite{rybakov2019}, and advanced tomography techniques capable of imaging 3D spin textures are being developed \cite{Donnelly2017}.

%%%%%%%%%%%%%%%%%%%%%
%%%%%%%%%%%%%%%%%

\subsection{Concluding remarks}
%Include brief concluding remarks. This should not be longer than a short paragraph. (150 words max)

As a conclusion, skyrmions exhibit novel properties in confined geometries. Skyrmions therein can exist at extended phase diagrams and experience nontrivial interactions with the sample boundary. Study of skyrmions in confined geometry is moving from low dimensions to three dimensions. Advanced magnetic imaging techniques with high spatial and temporal resolutions are called for. For the purpose of future device applications, it is important to investigate the footprints of skyrmions in nanoribbons using all-electric detections.

\subsection{Acknowledgements}
%Please include any acknowledgements and funding information as appropriate.

JZ was supported by the U.S. Department of Energy (DOE), Office of Science, Basic Energy Sciences (BES) under Award No. DE-SC0016424 and DE-SC0020221.
%%%%%%%%%%%%%%%%%%%%%%%%%%%%%%%%%%%%%%%%%%%%%%%%%
%%%%%%%%%%%%%%%%%%%%%%%%%%%%%%%%%%%%%%%%%%%%%%%%%
%%%%%%%%%%%%%%%%%%%%%%%%%%%%%%%%%%%%%%%%%%%%%%%%%

%\clearpage
%\newpage

%\input{_contributions/Parkin}

%%%%%%%%%%%%%%%%%%%%%%%%%%%%%%%%%%%%%%%%%%%%%%%%%
%%%%%%%%%%%%%%%%%%%%%%%%%%%%%%%%%%%%%%%%%%%%%%%%%
%%%%%%%%%%%%%%%%%%%%%%%%%%%%%%%%%%%%%%%%%%%%%%%%%

\clearpage
\newpage

\section[Skyrmion-based hybrid systems \\ {\normalfont Naoto Nagaosa}]{Skyrmion-based hybrid systems}
{\it Naoto Nagaosa$^{1, 2}$}
\label{nagaosa}
$^1$RIKEN Center for Emergent Matter Science, Wako 351-0198, Japan,
$^2$Department of Applied Physics and Quantum Phase Electronics Center, University of Tokyo, Tokyo 113-8656, Japan

\subsection{Status}

Skyrmion was originally proposed in nuclear physics as a 
model for hadrons, and is ubiquitous for 
various physical systems besides the magnets such as 
quantum Hall system, liquid crystal, and cold atoms. 
Especially, magnetic skyrmions are closely related to the relativistic 
spin-orbit interaction in crystals or at interfaces with inversion symmetry breaking. 
Therefore, they have the common features with other physical 
phenomena driven by spin-orbit interaction with broken inversion symmetry
such as the surface states of topological insulators and noncentrosymmetric 
superconductors, and hence it makes sense to consider the hybrid 
of skyrmions and these other systems \cite{Fert}.
However, at present, the researches on skyrmions, 
topological insulators, and noncentrosymmetric superconductors 
are rather independently developed as described below.

The skyrmion is characterized by the topological index, called skyrmion 
number, which is given by the integral of the emergent magnetic flux
defined by the spin Berry curvature in real space. Therefore, there are 
many intriguing physical phenomena related to the spin Berry phase 
such as the topological Hall effect, and emergent electromagnetic induction
\cite{Nagaosa}. 

On the other hand, topological insulators (TIs) are characterized by the 
Berry curvature of the Bloch wavefunctions in momentum space. 
It is symmetry protected topological state (SPT)
by the time-reversal symmetry T.
%as well as the U(1) symmetry corresponding to the conservation of charge.
The hallmark of TI appears at the surface where the inversion 
symmetry is broken, i.e., the 
robust gapless surface states with spin-momentum locking.
This surface states are proven to be useful for various spintronics 
applications, e.g., highly efficient spin-charge conversion \cite{Fert}.   
As for the noncentrosymmetric superconductors, the mixing of 
spin singlet and spin triplet pairings is inevitable, i.e., the spin 
degrees of freedom are (partially) active, which results in
novel phenomena such as the huge upper critical limit
beyond the Pauli limit. A remarkable noncentrosymmetric 
topological superconductor appears in the surface state of TI,
where the Majorana bound state is predicted and partially 
observed experimentally at the core of the vortex.

\subsection{Current and future challenges}

The current research front is the interplay among the 
spin-orbit driven phenomena described above, but 
this issue has been mostly explored theoretically.

\begin{figure}[b]
\begin{center}
\includegraphics [width=1.0\columnwidth]{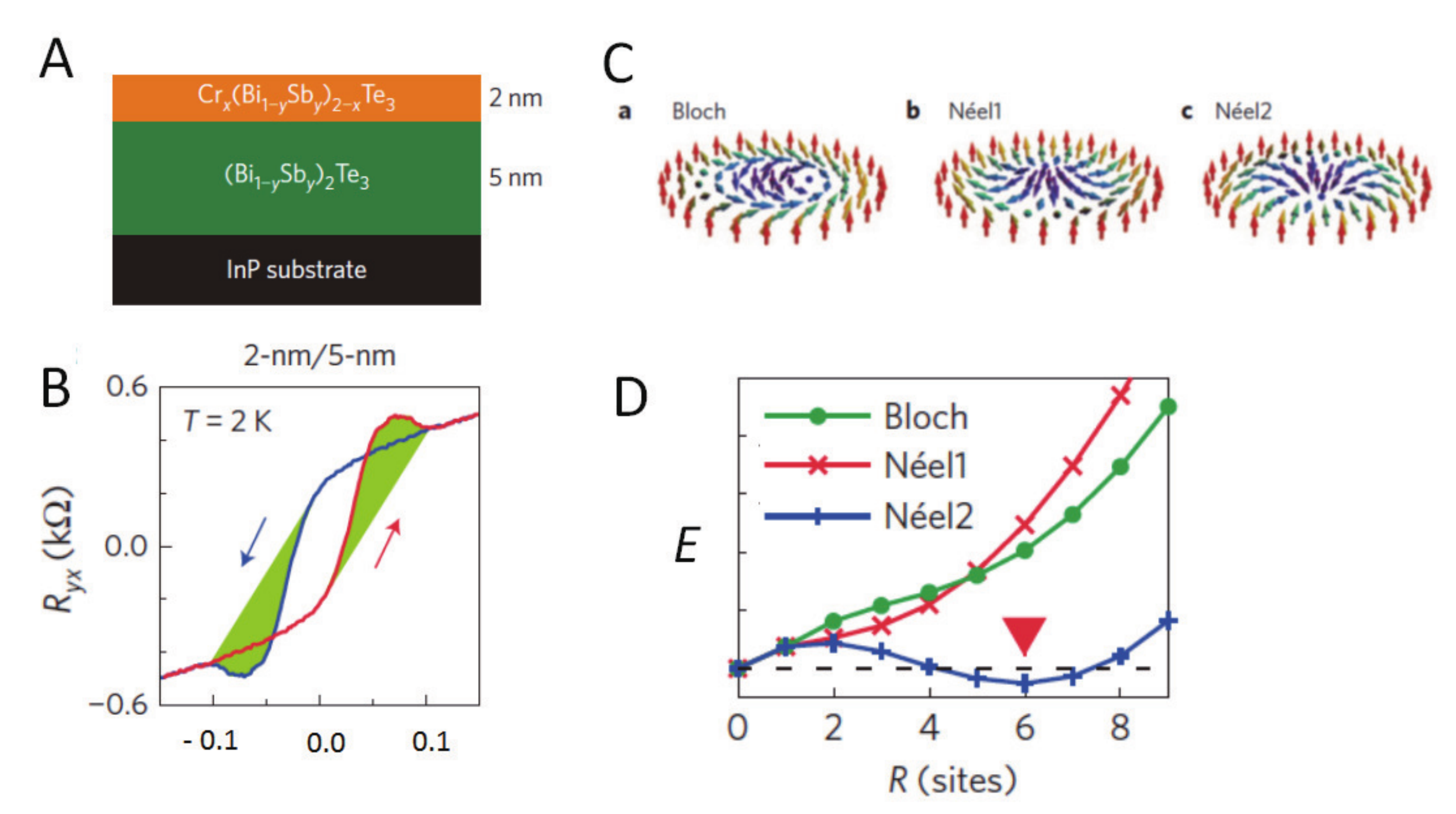}
\end{center}
\caption{(A) The schematic structure of InP/TI/magnetic TI.
(B) The experimental observation of  the Hall resistance as a function of the magnetic field. 
The green shaded part corresponds to the topological Hall effect due to the skyrmion formation.
(C) Three types of skyrmion.
(D) Theoretical calculation results of the energy for each type of skyrmion. 
Reproduced from Ref.\,\cite{Yasuda} with modification.}
\label{fig_image}
\end{figure}

The skyrnions on TI surface is an interesting topics
since the coupling to the surface state produces the 
Dzyaloshinskii-Moriya interaction (DMI), which 
can cause the skyrmion formation. 
A numerical calculation has been done for a model
system aiming at Bi$_2$Se$_3$ doped with Cr, and 
it is concluded that skyrmions of Neel type is stabilized
in the hole doped case. Experimentally, this prediction has been
tested by the measurement of topological Hall effect in the 
heterostructure of 
Cr$_x$(Bi$_{1-y}$Sb$_y$)$_{2-x}$Te$_3$/(Bi$_{1-y}$Sb$_y$)$_2$Te$_3$
\cite{Yasuda}. ( See Fig. \ref{fig_image}.)
Furthermore, the thickness dependence of the topological Hall effect in 
Mn-doped Bi$_2$Te$_3$ topological insulator (TI) films 
grown by molecular beam epitaxy is consistently interpreted as the 
skyrmion formation compared with the theoretical calculation \cite{Yayu}.
A related theoretical work has been done on the anomalous topological Hall effect
due to skyrmions on topological surface state.  
The next target in this system is the skyrmion motion driven by the 
current. Because of the spin-momentum locking in the surface state, 
the spin transfer torque and spin-orbit torque are expected to be
very different from the usual metals. The detailed studies changing 
the current density and chemical potential position by gating 
are highly desired to identify the mechanisms.

\begin{figure}[b]
\begin{center}
\includegraphics [width=1.0\columnwidth]{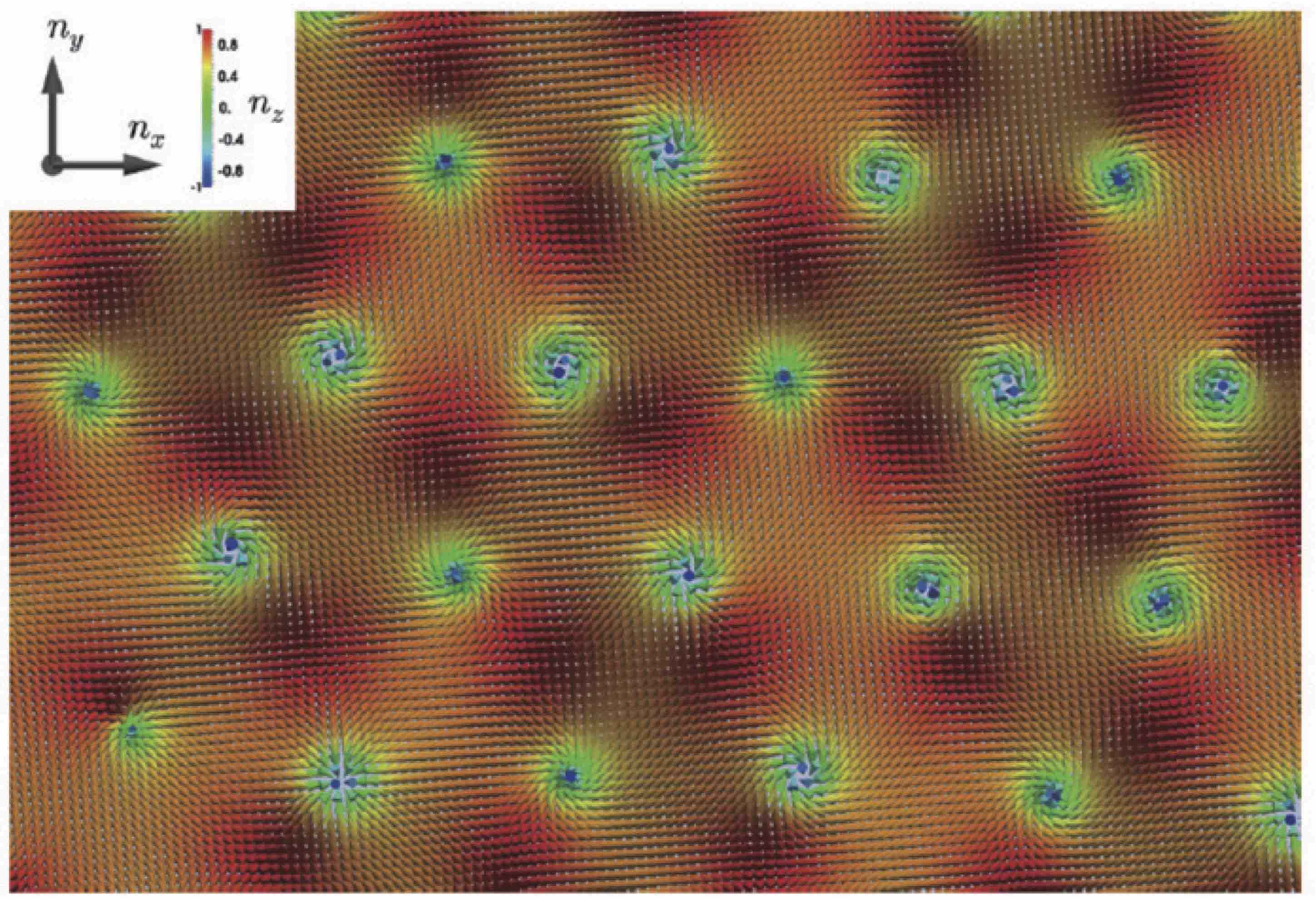}
\end{center}

\caption{The skyrmion lattice structure of the pseudospin predicted for the clean interface superconductor. 
Reproduced from Ref.\,\cite{LAO}.}
\label{fig_image2}
\end{figure}

The interplay between the skyrmion and superconductivity is another 
interesting issue. As for the Majorana fermions, it has been theoretically 
predicted that the $p+ip$ superconductivity and associated 
chiral Majorana edge channel are realized as the in-gap superconductivity 
induced by the Shiba states \cite{Nakosai}. Also the Majorana bound state 
is predicted theoretically at skyrmion \cite{Loss}. 
It is also proposed that the magnetic skyrmions can control the 
Josephson effect where the critical current is changed by several 
orders of magnitude simply by reversing the helicity of a magnetic 
skyrmion \cite{Yokoyama}. 
It is expected that in iron-based superconductors (FeSCs)
when inversion symmetry is broken, e.g., monolayer iron selenide (FeSe) 
on a strontium titanate (SrTiO3(001)) substrate, the noncollinar spin structure 
essentially equivalent to the skyrmion crystal is realized \cite{FeSe}.
Also similar skyrmion lattice texture is predicted for the 
superconductor at LAO/STO interface as shown in Fig. \ref{fig_image2} \cite{LAO}.
The remaining challenges are the experimental test of these theoretical predictions.
Furthermore, the Majorana bound states associated with the magnetic textures 
will have applications to the quantum information technology as the topologically protected 
qubits.

\subsection{Advances in science and technology to meet challenges}

Combining skyrmions with topological insulators and 
superconductors is a big challenge especially for experimentalists.
The most important step is to prepare the sample in a well
controlled way. First the homogeneity of proximity effect
of the magnets, superconductors, and topological surface is 
the crucial issue. For example, the spatially varying exchange gap observed 
in magnetic topological insulator surface is regarded as the
main reason to suppress the temperature where the quantized 
anomalous Hall effect is observed. Superlattice formation and modulation 
doping can be a method to improve this problem.
Second the direct real-space observation of the
spin and electronic structures is desirable in addition to the transport measurements.
Lorentz microscopy and spin-resolved STM have been successful for this purpose, 
and the simultaneous observation of the spin and electrons by combining these tools
would provide rich information. Thirdly, the fine tuning of the parameter such as 
the size of the skyrmions, chemical potential $\mu$ for electrons, thickness of the films, etc. is an 
important tool to identify the physical mechanisms behind the phenomenon. 
Lastly, the quantum dynamics of the composite system is 
a mostly unexplored problem, which will reveal the excitation spectrum 
of the system in addition to the ground state and low-lying excitations.
Accurate optical spectroscopy including real-time resolution would be 
a powerful tool for this purpose. For example, the rapid change in the 
magnetic structure would produce the gigantic emergent electric field,
which might induce novel phenomena unique to the skyrmionic systems.  
Although most of the hybrid systems discussed above are synthetic layered systems,
it would be an interesting direction to pursue the bulk hybrid systems. In this case,
the coexistence of the different orders is required in the bulk. 
Magnetic noncentrosymmetric superconductors, it they exist, 
can be a candidate from this viewpoint. 
Topological spin textures in three-dimensional magnetic topological insulators are another 
interesting system. 

Theoretically, the unified treatment of the skyrmion structures in real 
and momentum spaces is a challenging issue. The spin-orbit interaction 
produces the spin structure in the crystal momentum space in 
noncentrosymmetric system, i.e., the spin splitting occurs at generic 
k-point except the time-reversal symmetric momenta to produce 
the spin texture. This can be often topological, e.g., skyrmion and monopole.
The interplay between these real-space and momentum-space emergent 
electromagnetic field will be a most interesting theoretical issue including 
their dynamics.

\subsection{Concluding remarks}
%\begin{comment}
Inversion symmetry breaking combined with the relativistic spin-orbit
interaction offers a rich variety of physical phenomena which might 
leads to novel functions of materials. 
Spintronics application of TI surface, singlet-triplet mixing in 
noncentrosymmetric superconductors, skyrmion formation in chiral 
magnets are the representative examples, which 
have been studied rather independently. These apparently different
phenomena have a basic principle in common, i.e., the emergent gauge 
field or emergent electromagnetic field. Therefore, it is most
natural to consider the combination of these by fabricating the
hybrid systems of skyrmions, topological insulator, and superconductors,
and explore their interplay. From the point of view of applications, 
the nonreciprocal responses including the directional propagation
of quasi-particles and nonlinear rectification effect 
unique to the noncentrosymmetric systems with broken
time-reversal symmetry are the promising direction \cite{Tokura}.
Generalization of the concept of skyrmions will find the more progress
in the future. 
%\end{comment}

\subsection{Acknowledgements}

N.N. was supported by Ministry
of Education, Culture, Sports, Science, and Technology
Nos. JP24224009 and JP26103006, the Impulsing Paradigm
Change through Disruptive Technologies Program of Council
for Science, Technology and Innovation (Cabinet Office,
Government of Japan), and Core Research for Evolutionary
Science and Technology (CREST) No. JPMJCR16F1 and No. JPMJCR1874.

%%%%%%%%%%%%%%%%%%%%%%%%%%%%%%%%%%%%%%%%%%%%%%%%%
%%%%%%%%%%%%%%%%%%%%%%%%%%%%%%%%%%%%%%%%%%%%%%%%%
%%%%%%%%%%%%%%%%%%%%%%%%%%%%%%%%%%%%%%%%%%%%%%%%%

\clearpage
\newpage

\subsection{References}
%\bibliographystyle{iopart-num}
%\bibliography{roadmap}

\providecommand{\newblock}{}

\balance
\end{document}